\documentclass[aps,prx,twocolumn,showpacs,superscriptaddress]{revtex4-1}
\usepackage{amsmath,braket,amssymb,graphicx,hyperref,color,bbm,tabularx}
\usepackage[normalem]{ulem} 
\hypersetup{colorlinks, breaklinks, linkcolor={blue}, citecolor={blue}, urlcolor=blue}

\newcolumntype{Y}{>{\centering\arraybackslash}X}
\newcommand{\ignore}[1]{}

\begin{document}

\title{Cumulants and Scaling Functions of Infinite Matrix Product States}

\author{Jason C. Pillay}
\email{j.pillay@uq.edu.au}
\affiliation{School of Mathematics and Physics, The University of Queensland, St Lucia, QLD 4072, Australia}

\author{Ian P. McCulloch}
\email{ianmcc@physics.uq.edu.au}
\affiliation{School of Mathematics and Physics, The University of Queensland, St Lucia, QLD 4072, Australia}

\date{\today}


\begin{abstract}
The order parameter cumulants of infinite matrix product ground states are evaluated across a quantum phase transition. A scheme using the Binder cumulant, finite-entanglement scaling and scaling functions to obtain the critical point and exponents of the correlation length and cumulants is presented. Analogous to the scaling relations that relate the exponents of various thermodynamic quantities, a cumulant exponent relation is derived and used to check the consistency and relationship between the cumulant exponents. This scheme gives a numerically economical way of accurately obtaining the critical exponents. Examples of this scheme are shown for four one-dimensional models -- the transverse field Ising model, the topological Kondo insulator, the $S=1$ Heisenberg chain with single-ion anisotropy and the Bose-Hubbard model. A two-dimensional model is also exemplified in the square lattice transverse field Ising model on an infinite cylinder. These exemplary systems portray a variety of local and string order parameters as well as phase transition classes that can be studied with the scaling functions and infinite matrix product states.
\end{abstract}

\maketitle


\section{Introduction}
One of the cornerstones of quantum mechanics is the postulate that all information of a quantum system is contained in its wavefunction. From a practical perspective, storing and manipulating quantum wavefunctions can be a costly ordeal. This is due to the inherent nature of the Hilbert space of quantum systems, specifically, the size of the Hilbert space and its scaling with the number $N$ of particles of $d$ degrees of freedom is $d^N$, i.e. it scales exponentially with system size. Thus, capturing the information of merely 300 two-state particles would require $2^{300}$ bits, a number so large that it would bankrupt the visible universe of its information-storage capacity. This motivates the necessity to concisely represent a wavefunction while revealing the important physics in question.

In the field of low-dimensional many-body quantum physics, a class of ansatz known as tensor network states has been successful in faithfully representing the states of various gapped and gapless phases described by Hamiltonians with local interactions. This success is attributed to the entanglement structure that these physical systems possess, namely where the amount of entanglement entropy is constrained by the system's physical dimension $D$ \cite{Eisert}. For example, in gapped systems of size $L$, the entanglement entropy $S$ of the ground state is proportional to $L^{D-1}$. This is known as the ``area law of entanglement entropy" and it holds true as long as the system possesses a nonzero energy gap. At a critical point however, the system's energy gap vanishes and the entanglement entropy violates the area law. In 1D, Ref.~\cite{Vidal} has shown that $S$ diverges logarithmically with $L$. This drastic change of behavior in the entanglement entropy and its ease of computation via tensor network states has made it a viable tool in identifying critical points - by continuously tuning a Hamiltonian parameter, the entanglement entropy would gradually increase approaching the critical point and peaks at the critical point.

Unlike a finite system of size $L$, an infinite translationally-invariant system has no notion of boundaries that allows one to specify the size of the system. Thus, the common notion of length in such systems is the correlation length $\xi$ of some quasiparticle excitation. Being inversely proportional to the energy gap, $\xi$ diverges at a critical point. The tensor network state used to represent an infinite translationally-invariant 1D systems is known as an infinite matrix product state (iMPS) given as
\begin{eqnarray}
\ket{\psi} = \sum_{\{s_i\}} \left( \ldots A^{s_n} A^{s_{n+1}} \ldots \right) \ket{\ldots s_n s_{n+1} \ldots}
\label{eqn:mps}
\end{eqnarray}
for a one-site unit cell. $A^{s_i}$ is an $m \times m$ matrix and the superscript $s_i$ represents an element of the $d$-dimensional local Hilbert space at site $i$. The quantity $m$ is called the bond dimension, and it is a controllable parameter that governs the maximum amount of entanglement that the state can possess. Since $m$ is finite, the peak of $S$ at the critical point does not diverge to infinity, but instead saturates to a value that depends on $m$. The parameter value that corresponds to the maximum $S$ is known as a ``pseudocritical point" and has been thoroughly studied in Ref.~\cite{Tagliacozzo} where it was shown that $S \propto \log m^\kappa$. $\kappa$ is the finite-entanglement scaling exponent - a quantity related to the central charge of the critical point \cite{Pollmann}. More importantly, by using finite-entanglement scaling, the authors of Ref.~\cite{Tagliacozzo} presented a scheme to extract $\kappa$ from various universal quantities such as the correlation length, entropy, magnetization order parameter, etc. Analogous to finite-size scaling in finite systems, where critical exponents are extracted by scaling data of universal quantities at the critical point with respect to $L$, the scheme describes extracting the critical exponents by scaling data with respect to $m$. In addition to this, the scaling function of the magnetization order parameter relates the magnetization at different values of $m$ to each other across the critical point. This is again analogous to scaling functions in finite systems where such functions relate universal data of different system sizes to each other. Using the known value of the critical point, the authors used this scaling function to determine $\kappa$ by selecting the value of $\kappa$ that gave the best data collapse of the magnetization of several different values of $m$. Though not done in Ref.~\cite{Tagliacozzo}, the scaling functions could also be used to locate and fine-tune both the critical point and $\kappa$ simultaneously. This is motivated by the fact that the magnetization scales as a power law of $m$ at the critical point and thus its ability to detect the critical point is more pronounced than $S$ which scales logarithmically with $m$ at the critical point. More importantly, generalizing and applying these scaling functions to other universal quantities allows one to determine other critical exponents such as the correlation length exponent $\nu$ and higher order cumulant exponents - all of which are presented in this work. 

While an order parameter $M$ is an invaluable tool in discerning phases and locating critical points, there is more information available in its distribution function. Even though this distribution is typically difficult to obtain, it is still possible to gain information about the distribution from the cumulants $\kappa_n$ and the higher moments $\mu_n = \braket{M^n}$ of the order parameter. In probability theory, the cumulants specify the shape of a given distribution. The first cumulant is the mean value
\begin{eqnarray}
\kappa_1 = \braket{M},
\label{eqn:first_cumulant}
\end{eqnarray}
the second cumulant is the variance 
\begin{eqnarray}
\kappa_2 = \braket{M^2} - \braket{M}^2,
\label{eqn:second_cumulant}
\end{eqnarray}
the third cumulant is the skewness
\begin{eqnarray}
\kappa_3 = \braket{M^3} - 3\braket{M^2}\braket{M} + 2\braket{M}^3,
\label{eqn:third_cumulant}
\end{eqnarray}
and the fourth cumulant is the kurtosis
\begin{eqnarray}
\kappa_4 = \braket{M^4} - 4\braket{M^3}\braket{M} - 3\braket{M^2}^2 \nonumber \\
+ 12\braket{M^2}\braket{M}^2 - 6\braket{M}^4.
\label{eqn:fourth_cumulant}
\end{eqnarray}
Besides probability theory, a modification of the fourth cumulant has found practical usage in the field of phase transitions in what has been known as the Binder cumulant \cite{Binder}
\begin{eqnarray}
U_4 = 1 - \frac{\braket{M^4}}{3\braket{M^2}^2}.
\label{eqn:binder_cumulant}
\end{eqnarray}
By tabulating data the Binder cumulant for different system sizes $L$ across a phase transition, the critical point is read off the point where the Binder cumulant of different system sizes cross each other. The benefit of using the Binder cumulant is that by finding the crossing point between different systems sizes and using two higher order cumulants simultaneously, finite-size effects are drastically reduced. Hence, the critical point obtained from it is much more precise than using solely the order parameter. The occurrence of a crossing point in the Binder cumulant is also observed in infinite translationally-invariant systems represented by an iMPS, where the crossing occurs between $U_4$ of different bond dimension $m$ instead of $L$ \cite{West}.

A true phase transition only occurs when a system is in the thermodynamic limit $L \rightarrow \infty$ where the correlation length $\xi$ diverges. In a finite-sized system, $\xi$ is upper-bounded by the system size $L$. The conventional method of studying a phase transition with a finite MPS is to obtain data in the vicinity of the critical point for several system sizes. Since $\xi$ also depends on $m$, $m$ must be chosen such that the criteria $\xi \approx L$ is fulfilled for each system size. By employing finite-size scaling, the data is extrapolated with respect to $L$ to obtain the quantity of interest in the thermodynamic limit \cite{Nishino}. This dependence of $m$ on $L$ complicates the procedure of obtaining data since $m$ has to be obtained to satisfy the condition $\xi \approx L$ for each system size. A simpler and more direct way to probe phase transitions would thus be to use an iMPS. There are two advantages of this over a finite MPS. First, since there is no notion of a system size in an iMPS, one does not have to worry about determining the value of $m$ that satisfies the criteria $\xi \approx L$. Hence one only has to extrapolate data with respect to $m$ in order to obtain the data in the $m \rightarrow \infty$ limit. Second, the absence of boundaries completely removes any Friedel oscillations that affects quantities that are dependent on spatial properties of the system.

In the spirit of determining the critical point and exponents via finite-entanglement scaling and scaling functions of the order parameter, this work extends the scheme of Ref.~\cite{Tagliacozzo} through the addition of the order parameter cumulants, the Binder cumulant, and the cumulant scaling functions. The advantage this has over the previous schemes is that this scheme requires a much smaller $m$ in order to determine the critical point and exponents with a higher accuracy. To this end, several 1D examples are given, namely the transverse field Ising model, the topological Kondo insulator, the $S = 1$ Heisenberg chain with single-ion anisotropy and the Bose-Hubbard model. A 2D square lattice transverse field Ising model on an infinite cylinder is also investigated. These examples show the capacity of the scaling functions of higher order cumulants in determining the critical point and exponents of a variety of order parameters and phase transitions classes.


\section{Higher-order Moments and Cumulants in Tensor Networks}
\label{higher_order_moments}
To obtain moments and cumulants of an observable, one has to compute the expectation value of the observable operator to different orders. This can be done efficiently through the use of a triangular matrix-product operator (MPO) and its fixed-point equation. The latter is a recursive formula that relates the different elements of the environment matrix $E$ for a given triangular MPO. The advantage this recursive method has over other tensor network methods of calculating higher-order moments and cumulants such as that in Ref.~\cite{West} is that it unifies the method of computing the expectation values of local and string operators whereby the latter ends up being treated as a second order operator (i.e. a two-point correlator). For an upper triangular MPO, this recursive formula is given by \cite{Michel}
\begin{eqnarray}
E_i(L+1) = T_{W_{ii}} \left(E_i(L)\right) + \sum_{j<i} T_{W_{ji}} \left(E_j(L)\right),
\label{eqn:fixed_pt}
\end{eqnarray}
where $E_i$ is the $i$th element of $E$, $L$ are the number of sites, and $W_{ii}$ ($W_{ji}$) are the diagonal (off-diagonal) elements of the triangular MPO. For a lower triangular MPO, the index in the second sum is simply swapped, i.e. $j > i$. $T_{X}$ is the transfer operator that acts the matrix-valued $X$ operator element of the MPO:
\begin{eqnarray}
T_X \left(E(L)\right) = \sum_{ss'} \bra{s'} X \ket{s} A^{s' \dagger} E(L) A^s .
\label{eqn:transfer_op}
\end{eqnarray}
Eq.~\ref{eqn:fixed_pt} specifies the action of adding one site to the expectation value in terms of the polynomial form for the $d$ different matrices $E_i$ ($1 \leq i \leq d$), for a $d \times d$ dimensional MPO \cite{McCulloch1, McCulloch2, Michel}. A triangular MPO with zero momentum is characterized by diagonal elements $W_{ii}$ that are proportional to the identity operator, $W_{ii} = xI$, with prefactor $x$ satisfying either $x = 1$ or $|x| < 1$. As such, the expectation value is a polynomial function of $L$ with matrix-valued coefficients \cite{Michel}:
\begin{eqnarray}
E_i(L) = \sum_{m=0}^p E_{i,m} L^m ,
\label{eqn:E_poly_L}
\end{eqnarray}
where $p$ is the polynomial degree of $E_i(L)$.

The expectation value of operator $M$ written in the form of an upper triangular MPO is \ignore{$\braket{M} = \text{Tr} \left(E_i(L+1) \rho_R \right)$} $\braket{M} = \text{Tr} \left(E_d(L) \rho_R \right)$, where $\rho_R$ is the reduced density-matrix of the right bipartition of the state, and \ignore{$E_i(L+1)$}$E_d(L)$ is the component of the left environment matrix that contains the matrix-valued operator that represents the accumulation of all terms of the MPO, corresponding to the $d^\text{th}$ column of an upper triangular MPO. For a lower triangular MPO, the expectation value is $\braket{M} = {Tr} \left(E_1(L) \rho_R \right)$ where $E_1(L)$ is the component of the left environment matrix that contains the matrix-valued operator accumulated from the first column of the lower triangular MPO. The choice of left (right) environment matrix, left (right) transfer operator and right (left) reduced density matrix is arbitrary, and one could use either choice. As an example, the magnetization order parameter is shown here. This order parameter is given by the upper triangular MPO
\begin{eqnarray}
W = \left( \begin{array}{cc} I & Z \\ & I \end{array} \right) ,
\label{eqn:magnetization_mpo}
\end{eqnarray}
where $Z \equiv S^z$ is the $z$ component of the spin operator, and $I$ is the identity. For this triangular MPO, the operator representing the observable is $E_2(L)$ which has the expectation value of $\braket{M} = \text{Tr} (E_2(L) \rho_R)$. This quantity is actually trivial to compute via conventional tensor network methods since it is a local expectation value and the MPO is a single-site operator. However, its evaluation from first principles is shown here since, in contrast to the conventional tensor network approach, this method is applicable to more complicated MPO's such as higher-order moments and string order parameters. The goal here is to show that the correct polynomial degree matrix-valued coefficient Eq.~\ref{eqn:E_poly_L} is related to $\text{Tr} (E_2(L) \rho_R)$ -- this is done as follows. $W$ has dimension $d = 2$, so Eq.~\ref{eqn:fixed_pt} gives two terms. First,
\begin{eqnarray}
E_1(L+1) &=& T_{W_{11}}(E_1(L)) \nonumber \\
&=& T_I(E_1(L)) .
\label{eqn:E1}
\end{eqnarray}
The only operator acting here is transfer operator $T_I$ containing the identity operator $I$. As a result, its operation on $E_1(L)$ is trivial and therefore independent of $L$. Hence, the polynomial degree $p$ in Eq.~\ref{eqn:E_poly_L} for $E_1(L)$ is $p = 0$ and thus $E_1(L)$ can be written as
\begin{eqnarray}
E_1(L) &=& E_{1,0} L^0 \nonumber \\
&=& E_{1,0} .
\label{eqn:E1_2}
\end{eqnarray}
Inserting this into Eq.~\ref{eqn:E1} gives
\begin{eqnarray}
E_{1,0} = T_{I}(E_{1,0}) ,
\label{eqn:E1_I}
\end{eqnarray}
which implies that $E_{1,0}$ is an eigenvector of operator $T_I$ with eigenvalue equal to 1. If the iMPS is in left-canonical form, then $E_{1,0} \propto \tilde{I}$ where $\tilde{I}$ is the $m \times m$ identity matrix. An obvious choice of proportionality factor is to fix $\text{Tr} (E_{1,0} \rho_R) = 1$, which implies
\begin{eqnarray}
E_1(L) = E_{1,0} = \tilde{I} .
\label{eqn:E1_3}
\end{eqnarray}

The second term of the recursion formula is
\begin{eqnarray}
E_2(L+1) &=& T_{W_{22}}(E_2(L)) + T_{W_{12}}(E_1(L)) .
\end{eqnarray}
Using Eq.~\ref{eqn:E1_3} and the elements of the MPO from Eq.~\ref{eqn:magnetization_mpo}, this becomes
\begin{eqnarray}
E_2(L+1) &=& T_I(E_2(L)) + T_{Z}(\tilde{I}) \nonumber \\
		 &=& T_I(E_2(L)) + C_Z ,
\label{eqn:E2}
\end{eqnarray}
where in the last step, $C_Z \equiv T_{Z}(\tilde{I})$ is a constant matrix, i.e. it has no dependence on any $E_i$'s and thus its value can be calculated beforehand. With this, one can now proceed to find an ansatz for the form of $E_2(L)$ in the asymptotic large-$L$ limit. This limit shouldn't depend on the ``boundary" form $E_2(0)$, hence one can choose $E_2(0) = 0$ (this choice has no effect on the final solution, up to an irrelevant constant), which leads to
\begin{eqnarray}
E_2(1) &=& C_Z \nonumber \\
E_2(2) &=& T_I(C_Z) + C_Z \nonumber \\
E_2(3) &=& T_I^2(C_Z) + T_I(C_Z) + C_Z \nonumber \\
\vdots & & \nonumber \\
E_2(L) &=& \sum_{n=0}^{L-1} T_I^{n}(C_Z)
\label{eqn:E2_geo_series}
\end{eqnarray}
where $T^n_I(C_Z)$ means $n$ repeated applications of the transfer operator, $T_I( T_I( \ldots (C_Z) \ldots ))$. Thus $E_2(L)$ is the form of a geometric series, and the large $L$ limit depends on the
nature of the spectral decomposition of $C_Z$. Since $T_I$ only has a single eigenvalue
equal to 1 and all other eigenvalues are strictly less than 1, it follows that
$E_2(L)$ can diverge at most linearly with $L$, hence the polynomial degree $p$ of Eq.~\ref{eqn:E_poly_L} for $E_2(L)$ is $p = 1$ and thus $E_2(L)$ can be written as
\begin{eqnarray}
E_2(L) = E_{2,0} + E_{2,1} L .
\label{eqn:poly_expansion_E_2}
\end{eqnarray}
Inserting this into Eq.~\ref{eqn:E2} gives
\begin{eqnarray}
E_{2,0} + E_{2,1} (L+1) &=& T_I(E_{2,0} + E_{2,1}L) + C_Z \nonumber \\
E_{2,0} + E_{2,1} + E_{2,1} L &=& T_I(E_{2,0}) + T_I(E_{2,1})L \nonumber \\
&&+ C_Z .
\end{eqnarray}
Equating powers of $L$ gives
\begin{eqnarray}
L^0 &:& E_{2,0} + E_{2,1} = T_I(E_{2,0}) + C_Z \label{eqn:L0} \\
L^1 &:& E_{2,1} = T_I(E_{2,1}) \label{eqn:L1} .
\end{eqnarray}
Similar to Eq.~\ref{eqn:E1_I}, Eq.~\ref{eqn:L1} implies $E_{2,1} \propto \tilde{I} = e_{2,1,1} \tilde{I}$, where $e_{2,1,1}$ is a proportionality constant. The first two indices in the subscript of $e_{2,1,1}$ denotes that it corresponds to $E_{2,1}$ (i.e. same subscript indices for clarity), while the third subscript denotes the eigenvalue number of the transfer operator. Since the first eigenvector of the transfer operator $T_I$ is $\tilde{I}$, the third index in the subscript of $e_{2,1,1}$ is $1$. The transfer operator $T_I$ can be further decomposed into its components through an eigenvalue decomposition $T_I = \sum_{i=1}^{m^2} \lambda_i \ket{\lambda_i}\bra{\lambda_i}$, where $\ket{\lambda_i}$ are the eigenvectors of $T_I$. $E_{2,0}$ can also be expanded in terms of the basis $\ket{\lambda_i}$, i.e. $E_{2,0} = \sum_{i=1}^{m^2} e_{2,0,i} \ket{\lambda_i}$. The constant matrix $C_Z$ on the other hand is expanded as $C_Z \equiv T_Z(\tilde{I}) = \sum_{i=1}^{m^2} C_{Z,i} \ket{\lambda_i}$ where $C_{Z,i}$ are elements of the constant matrix $C_Z$. Using these in Eq.~\ref{eqn:L0} gives
\begin{eqnarray}
\sum_{i=1}^{m^2} e_{2,0,i} \ket{\lambda_i} + e_{2,1,1} \tilde{I} &=& \sum_{i=1}^{m^2} \lambda_i \ket{\lambda_i}\bra{\lambda_i} \left( \sum_{i'=1}^{m^2} e_{2,0,i'} \ket{\lambda_{i'}} \right) \nonumber \\
&& + \sum_{i=1}^{m^2} C_{Z,i} \ket{\lambda_i} \nonumber \\
&=& \sum_{ii'=1}^{m^2} \lambda_i e_{2,0,i'} \ket{\lambda_i} \left\langle \lambda_i | \lambda_{i'} \right\rangle \nonumber \\
&&+ \sum_{i=1}^{m^2} C_{Z,i} \ket{\lambda_i} \nonumber \\
&=& \sum_{i=1}^{m^2} \lambda_i e_{2,0,i} \ket{\lambda_i} + \sum_{i=1}^{m^2} C_{Z,i} \ket{\lambda_i} , \nonumber \\
\label{eqn:L1_2}
\end{eqnarray}
where the orthogonality relation $\sum_{i'} \left\langle \lambda_i | \lambda_{i'} \right\rangle = \delta_{ii'}$ was used in the last step. By construction, the spectrum of $T_I$ contains the eigenvalue $1$ with corresponding eigenvector $\tilde{I}$. Thus, the eigenvalue decomposition of $T_I$ and $E_{2,0}$ can be separated into a part parallel to the identity $\tilde{I}$ and remaining parts that are perpendicular to $\tilde{I}$, i.e.
\begin{eqnarray}
\sum_{i=1}^{m^2} e_{2,0,i} \ket{\lambda_i} = e_{2,0,1} \ket{\lambda_1} + \sum_{i=2}^{m^2} e_{2,0,i} \ket{\lambda_i} ,
\end{eqnarray}
with $\ket{\lambda_1} = \tilde{I}$, and similarly for the decomposition of $\sum_{i=1}^{m^2} \lambda_i e_{2,0,i} \ket{\lambda_i} = e_{2,0,1} \ket{\lambda_1} + \sum_{i=2}^{m^2} \lambda_i e_{2,0,i} \ket{\lambda_i}$, with $\lambda_1 = 1$. Eq.~\ref{eqn:L1_2} then becomes
\begin{eqnarray}
e_{2,0,1} \tilde{I} + \sum_{i=2}^{m^2} e_{2,0,i} \ket{\lambda_i} &+& e_{2,1,1} \tilde{I} \nonumber \\
&=& e_{2,0,1} \tilde{I} + \sum_{i=2}^{m^2} \lambda_i e_{2,0,i} \ket{\lambda_i} \nonumber \\
&&+ C_{Z,1} \tilde{I} + \sum_{i=2}^{m^2} C_{Z,i} \ket{\lambda_i} \nonumber \\
\sum_{i=2}^{m^2} e_{2,0,i} \ket{\lambda_i} + e_{2,1,1} \tilde{I} &=& \sum_{i=2}^{m^2} \lambda_i e_{2,0,i} \ket{\lambda_i} + C_{Z,1} \tilde{I} \nonumber \\
&&+ \sum_{i=2}^{m^2} C_{Z,i} \ket{\lambda_i}.
\label{eqn:L1_3}
\end{eqnarray}
As stated above, the goal is to show that the correct polynomial degree of Eq.~\ref{eqn:E_poly_L} is related to $\text{Tr} (E_2(L) \rho_R)$. In Eq.~\ref{eqn:poly_expansion_E_2} the polynomial degree $p = 1$ is that corresponding to the coefficient $E_{2,1}$. The latter's proportionality constant is $e_{2,1,1}$ and this can be related to $\text{Tr} (E_2(L) \rho_R)$ via Eq.~\ref{eqn:L1_3} by multiplying $\rho_R$ on the left and right hand sides and taking the trace:
\begin{eqnarray}
\text{Tr} \left( \sum_{i=2}^{m^2} e_{2,0,i} \ket{\lambda_i} \rho_R \right) &+& \text{Tr} \left( e_{2,1,1} \tilde{I} \rho_R \right) \nonumber \\
&=& \text{Tr} \left(\sum_{i=2}^{m^2} \lambda_i e_{2,0,i} \ket{\lambda_i} \rho_R \right) \nonumber \\
&&+ \text{Tr} \left( C_{Z,1} \tilde{I} \rho_R \right) \nonumber \\
&&+ \text{Tr} \left( \sum_{i=2}^{m^2} C_{Z,i} \ket{\lambda_i} \rho_R \right)
\label{eqn:L1_4}
\end{eqnarray}
The spectrum of the left transfer operator $T_I$ contains the eigenvalue $\lambda_1 = 1$ corresponding to its left eigenvector $\tilde{I}$ and right eigenvector $\rho_R$ (the right transfer operator on the other hand has eigenvalue 1 corresponding to its left eigenvector $\rho_L$, which is the reduced density matrix of the left bipartition, and its right eigenvector $\tilde{I}$). As such, the product between the term parallel to the identity and $\rho_R$ in Eq.~\ref{eqn:L1_4} is $\tilde{I}\rho_R = 1$, while the products of the terms perpendicular to the identity with $\rho_R$ give $\ket{\lambda_i} \rho_R = 0$. Eq.~\ref{eqn:L1_4} thus reduces to
\begin{eqnarray}
\text{Tr} \left( e_{2,1,1} \tilde{I} \rho_R \right) &=& \text{Tr} \left( C_{Z,1} \tilde{I} \rho_R \right) \nonumber \\
e_{2,1,1} &=& C_{Z,1} ,
\label{eqn:L1_5}
\end{eqnarray}
where from the first to second line of Eq.~\ref{eqn:L1_5}, $e_{2,1,1}$ and $C_{Z,1}$ are just numbers, so their trace are the numbers themselves. Eq.~\ref{eqn:L1_5} states that the polynomial degree $p = 1$, i.e. the degree of $L^1$ with coefficient $E_{2,1}$, is related to the desired expectation value $\text{Tr} (E_2(L) \rho_R)$. Though the terms perpendicular to the identity vanished upon multiplication of $\rho_R$, they can be evaluated separately when they are needed for the evaluation of other terms such as in the case of where higher order moments are needed. This derivation can be generalized for the higher order moments and a detailed derivation of $\braket{M^2}$ of the MPO in Eq.~\ref{eqn:magnetization_mpo} is shown in Section \ref{appendix_variance}. Ultimately, the moments are expressed as an $n$-degree polynomial in $L$ \cite{Michel}. For example, the first four moments in terms of cumulants per site are
\begin{eqnarray}
\braket{M} &=& \kappa_1 L,
\label{eqn:first_cumulant_per_site}
\\
\braket{M^2} &=& \kappa_2 L + \kappa_1^2 L^2,
\label{eqn:second_cumulant_per_site}
\\
\braket{M^3} &=& \kappa_3 L + 3\kappa_2\kappa_1 L^2 + \kappa_1^3 L^3,
\label{eqn:third_cumulant_per_site}
\\
\braket{M^4} &=& \kappa_4 L + (4\kappa_3\kappa_1 + 3\kappa_2^2)L^2 \nonumber \\
&& + 6\kappa_2\kappa_1^2 L^3 + \kappa_1^4 L^4,
\label{eqn:fourth_cumulant_per_site}
\end{eqnarray}
Comparing Eq.~\ref{eqn:first_cumulant_per_site} and the result of Eq.~\ref{eqn:L1_5} reveals that the coefficient $e_{2,1,1}$ is indeed the first cumulant $\kappa_1$.


\section{Relation between Cumulant Exponents}
The fact that the Binder cumulant of different system sizes cross one another at the critical point implies that $U_4$ is independent of $L$ at the critical point. This sets a constraint or relationship between the cumulant exponents $\alpha_i$ since varying them should not change the value of $U_4$ at the critical point. This relationship is revealed by equating the exponent of the bond dimension $m$ in the $n$th order derivative of the singular part of the free energy density $f$ to the exponent of $m$ in the $n$th order cumulant written as a power law function of $m$.

The singular part of the free energy density $f$ introduced in Ref.~\cite{Privman} for a finite system is given by:
\begin{eqnarray}
f \approx L^{-d} Y \left( C_1 t L^{1/\nu} , C_2 h L^{ \left( \beta + \gamma \right)/\nu} \right),
\label{eqn:free_en_uni_scaling}
\end{eqnarray}
where $L$ is the system size, $d$ is the spatial dimension, $t$ is the reduced temperature given as $t \equiv \frac{T - T_c}{T_c}$, and $h$ is the scaled applied field $h \equiv H/k_B T$. $Y$ is the universal scaling function which even though is universal, depends on system specific properties such as the boundary conditions, lattice geometry, coupling constants, etc, which are captured in the nonuniversal metric factors $C_1$ and $C_2$. In an infinite system, there is no notion of a system size $L$. Instead, any need for a length is replaced by the correlation length $\xi$, i.e. $L \propto \xi$. This in turn is related to the bond dimension $m$ through the expression $\xi \sim m^\kappa$ \cite{Tagliacozzo, Pollmann} which will be discussed in the Section \ref{finite-entanglement}. One now has $L \propto m^\kappa$ and by substituting this into Eq.~\ref{eqn:free_en_uni_scaling} gives
\begin{eqnarray}
f \approx m^{-\kappa d} Y \left( C_1 t m^{\kappa/\nu} , C_2 h m^{ \left( \beta + \gamma \right)\kappa/\nu} \right).
\label{eqn:free_en_uni_scaling_m}
\end{eqnarray}

From the rules of thermodynamics, the change of the free energy density with respect to a parameter of interest gives the measure of that parameter i.e. its thermodynamic observable or order parameter. Therefore, the first order derivative of $f$ with respect to $h$ gives the first cumulant:
\begin{eqnarray}
\kappa_1 &=& - \frac{\partial f}{\partial h} \nonumber \\
		 &=& C_2 m^{ \left( \beta + \gamma - \nu d \right) \kappa / \nu} Y^{(1)} \left( C_1 t m^{\kappa/\nu} , C_2 h m^{ \left( \beta + \gamma \right)\kappa/\nu} \right) \nonumber \\
		 &=& C_2 m^{ \left( \beta + \gamma - \left( 2\beta + \gamma \right) \right) \kappa / \nu} Y^{(1)} \left( C_1 t m^{\kappa/\nu} , C_2 h m^{ \left( \beta + \gamma \right)\kappa/\nu} \right) \nonumber \\
		 &=& C_2 m^{-\beta\kappa/\nu} Y^{(1)} \left( C_1 t m^{\kappa/\nu} , C_2 h m^{ \left( \beta + \gamma \right)\kappa/\nu} \right) ,
\label{eqn:kappa1_uni_scaling}
\end{eqnarray}
where from the second to third lines, the Josephson relation $\nu d = 2 - \alpha$, and the Rushbrooke relation $\alpha + 2\beta + \gamma = 2$, were combined to eliminate $\alpha$ and give $\nu d = 2\beta + \gamma$. The superscript of $Y$ marks the order of the derivative. Proceeding in the similarly fashion for the higher order derivatives of $f$:
\begin{eqnarray}
\kappa_2 &=& - \frac{\partial^2 f}{\partial h^2} \nonumber \\
		 &=& C_2^2 m^{\gamma\kappa/\nu} Y^{(2)} \left( C_1 t m^{\kappa/\nu} , C_2 h m^{ \left( \beta + \gamma \right)\kappa/\nu} \right) , \label{eqn:kappa2_uni_scaling} \\
\kappa_3 &=& - \frac{\partial^3 f}{\partial h^3} \nonumber \\
		 &=& C_2^3 m^{(\beta + 2\gamma)\kappa/\nu} Y^{(3)} \left( C_1 t m^{\kappa/\nu} , C_2 h m^{ \left( \beta + \gamma \right)\kappa/\nu} \right) , \label{eqn:kappa3_uni_scaling} \\
\kappa_4 &=& - \frac{\partial^4 f}{\partial h^4} \nonumber \\
		 &=& C_2^4 m^{(2\beta + 3\gamma)\kappa/\nu} Y^{(4)} \left( C_1 t m^{\kappa/\nu} , C_2 h m^{ \left( \beta + \gamma \right)\kappa/\nu} \right) , \label{eqn:kappa4_uni_scaling} \\
		 &\vdots \nonumber \\
\kappa_n &=& - \frac{\partial^n f}{\partial h^n} \nonumber \\
		 &=& C_2^n m^{((n-2)\beta + (n-1)\gamma)\kappa/\nu} \nonumber \\
		 &&\times Y^{(n)} \left( C_1 t m^{\kappa/\nu} , C_2 h m^{ \left( \beta + \gamma \right)\kappa/\nu} \right) . \label{eqn:kappa_n_uni_scaling}
\end{eqnarray}
This shows that the exponent of $m$ of any $n$th order cumulant can be obtained from the exponents of just the two first cumulants $\beta$ and $\gamma$.

Cumulants of an order parameter show singular behavior at the critical point, i.e. they either vanish or diverge at the critical point. For an iMPS with a fixed $m$ these divergences still occur, at least in the limit of perfectly converged numerics, although sufficiently close to a pseudo-critical point the iMPS exhibits mean-field-like behavior \cite{Liu}, i.e. the exponents diverge from their true values. Nevertheless, finite-entanglement scaling of the cumulants at the true critical point follows the correct power-law scaling behavior with the basis size,
\begin{eqnarray}
\kappa_i \sim m^{\alpha_i},
\label{eqn:cumulant_power_law}
\end{eqnarray}
where $\alpha_i$ is the cumulant critical exponent. Equating the exponents of $m$ of Eq.~\ref{eqn:kappa_n_uni_scaling} and \ref{eqn:cumulant_power_law} gives a relation between the exponent of the $n$th order cumulant and the first two cumulant exponents:
\begin{equation}
\alpha_n = \left[(n-2) \beta + (n-1) \gamma \right] \frac{\kappa}{\nu}.
\end{equation}
Equivalently, a self-contained expression for all $\alpha_n$'s in terms of $\alpha_1$ and $\alpha_2$ can be written as
\begin{eqnarray}
\alpha_n = -(n-2)\alpha_1 + (n-1)\alpha_2 ,
\label{eqn:higher_cumulant_exponent_relation}
\end{eqnarray}
where $\alpha_1 = -\beta\kappa/\nu$ and $\alpha_2 = \gamma\kappa/\nu$. Eq.~\ref{eqn:higher_cumulant_exponent_relation} is dubbed the cumulant exponent relation and it will be tested for the various models in Section \ref{results}.

An especially useful application of this relation is in the calculation of certain cumulant exponents whose cumulants are not directly accessible, from those that are readily obtained. For example, in the case when certain symmetries are enforced on a wave function, the odd $n$ order cumulants may turn out to be identically zero, whereas the even $n$ order cumulants are nonzero. In such a circumstance, Eq.~\ref{eqn:higher_cumulant_exponent_relation} can be used to calculate the odd order cumulant exponents from the even ones. To illustrate this, supposing $\alpha_2$ and $\alpha_4$ have be determined, then by setting $n = 4$, Eq.~\ref{eqn:higher_cumulant_exponent_relation} gives
\begin{eqnarray}
\alpha_4 = 2\alpha_1 + 3\alpha_2 \nonumber
\end{eqnarray}
which gives the first cumulant exponent:
\begin{eqnarray}
\alpha_1 = \frac{1}{2} (-3\alpha_2 + \alpha_4) . \nonumber
\end{eqnarray}
Subsequently, by setting $n = 3$, Eq.~\ref{eqn:higher_cumulant_exponent_relation} gives
\begin{eqnarray}
\alpha_3 &=& \alpha_1 + 2\alpha_2 , \nonumber
\end{eqnarray}
which upon substitution of $\alpha_1$ above gives
\begin{eqnarray}
\alpha_3 &=& \left( \frac{1}{2} (-3\alpha_2 + \alpha_4) \right) + 2\alpha_2 \nonumber \\
&=& \frac{1}{2} (\alpha_2 + \alpha_4) . \nonumber
\end{eqnarray}


\section{Finite-entanglement scaling and scaling functions}
\label{finite-entanglement}
The idea of finite-entanglement scaling (FES) for infinite systems represented by iMPS was directly adopted from finite-size scaling (FSS) of finite systems \cite{Tagliacozzo}. The need for FES can be appreciated by understanding the role $m$ plays in the iMPS. To do so, one has to apply a Schmidt decomposition of a state $\ket{\psi}$ which will allow one to observe the entanglement within the bipartition of the state through means of the Schmidt values $\lambda_n$. Splitting a state into two parts $A$ and $B$, the Schmidt decomposition is a basis choice that expresses $\ket{\psi}$ as a sum of product between the two parts of the wave function:
\begin{eqnarray}
\ket{\psi} = \sum_{n=1}^\infty \lambda_n \ket{\phi_n^A} \ket{\phi_n^B} ,
\end{eqnarray}
where $\ket{\phi_n^A}$ and $\ket{\phi_n^B}$ are orthonormal bases of the Hilbert spaces of the two subsystems $A$ and $B$. The entanglement entropy $S$ between the two subsystems is given as
\begin{eqnarray}
S = - \sum_n \lambda_n^2 \log \lambda_n^2 .
\end{eqnarray}
Since the size of each matrix $A^{s_i}$ in an iMPS is bounded by the bond dimension $m$, all properties of the state at and away from a critical point is dictated by $m$. Away from the critical point, the entanglement entropy $S$ of a ground state is finite, therefore a finite $m$ iMPS would be able to represents the state well. At the critical point, the maximum $S$ for $m$ Schmidt eigenvalues is $\log(m)$. As a result, the finite $m$ dimensional matrices $A^{s_i}$ cannot approximate the ground state well and all singular behavior of observables or thermodynamic quantities are blurred out. This is further corroborated by the fact that all correlation functions of MPS decay exponentially, implying that they have finite correlation lengths \cite{Ostlund, Rommer}. In spite of this, it is possible to quantitatively measure how observables and thermodynamic quantities behave at and in the vicinity of the critical point through means of FES.

The nature of the transition in a finite and infinite system can be appreciated from their energy landscapes and magnetization order parameter. In a finite system, the energy landscape contains minimums which cross each other at the $m$ and $L$ dependent pseudocritical point $B_c(m,L)$. This causes a discontinuous change of the magnetization at $B_c(m,L)$ which is a first order transition. As $L$ is increased, these energy minimums move closer to each other in parameter space and finally coincide at $B_c(m,L \rightarrow \infty)$ when $L \rightarrow \infty$. This causes the magnetization to vanish smoothly which is a continuous (second-order) phase transition. In an infinite system, the behavior of the order parameter in the vicinity of the critical point is always mean-field in nature. Increasing $m$ shrinks this mean-field behavior around the critical point and the true transition type is recovered in the region left behind by the mean-field behavior. In the case of the transverse field Ising (TFI) model, Ref.~\cite{Liu} has shown this change of the magnetization exponent which is $\beta = 1/2$ when $m$ is small whereas a region of $\beta = 1/8$ increases in size as $m$ is increased. The former is the well-known magnetization exponent of the TFI model under a mean-field treatment whereas the latter is the true magnetization exponent of the Ising class.

In a finite system, all observables do not form singularities. Taking the correlation length as an example, $\xi$ cannot exceed the system size $L$. Thus, as $\xi$ approaches its maximum value of $\sim L$, it forms a smooth, rounded peak at the $L$-dependent pseudocritical point. This pseudocritical point is located some distance from the true critical point and approaches it as $L$ is increased. This smoothness ensures that $\xi$ is always continuous i.e. no singularity. In contrast to this, the infinite system's observables always forms a singularity because there is no system-size restriction. Hence, at its $m$-dependent pseudocritical point, $\xi$ forms a sharp, discontinuous peak whose height and pseudocritical point is dependent on $m$. Using these facts, one is able to determine the true critical point by tabulating the singular part of the observable with respect to $m$ and extrapolating the observable or parameter value to the $m \rightarrow \infty$ limit in order to obtain the true value of that observable and the true critical point.

To draw parallels between FSS and FES, let's first look at FSS. In the thermodynamic limit, the correlation length $\xi(B)$ in the vicinity of the critical point $B_c$ diverges as a power law function
\begin{eqnarray}
\xi(B) \propto \frac{1}{|B - B_c|^\nu} ,
\label{eqn:cor_len_thermo_limit}
\end{eqnarray}
where $\nu$ is the critical exponent of $\xi$. Similarly, suppose there is some observable $\kappa_n(B)$ that also diverges at the critical point $B_c$ as a power law function:
\begin{eqnarray}
\kappa_n(B) &\propto& \frac{1}{|B - B_c|^q} \nonumber \\
		&\propto& \xi(B)^{q / \nu} ,
\label{eqn:kappa_n_xi}
\end{eqnarray}
where $q$ is the critical exponent of $\kappa_n$. In a finite system of size $L$, both $\xi$ and $\kappa_n$ are now a functions of $B$ and $L$ i.e. $\xi(L,B)$ and $\kappa_n(L,B)$, and they both form peaks at the pseudocritical point $B^*(L)$. Since the system is finite, the ground state is effectively ordered when $\xi(L,B^*(L)) \approx L$ since $\xi(L,B^*(L))$ cannot grow larger than $L$. Thus Eqs.~\ref{eqn:cor_len_thermo_limit} for a finite system is written as
\begin{eqnarray}
\xi(L,B^*(L)) &\propto& \frac{1}{|B^*(L) - B_c|^\nu} \nonumber \\
L &=& \frac{g}{|B^*(L) - B_c|^\nu} ,
\label{eqn:finite_size_effect}
\end{eqnarray}
where $g$ is the proportionality constant. Eq.~\ref{eqn:finite_size_effect} states that the effective distance between the pseudocritical point $B^*(L)$ and the true critical point $B_c$ is determined by the system size $L$. This equation can be re-expressed in two useful ways. First, as
\begin{eqnarray}
|B^*(L) - B_c| L^{1 / \nu} = g' ,
\label{eqn:L_indep_finite_size_effect}
\end{eqnarray}
where $g' = g^{1 / \nu}$. This equation states that all $L$-dependent terms on the left hand side must give an overall $L$-independent constant which is on the right hand side. The second expression is
\begin{eqnarray}
B^*(L) = \frac{g'}{L^{1 / \nu}} + B_c .
\label{eqn:finite_size_scaling}
\end{eqnarray}
This is the FSS equation. It enables one to determine $B_c$ and $\nu$ from data of an observable e.g. the location of $B^*(L)$ corresponding to the peak of $\kappa_n(L,B^*(L))$, of different values of $L$. Similarly, one can also apply these steps to $\kappa_n(B)$ as follows. For a finite system, the peaks of $\kappa_n(L,B^*(L))$ for a given $L$ occurs $B^*(L)$. Using the fact that $\xi(L,B^*(L)) \approx L$ in the finite system, Eq.~\ref{eqn:kappa_n_xi} becomes
\begin{eqnarray}
\kappa_n(L,B^*(L)) &\propto& \xi(L,B^*(L))^{q / \nu} \nonumber \\
					&=& \tilde{g} L^{q / \nu} ,
\end{eqnarray}
where $\tilde{g}$ is the proportionality constant. This equation can be re-expressed as
\begin{eqnarray}
\frac{\kappa_n(L,B^*(L))}{L^{q / \nu}} = \tilde{g} ,
\label{eqn:scaling_function}
\end{eqnarray}
which states that all the $L$-dependent terms on the left must give an overall $L$-independent constant on the right. This is called the scaling function of $\kappa_n(L,B)$. By comparing Eqs.~\ref{eqn:L_indep_finite_size_effect} and \ref{eqn:scaling_function}, one notices that the former appears to be a form of scaling of the parameter $B$ while the latter appears to be a scaled form of $\kappa_n(L,B)$. Both equations are scaled such that there is no overall dependence of $L$. Therefore by plotting the data of $\kappa_n(L,B^*(L)) / L^{q / \nu}$ versus $|B^*(L) - B_c| L^{1 / \nu}$ for several values of $L$, and by treating the $B_c$, $q$ and $\nu$ as tuning parameters, one would find that the data for the different values of $L$ will collapse onto a single curve when the suitable values of the tuning parameters are chosen.

In an infinite system, there is no notion of a system size $L$. Instead, any need of a length is replaced by the correlation length $\xi$. If this system is described by an iMPS, then $\xi$ is related to $m$ by
\begin{eqnarray}
\xi \propto m^\kappa,
\label{eqn:cor_len_m_kappa}
\end{eqnarray}
where $\kappa$ is the finite-entanglement scaling exponent. This relation was found empirically in Refs.~\cite{Tagliacozzo, Andersson}, and later derived by Pollmann \textit{et al} \cite{Pollmann}. The latter was done by relating the distribution of Schmidt values $\lambda_n$ of the bipartition of the wave function, the central charge $c$ of the transition described by the associated conformal field theory, $m$ and $\xi$ at the critical point. This revealed that $\kappa$ is intimately related to the central charge $c$ by:
\begin{eqnarray}
\kappa = \frac{6}{c \left( \sqrt{\frac{12}{c}} + 1 \right)} .
\label{eqn:kappa_central_charge}
\end{eqnarray}
The significance of Eq.~\ref{eqn:kappa_central_charge} is that since only a handful of transition classes, and thus $c$'s, are known in 1D, this constraints the possible values of $\kappa$. By using $L \propto \xi \propto m^\kappa$, the three important equations \ref{eqn:L_indep_finite_size_effect}, \ref{eqn:finite_size_scaling} and \ref{eqn:scaling_function} become:
\begin{eqnarray}
|B^*(m) - B_c| m^{\kappa / \nu} = g' ,
\label{eqn:m_indep_finite_size_effect}
\end{eqnarray}
\begin{eqnarray}
B^*(m) = \frac{g'}{m^{\kappa / \nu}} + B_c ,
\label{eqn:finite_m_scaling}
\end{eqnarray}
and
\begin{eqnarray}
\frac{\kappa_n(m,B^*(m))}{m^{\kappa q / \nu}} = \tilde{g} .
\label{eqn:scaling_function_m}
\end{eqnarray}
Just as in the case of finite $L$, Eqs.~\ref{eqn:m_indep_finite_size_effect} and \ref{eqn:scaling_function_m} describe a scaled parameter $B$ and scaled function $\kappa_n(m,B)$ that are overall independent of $m$ respectively. Hence by plotting data of $\kappa_n(m,B^*(m)) / m^{\kappa q / \nu}$ versus $|B^*(m) - B_c| m^{\kappa / \nu}$ for several values of $m$, one can expect a data collapse when the suitable values of $B_c$, $\gamma$, $\nu$ and $\kappa$ are chosen. Besides the scaling function of the cumulants, Eq.~\ref{eqn:cor_len_m_kappa} gives a scaling function of $\xi$ that does not have an analog in finite-size scaling. This is written as
\begin{eqnarray}
\frac{\xi(m,B^*(m))}{m^\kappa} = g ,
\label{eqn:scaling_function_cor_len_m}
\end{eqnarray}
where $g$ is a proportionality constant. Thus plotting this against Eq.~\ref{eqn:m_indep_finite_size_effect} would give a data collapse of $\xi(m,B^*(m))$ with tuning parameters $B_c$, $\nu$ and $\kappa$.


\section{Results}
\label{results}
All ground state wave functions in this work were variationally optimized using the infinite density-matrix renormalization group (iDMRG) algorithm with single-site optimization \cite{McCulloch2, Schollwock, Hubig2}. Wave functions of several different bond dimensions $m$ were generated to demonstrate the finite-entanglement scaling of the cumulants $\kappa_i$ and correlation length $\xi$, as well as to be used to locate the critical point through means of the Binder cumulant $U_4(m)$.


\subsection{One-dimensional transverse field Ising model}
\label{result_tfi}
\begin{figure}[h!]
  \centering\includegraphics[width=0.475\textwidth]{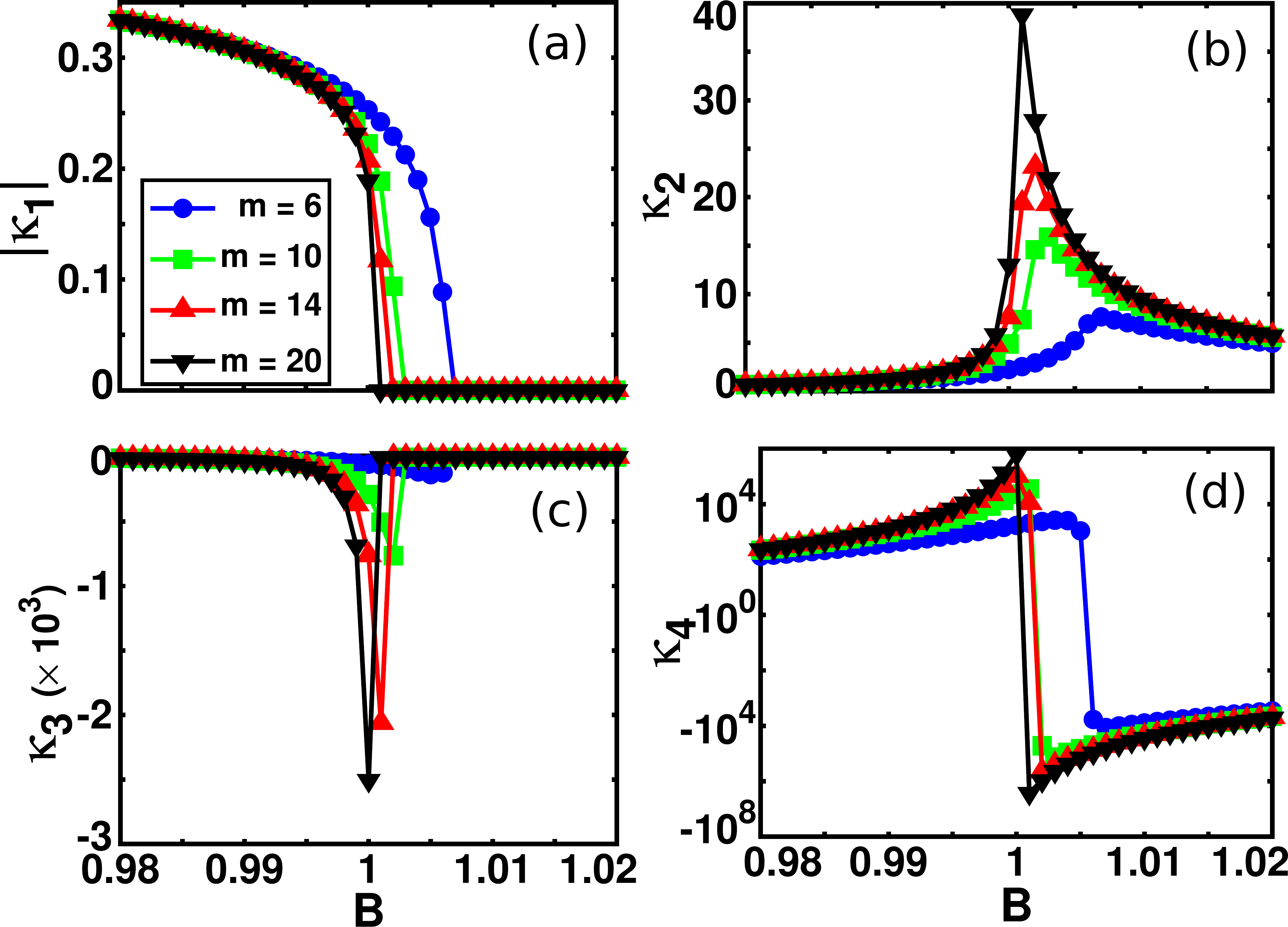}
  \caption{(Colour online) 1D TFI model. First four cumulants of the order parameter $\braket{M} = \sum_i \sigma^z_i$ as a function of transverse field $B$ for several values of $m$. (a) First cumulant $\kappa_1$ is the order parameter itself which is nonzero when $B < B_c$ and zero when $B > B_c$. (b) Second cumulant $\kappa_2$ is the variance of the order parameter, which diverges at the critical point. (c) Third cumulant $\kappa_3$ is the skewness. (d) Fourth cumulant $\kappa_4$ is the kurtosis. The point where the cumulants vanish/diverge shifts towards $B_c = 1$ as $m$ is increased.}
  \label{fig:cumulants_ising1d}
\end{figure}

The 1D transverse field Ising (TFI) model is the quintessential model for studying phase transitions. Its Hamiltonian is given by
\begin{eqnarray}
H = -\sum_i \sigma^z_i \sigma^z_{i+1} + B\sum_i \sigma^x_i,
\label{eqn:Hamiltonian_ising_1d}
\end{eqnarray}
where $B$ is the transverse field strength. The ground state of this model is ferromagnetic when $B < B_c$ and paramagnetic when $B > B_c$ with $B_c = 1$. Figure \ref{fig:cumulants_ising1d} shows the first four cumulants of the magnetization order parameter
\begin{eqnarray}
\braket{M} = \sum_i \sigma^z_i,
\label{eqn:magnetization_order_parameter}
\end{eqnarray}
as a function of the transverse field $B$. The ground state is ordered at $B < B_c$, giving a nonzero $\braket{M}$, and disordered when $B > B_c$, giving $\braket{M} = 0$. The variance, skewness and kurtosis on the other hand diverge at the critical point due to large fluctuations in the ground state. One can picture that as the variance diverges, the distribution function expands infinitely. As a result, the skewness and kurtosis of the distribution function also diverges. All four cumulants Figs.~\ref{fig:cumulants_ising1d}(a) - \ref{fig:cumulants_ising1d}(d) show the same dependence of $m$, i.e. the point where they vanish/diverge shifts towards the known value of $B_c$ as $m$ is increased. This indicates that the iMPS wave function better approximates the true wave function with increasing $m$.

\begin{figure}[h!]
  \centering\includegraphics[width=0.45\textwidth]{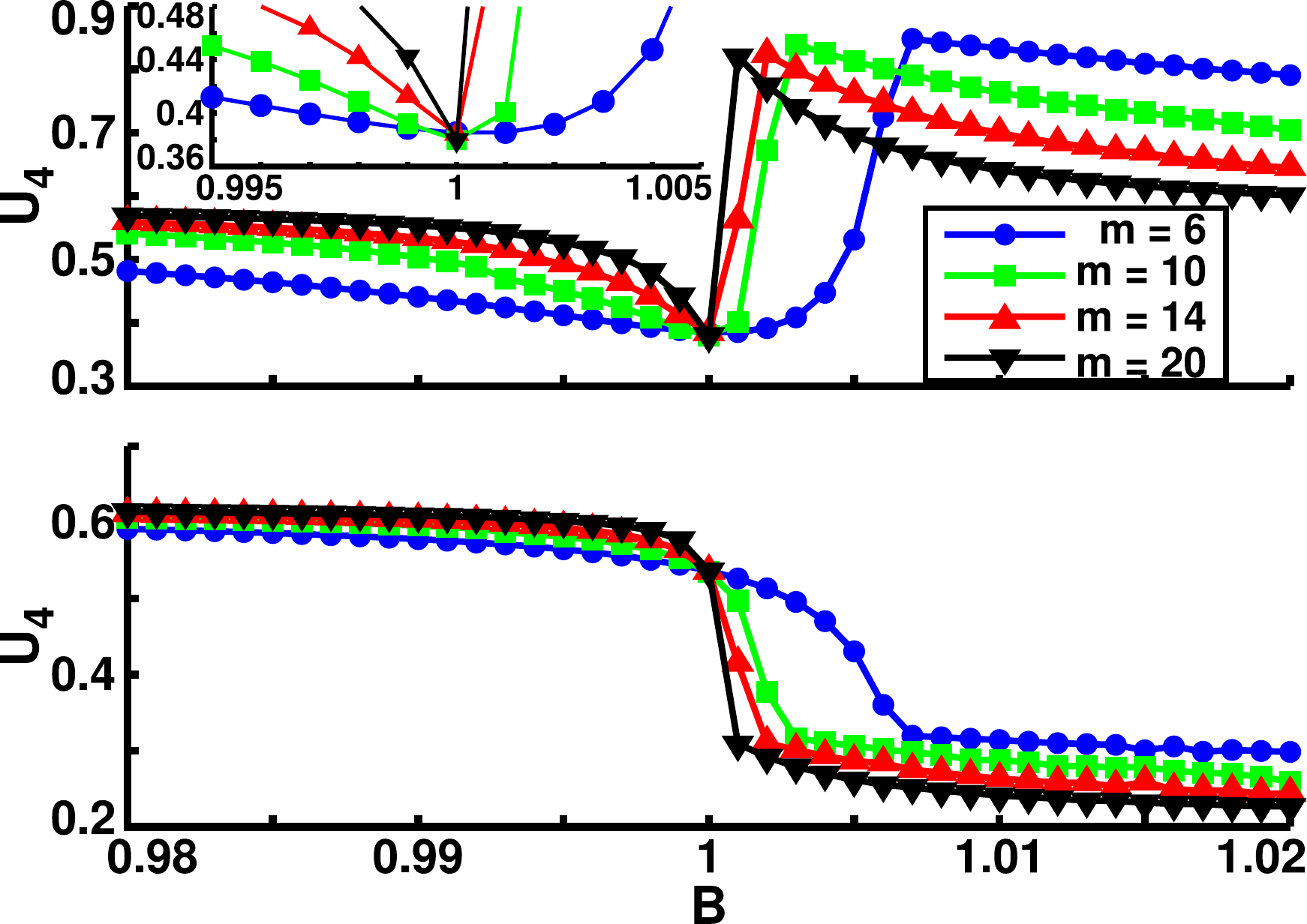}
  \caption{(Colour online) 1D TFI model. The Binder cumulant for $s = 2$ (top) and $s = s^* = 5.31$ (bottom) as a function of $B$ for various values of $m$. In the top figure, there is an intersection between the different values of $m$ at $B = 1$ and spurious crossing points in the region $1 < B < 1.007$. Increasing $s$ removes the spurious crossing points, leaving only one point with the largest number of crossings (bottom). This marks the critical point $B_c = 1$. Inset of top figure: Binder cumulant for $s = 2$ in the vicinity of $B = 1$ shows an imperfect intersection, i.e. $\frac{\partial U_4(m,B)}{\partial m}$ is a small, nonzero value. A larger version of this inset is shown in Fig.~\ref{fig:binder_cumulant_s2_s5_3075_inset_ising1d}.}
  \label{fig:binder_cumulant_s2_s5_ising1d}
\end{figure}

While $B_c$ of the 1D TFI model is known, it is good practice to extract its value directly from the cumulant data by employing the Binder cumulant. This also serves as practice for systems where the critical point is not known. In finite systems, the Binder cumulant of several system sizes $U_4(L,B)$ are tabulated as a function of a Hamiltonian parameter $B$, and the critical point $B_c$ is read off from the point where $U_4(L,B) \forall L$ cross or intersect each other, which is denoted $U_4(L,B_c)$. In an infinite system described by an iMPS, the length $L$ is replaced by the correlation length $\xi$ up to a factor $s$:
\begin{eqnarray}
L = s\xi,
\label{eqn:L_xi_relation}
\end{eqnarray}
and $\xi$ in turn is related to the bond dimension $m$ through Eq.~\ref{eqn:cor_len_m_kappa}. Thus, just as in the case of finite system sizes, the Binder cumulant of different bond dimensions, denoted $U_4(m,B)$, can be tabulated and the critical point read off where $U_4(m,B) \forall m$ cross or intersect each other. The factor $s$ is treated as a length scaling parameter, and it affects the Binder cumulant by shifting $U_4(m,B)$ at different rates that are dependent on the value of $m$, causing $U_4(m,B)$ to cross and/or intersect each other at different values of $B$. The optimal value $s^*$ in determining the critical point in this work is defined as the value $s$ that gives the crossing or intersection between $U_4(m,B) \forall m$, which is denoted $U_4(m,B_c)$. This will be important where there arises a need to distinguish between $U_4(m,B_c)$ and crossing/intersecting points that are formed between $U_4(m,B)$ of \emph{several} different values of $m$ but not \emph{all} the different values of $m$ - this will be referred to as ``spurious crossing points" and they are disregarded as critical point candidates. Since $s^*$ is defined to occur at a crossing/intersection of $U_4(m,B) \forall m$, it can be obtained by solving $\frac{\partial U_4(m,B)}{\partial m} = 0$ for $s$ with a linear solver over the range of $B$ - this is how $s^*$ is obtained throughout this work. The Binder cumulant for $s = 2$ (top) and $s = s^* = 5.31$ (bottom) are plotted in Fig.~\ref{fig:binder_cumulant_s2_s5_ising1d}, where the latter was obtained using a linear solver. In the top figure, $U_4(m,B) \forall m$ appears to intersect at $B = 1$. However, upon close inspection as shown in the inset of the top figure, this is not a true intersection because $\frac{\partial U_4(m,B)}{\partial m} \neq 0$, instead, it is a small but nonzero number. The fact that this occurs at the known critical point is possibly attributed to the simplicity of the 1D Ising model and further examples of other models will elucidate that this is not generic. Within the region of $1 < B < 1.007$, there are multiple spurious crossing points - each formed from the crossing between \emph{pairs} of different $m$'s, and can be disregarded as candidates of the critical point. By gradually increasing $s$, the Binder cumulant gradually shift at different rates. The overall effect is that $U_4(m,B)$ in the region of $B < 1$ moves upwards while $U_4(m)$ in the region of $B > 1$ moves downwards as can be seen by comparing the top and bottom figures. This causes the intersection point at $B_c$ to gradually change into a crossing point at $B_c$, all while maintaining the value of $B_c = 1$. By tuning $s$ to $s^*$, $U_4(m,B_c)$ is achieved at $B = 1$ which is taken as the critical point $B_c$. Increasing $s$ beyond $s^*$ does not shift the crossing point any further, however, it must be once again stressed that this observation of having a stable critical point when $s \neq s^*$ is special to the 1D Ising model. The latter will be elucidated in the remaining exemplary systems. The key difference of using the Binder cumulant in an iMPS and a finite MPS is the existence of the scaling parameter $s$ in the iMPS. Though it takes an extra step to determine $s = s^*$, this step does not require additional simulation time or computational cost. In fact, the overall simulation time and computational cost is greater in the case of finite MPS since data have to be produced for both $L$ and $m$. The utilization of the linear solver requires a guess input value of $s$. Within a guess range of $s$, the linear solver typically converges to a value of $s^*$ that gives the single crossing point of $U_4(m,B_c) \forall m$. Whereas far outside this range, the linear solver will either not converge or produce a value $s^*$ that does not give the single crossing point of $U_4(m,B)$.

\begin{figure}[h!]
  \centering\includegraphics[width=0.475\textwidth]{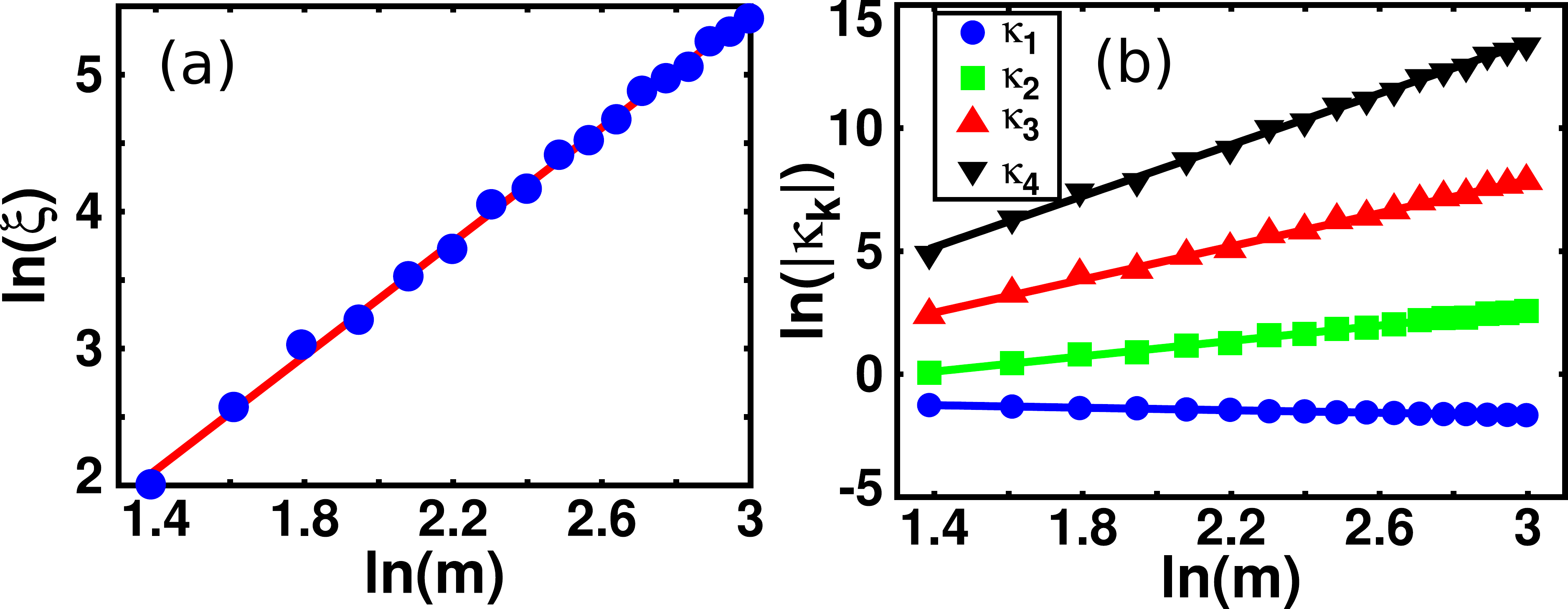}
  \caption{(Colour online) 1D TFI model. Log-log plot of (a) correlation length $\xi$ and (b) cumulants, at the critical point with respect to $m$. The symbols are the data points and the lines are the linear fits. The linear fit's gradient corresponds to the critical exponent.}
  \label{fig:cor_len_cumulant_m_scaling_ising1d}
\end{figure}

Now that $B_c$ has been obtained, the finite-entanglement scaling exponent $\kappa$ and the exponents of the cumulants $\alpha_i$ (where $i = 1,2,3,4$) have to be extracted before the correlation length exponent $\nu$. By taking the logarithm of Eq.~\ref{eqn:cor_len_m_kappa}, $\kappa$ can be extracted as the gradient of the linear equation $\ln \xi = \kappa \ln m$. This is plotted in Fig.~\ref{fig:cor_len_cumulant_m_scaling_ising1d}(a) where the blue circles are the value of the correlation length at the critical point and the red line is the linear fit whose gradient is $\sim 2.083 \pm 0.015$. This is rather close to the previously known value of $\kappa$ which is $\sim 2$ \cite{Tagliacozzo}. In a similar fashion, the cumulants at $B_c$ are also plotted as a function of $m$ to extract their critical exponents $\alpha_i$ in Fig.~\ref{fig:cor_len_cumulant_m_scaling_ising1d}(b). The respective critical exponents are $\alpha_1 = -0.257 \pm 0.004$, $\alpha_2 = 1.558 \pm 0.015$, $\alpha_3 = 3.37 \pm 0.03$ and $\alpha_4 = 5.20 \pm 0.04$. Eq.~\ref{eqn:higher_cumulant_exponent_relation} allows one to check the consistency and accuracy of the cumulant exponents $\alpha_i$'s. Using the obtained values of $\alpha_1$ and $\alpha_2$, one gets $\alpha_3 = -\alpha_1 + 2\alpha_2 = 0.257 + 2(1.558) = 3.373$, which differs from value obtained via linear fit in Fig.~\ref{fig:cor_len_cumulant_m_scaling_ising1d}(b) by $\sim 0.15\%$. This difference stems from the uncertainty in the linear fit used to determine the values of $\alpha_i$ in Fig.~\ref{fig:cor_len_cumulant_m_scaling_ising1d}(b). Similarly, for the fourth cumulant, $\alpha_4 = -2\alpha_1 + 3\alpha_2 = 2(0.257) + 3(1.558) = 5.188$, which differs from the value obtained via linear fit by $\sim 0.25\%$.

\begin{figure}[h!]
  \centering\includegraphics[width=0.475\textwidth]{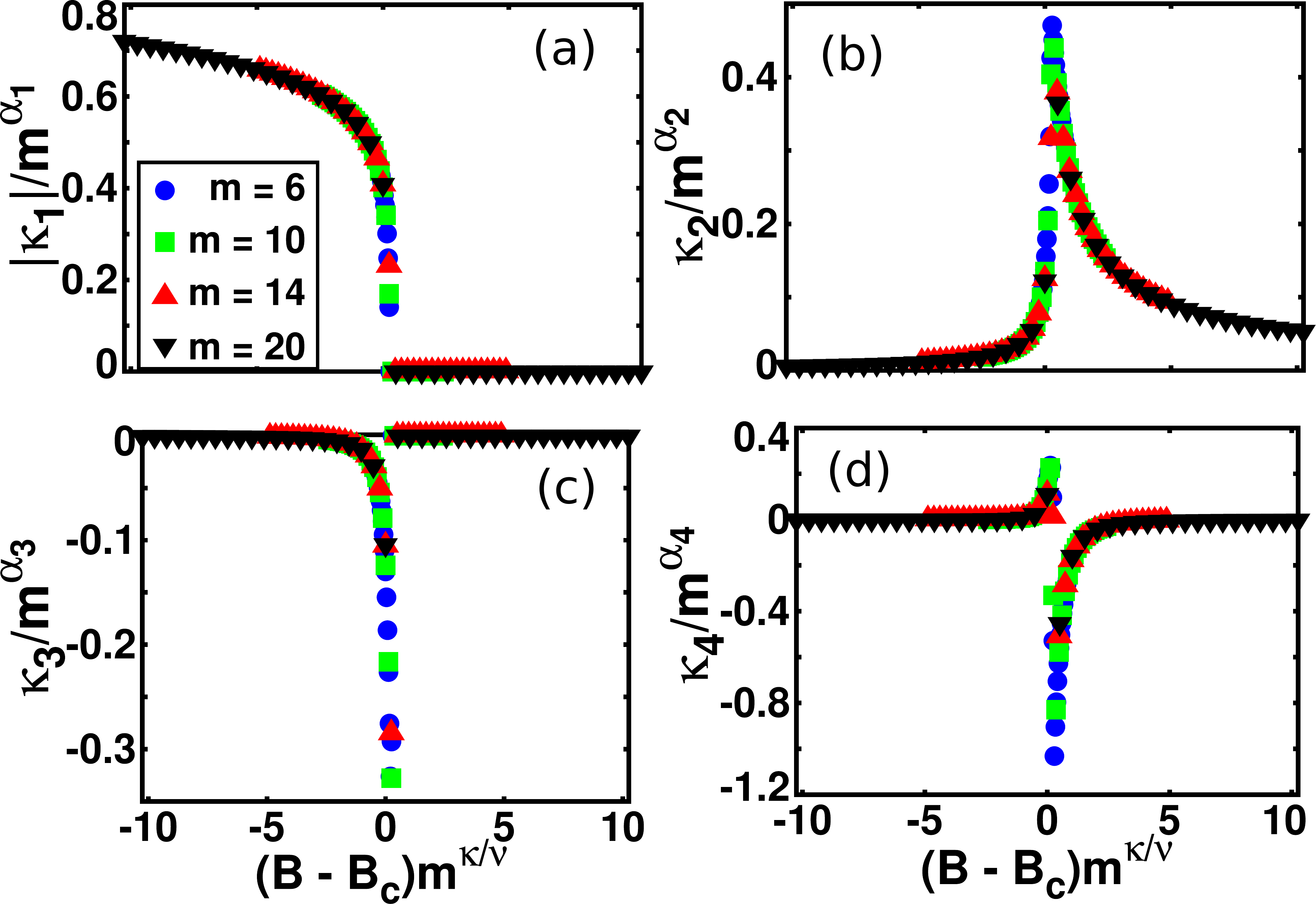}
  \caption{(Colour online) 1D TFI model. Cumulants scaled according to Eq.~\ref{eqn:scaling_function_m} for several values of $m$. (a) The order parameter, (b) variance, (c) skewness and (d) kurtosis. $B_c$, $\kappa$ and $\alpha_i$ used in the scaling function are those obtained from the Binder cumulant and the linear fits of $\ln(\xi)$ and $\ln(\kappa_i)$ versus $\ln(m)$ respectively. $\nu$ is tuned such that the cumulants' sum of residual of the different values of $m$ at $B_c$ is minimized, this gives $\nu = 1.000 \pm 0.005$.}
  \label{fig:cumulants_scale_ising1d}
\end{figure}

The final exponent left to obtain now is $\nu$. By plotting the cumulants according to the scaling function Eqs.~\ref{eqn:m_indep_finite_size_effect} and \ref{eqn:scaling_function_m} with the obtained values of $B_c$, $\kappa$ and the cumulant exponents $\alpha_i$, $\nu$ can be tuned to obtain the best data collapse e.g. by minimizing the cumulants' sum of residual of the different values of $m$ at $B_c$. This is shown in Fig.~\ref{fig:cumulants_scale_ising1d} with the result $\nu = 1.000 \pm 0.005$.

\begin{figure}[h!]
  \centering\includegraphics[width=0.45\textwidth]{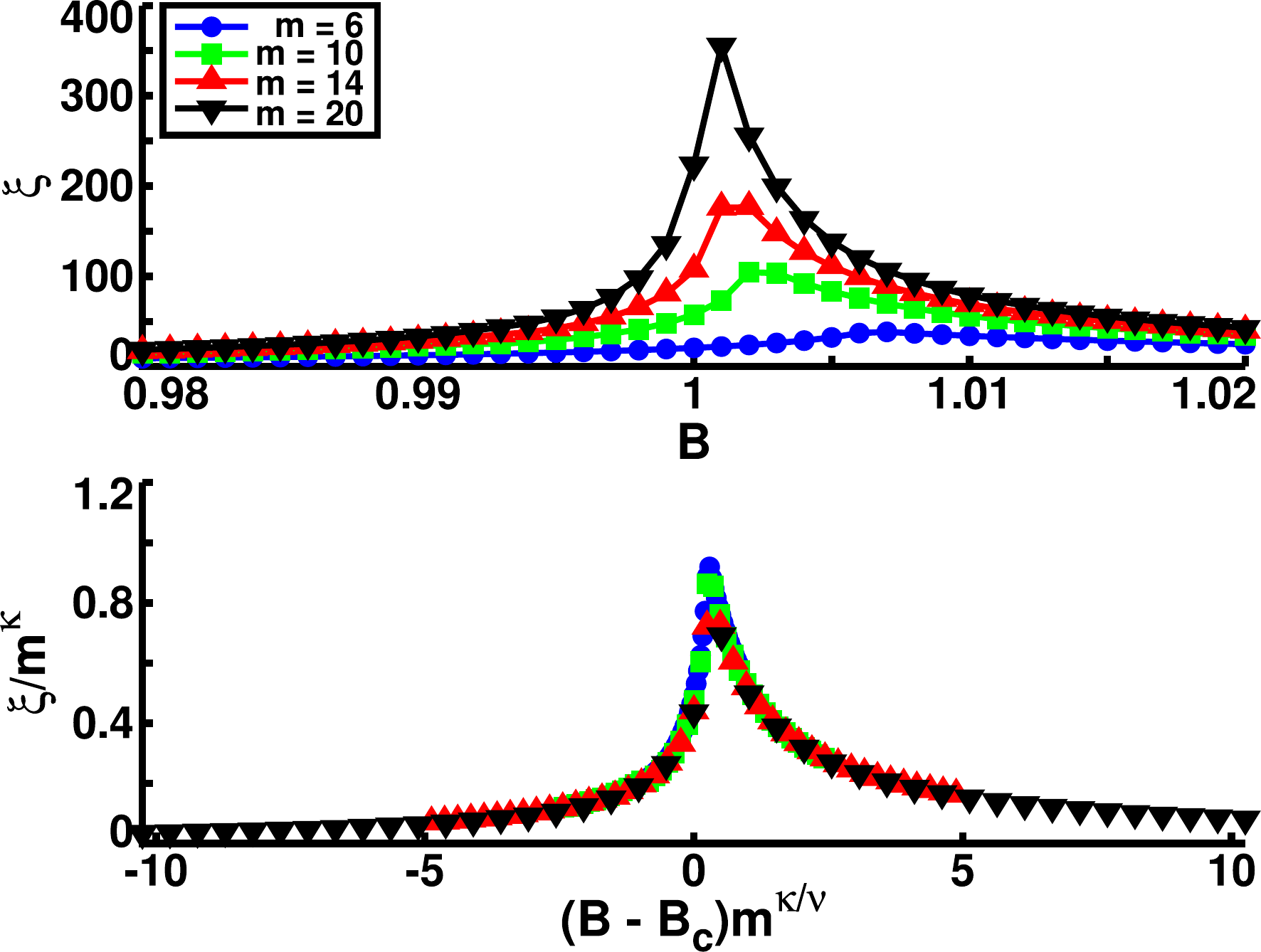}
  \caption{(Colour online) 1D TFI model. Correlation length versus $B$ (top) and correlation length scaled according to Eq.~\ref{eqn:scaling_function_cor_len_m} (bottom) for several values of $m$. $B_c$ and $\kappa$ used in the scaling function are those obtained from the Binder cumulant and the linear fit of $\ln(\xi)$ versus $\ln(m)$ respectively. $\nu$ is tuned such that the correlation length's sum of residual of the different values of $m$ at $B_c$ is minimized, this gives $\nu = 1.000 \pm 0.005$.}
  \label{fig:cor_len_and_cor_len_scale_vs_B_ising1d}
\end{figure}

Using Eqs.~\ref{eqn:m_indep_finite_size_effect} and \ref{eqn:scaling_function_cor_len_m}, the correlation length $\xi$ can also be used to determine $\nu$. Using the obtained values of $B_c$ and $\kappa$, $\nu$ is tuned so that the correlation length's sum of residual of the different values of $m$ at $B_c$ is minimized. This is shown in Fig.~\ref{fig:cor_len_and_cor_len_scale_vs_B_ising1d} with $\nu = 1.000 \pm 0.005$, which is in agreement with that obtained from the data collapse of the cumulants. Eqs.~\ref{eqn:m_indep_finite_size_effect}, \ref{eqn:scaling_function_m} and \ref{eqn:scaling_function_cor_len_m} allows one to additionally fine-tune $B_c$ to obtain a better data collapse. Since the value of $B_c = 1$ gave a sufficiently good data collapse, no further fine-tuning of $B_c$ was needed.

In order to make a connection between the more familiar magnetization order parameter exponent $\beta$, one can compare the term $m^{\alpha_1}$ from the cumulant scaling function Eq.~\ref{eqn:scaling_function_m}, to that of the conventional scaling function of the magnetization order parameter used in iMPS, $m^{-\beta\kappa/\nu}$ \cite{Tagliacozzo}. This comparison implies that $\alpha_1 = -\beta\kappa/\nu$. Using the obtained values of $\alpha_1$, $\kappa$ and $\nu$ gives $\beta = -\alpha_1\nu/\kappa = 0.1234$, which differs from the known value of $\beta = 1/8$ by $\sim 1.3\%$. Proceeding in the same way, the second cumulant's exponent $\alpha_2$ is related to the exponent of the variance $\gamma^*$ by $\alpha_2 = \gamma^*\kappa/\nu$. Using the obtained values of $\alpha_2$, $\kappa$ and $\nu$ gives $\gamma^* = \alpha_2\nu/\kappa = 0.748 \sim 3/4$. It is important to note that the second cumulant here is the variance of the order parameter but it has no direct relation to the susceptibility as in the case of the 2D classical Ising model in a longitudinal field. This is because there is no quantum analog of the fluctuation-dissipation theorem that relates the variance to the susceptibility as in the classical case. In the 2D classical Ising model, the susceptibility/variance exponent is $\gamma = 7/4$ which can be related to the 1D quantum Ising model with an applied \emph{longitudinal} field. The exponent $\gamma^* = 3/4$ obtained in this work is consistent with that found in Ref.~\cite{Pfeuty}.


\subsection{One-dimensional topological Kondo insulator}
\label{result_1d_tki}
\begin{figure}[h!]
  \centering\includegraphics[width=0.46\textwidth]{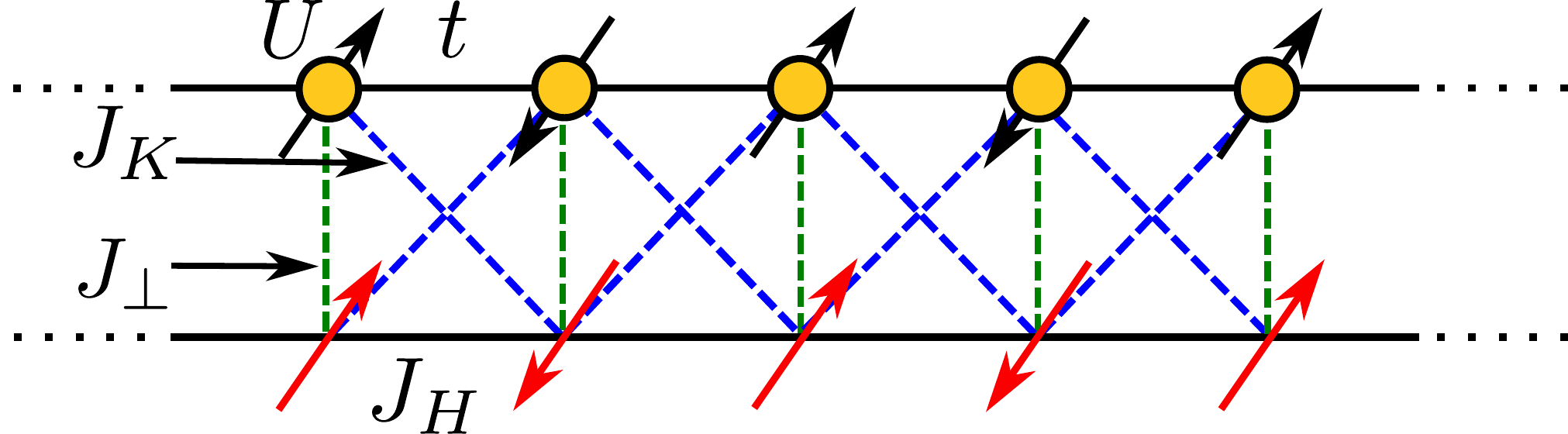}
  \caption{(Colour online) Depiction of a segment of the 1D \textit{p}-wave Kondo-Heisenberg lattice. The lattice consists a Hubbard chain (top) and an $S = \frac{1}{2}$ antiferromagnetic Heisenberg chain (bottom). The non-local Kondo exchange $J_K > 0$ couples a spin $S_j$ at site $j$ in the Heisenberg chain to its nearest-neighbour \textit{p}-wave spin density $\pi_j$ in the Hubbard chain. (Figure taken from Ref.~\cite{Pillay}.)}
  \label{fig:lattice}
\end{figure}

The 1D topological Kondo insulator (TKI) is an effective model introduced in Ref.~\cite{Alexandrov} to study the effects of the strong electron interaction on the topological properties of a 3D bulk insulator. The 1D model consists of a Hubbard chain coupled to a spin-$\frac{1}{2}$ Heisenberg chain by a nonlocal coupling as shown in Fig.~\ref{fig:lattice}. The Hamiltonian of this system is given by
\begin{eqnarray}
H = H_c + H_H + H_K + H_\perp,
\label{eqn_Hamiltonian_tot}
\end{eqnarray}
where
\begin{eqnarray}
H_c &=& -t \sum_{j,\sigma} \left( c^\dagger_{j+1, \sigma} c_{j, \sigma} + c^\dagger_{j, \sigma} c_{j+1, \sigma} \right) \nonumber
\\
&& + U \sum_j n_{j, \uparrow} n_{j, \downarrow}
\label{eqn_Hamiltonian_Hubbard}
\end{eqnarray}
is the 1D Hubbard Hamiltonian (top chain in Fig.~\ref{fig:lattice}) describing fermions hopping between sites $j$ and $j+1$ with amplitude $t$, and a Hubbard repulsion $U$ between fermions of opposite spins at site $j$. The second term
\begin{eqnarray}
H_H = J_H \sum_j \vec{S}_j \cdot \vec{S}_{j+1} ,
\label{eqn_Hamiltonian_Heisenberg}
\end{eqnarray}
is the 1D Heisenberg Hamiltonian (bottom chain in Fig.~\ref{fig:lattice}) that describes the spin exchange between nearest-neighbour $S = \frac{1}{2}$ localized spins. The third term $H_K$ represents a non-local Kondo coupling between the Heisenberg and Hubbard chains:
\begin{eqnarray}
H_K = J_K \sum_j \left[ \frac{1}{2} \left( S^+_j \pi^-_j + S^-_j \pi^+_j \right) + S^z_j \pi^z_j \right] ,
\label{eqn_Hamiltonian_Kondo}
\end{eqnarray}
where $S^\pm_j$ and $\pi^\pm_j$ ($S^z_j$ and $\pi^z_j$) are the ladder operators ($z$ components) of the spin $\vec{S}_j$ in the Heisenberg chain and the \textit{p}-wave spin density $\vec{\pi}_j$ in the Hubbard chain. The $\vec{\pi}_j$ operator is given as
\begin{eqnarray}
\vec{\pi}_j = \frac{1}{2} \sum_{\alpha, \beta} p^\dagger_{j,\alpha} \vec{\sigma}_{\alpha,\beta} p_{j,\beta},
\label{eqn_pi_operator}
\end{eqnarray}
where $\vec{\sigma}$ is the vector of Pauli matrices, and
\begin{eqnarray}
p_{j,\sigma} = \frac{1}{\sqrt{2}} \left( c_{j+1,\sigma} - c_{j-1,\sigma} \right).
\label{eqn_p_operator}
\end{eqnarray}
The last term in Eq.~(\ref{eqn_Hamiltonian_tot}) is the conventional Kondo coupling between a fermion and a localized spin at site $j$,
\begin{eqnarray}
H_\perp = J_\perp \sum_j \left[ \frac{1}{2} \left( S^+_j s^-_j + S^-_j s^+_j \right) + S^z_j s^z_j \right].
\label{eqn_Hamiltonian_Kondo_local}
\end{eqnarray}
In this work, the system is set at half-filling and Hamiltonian parameters that are held fixed are $U = 0$, $t = J_H = 1$, and $J_K = 2$. The iMPS ground state generated here is enforced with $U(1)$ particle number symmetry to conserve particle number and SU(2) spin rotation symmetry.

\begin{figure}[h!]
  \centering\includegraphics[width=0.475\textwidth]{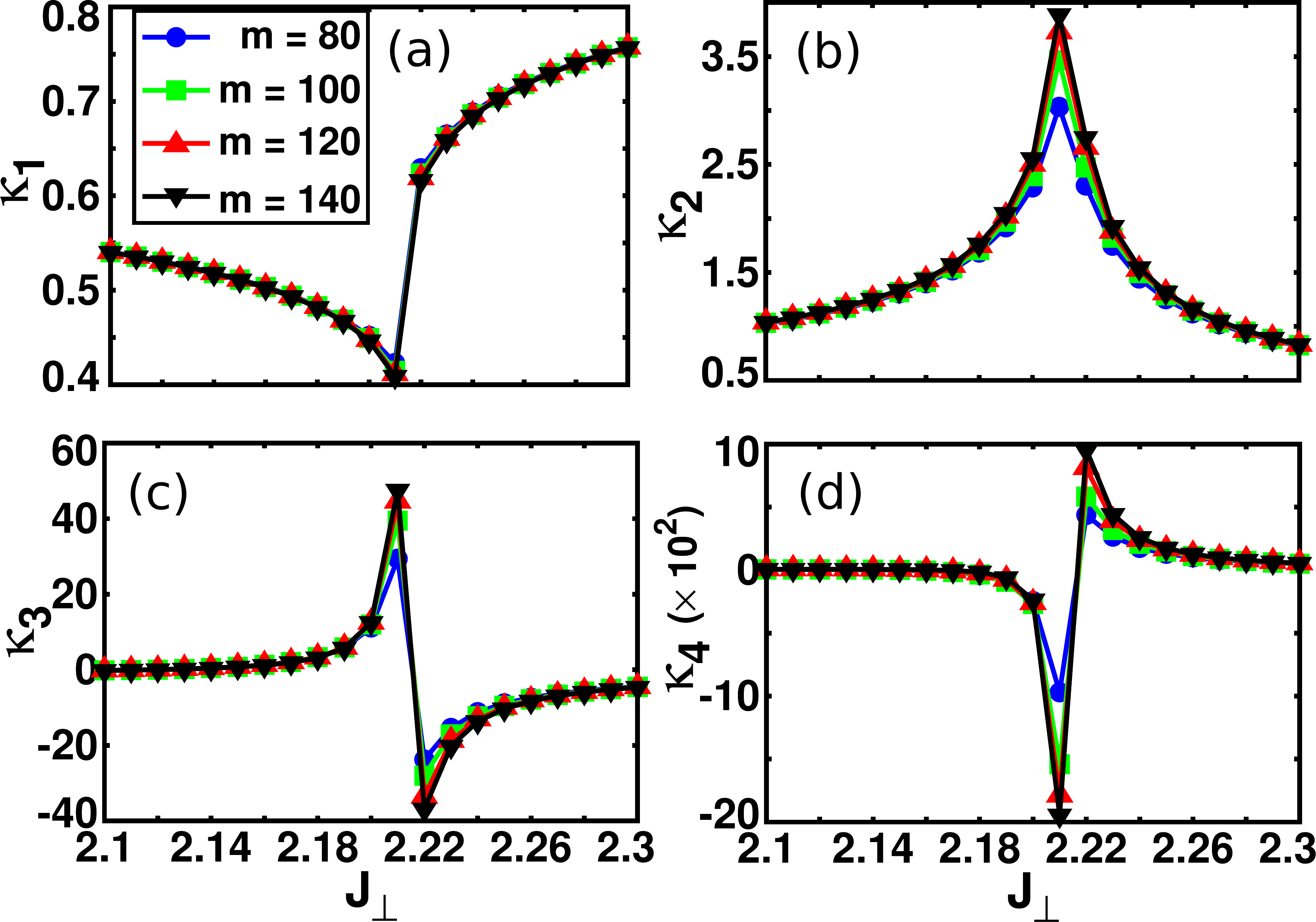}
  \caption{(Colour online) 1D TKI. First four cumulants of the string order parameter $O^2_\text{string}$ as a function of $J_\perp$ for several values of $m$. (a) First cumulant $\kappa_1$ is the order parameter which is zero only when the charge gap vanishes. (b) Second cumulant $\kappa_2$ is the variance of the order parameter, (c) the third cumulant $\kappa_3$ and (d) the fourth cumulant $\kappa_4$ all diverge at the critical point. The cumulants do not shift horizontally with increasing $m$, signifying the critical point is located in a narrow region of parameter space.}
  \label{fig:cumulants_tki}
\end{figure}

When $J_\perp < J^c_\perp$, the ground state is in a symmetry-protected topological (SPT) phase protected by inversion symmetry and undergoes a topological phase transition into a topologically trivial state consisting of local Kondo singlets when $J_\perp > J^c_\perp$ \cite{Pillay}. This phase transition occurs with the vanishing of the charge excitation gap while the spin excitation gap remains nonzero and finite. This coincides with the presence of a spinless two-particle excitation which is detected with the string order parameter
\begin{eqnarray}
O^2_\text{string} = \lim_{|j-k| \rightarrow \infty} \left\langle  \mathbbm{1}_j \text{exp} \left[ \frac{i \pi}{2} \sum^k_{l=j} \left( \hat{n}_l - 1 \right) \right] \mathbbm{1}_k \right\rangle ,
\label{eqn:string_order_parameter}
\end{eqnarray}
where $\hat{n} = \sum_\sigma c^\dagger_{l,\sigma} c^{l,\sigma}$, which is shown in the Fig.~\ref{fig:cumulants_tki}(a) as a function of $J_\perp$. As shown in Ref.~\cite{Pillay}, $O^2_\text{string}$ can be expressed as a correlation function $O^2_\text{string} = \lim_{|j-k| \rightarrow \infty} \braket{p(j)p(k)}$, where $p(j) = \prod_{i<j} (-1)^{\frac{n_i - 1}{2}}$ is the ``kink operator", and $\braket{p(j)p(k)} = \left\langle \prod_{i<j} \mathbbm{1}_i \prod_{i=j}^k (-1)^\frac{n_i-1}{2} \prod_{i>k} \mathbbm{1}_i \right\rangle$. $O^2_\text{string}$ thus appears similar to a local order parameter, e.g., $\braket{M^2}$, where $M$ is the magnetization operator. The calculation of $O^2_\text{string}$ is done by constructing an extensive order parameter $P = \sum_i p(i)$,  which is written in the matrix product operator (MPO) form as
\begin{eqnarray}
P_i = \left[ \begin{array}{cc} (-1)^\frac{n_i-1}{2} & I \\
								0 & I \end{array} \right] .
\end{eqnarray}
The square $\braket{P^2}$ is then taken and related to $O^2_\text{string}$ via $O^2_\text{string} = \braket{P^2}/L^2$, where $L = b\xi$, $\xi$ is the correlation length and $b$ is a scaling factor. As shown in Section \ref{higher_order_moments}, the expectation value of an $n$th power of an operator $P^n$ in an iMPS is obtained as a degree $n$ polynomial of the lattice size $L$, which is exact in the asymptotic large-$L$ limit. Thus $O^2_\text{string} = \braket{P^2}/L^2$ is evaluated as the coefficient of the degree 2 component of $\braket{P^2}$. The example of the evaluation of this second degree polynomial is shown in Section \ref{appendix_variance}.

Fig.~\ref{fig:cumulants_tki} shows the cumulants of $O^2_\text{string}$ as a function of $J_\perp$. From these figures, $O^2_\text{string}$ vanishes, while its variance, skewness and curtosis diverge at $J^c_\perp \sim 2.21$, consistent with previous results of $J^c_\perp = 2.214$ \cite{Pillay}. Unlike the case of the 1D TFI model where the magnetization order parameter $\braket{M}$ is finite in the ordered phase and zero in the disordered phase, $O^2_\text{string}$ remains finite in both SPT and topologically trivial phases and only vanishes at the critical point. This happens because $O^2_\text{string}$ is not an order parameter in the traditional sense where it discerns an ordered phase from a disordered one. On the contrary, $O^2_\text{string}$ was constructed specifically to detect a spinless even number-particle excitation. In the TKI model, this excitation is gapped in both phases but vanishes at the critical point. Compared to the cumulants of the 1D TFI model, the cumulants of the TKI in Fig.~\ref{fig:cumulants_tki} appear to only shift vertically and not horizontally with increasing $m$. This occurs because the critical point is located within a range of $J_\perp$ that is much narrower than the range of $J_\perp$ shown in Fig.~\ref{fig:cumulants_tki}. One can thus expect to see this horizontal shift if one were to use a finer $J_\perp$ grid and zoom-in closer around $J_\perp \approx 2.20-2.22$.

\begin{figure}[h!]
  \centering\includegraphics[width=0.45\textwidth]{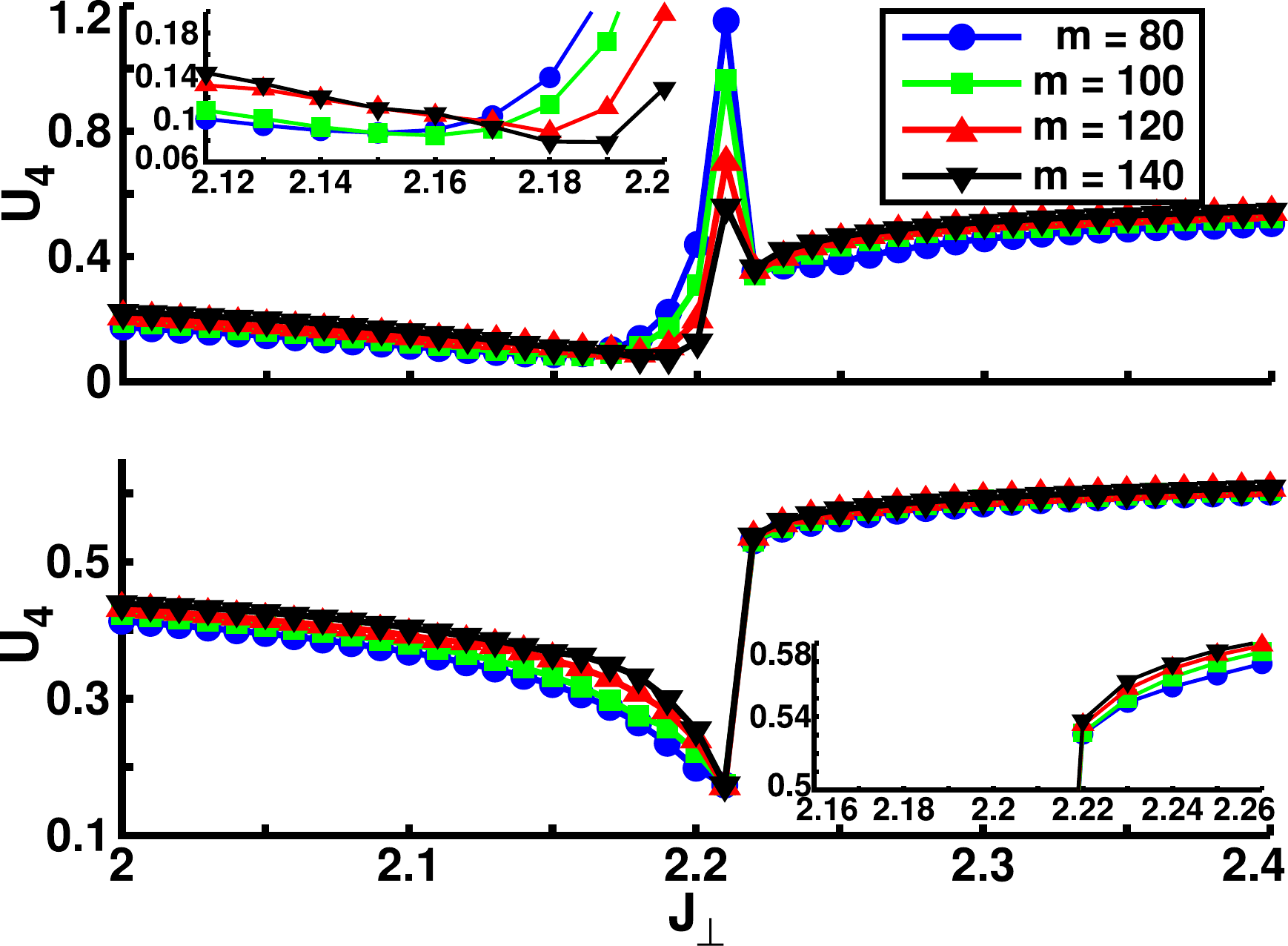}
  \caption{(Colour online) 1D TKI. The Binder cumulant with $s = 2$ (top) and $s = s^* = 5.29$ (bottom) as a function of $J_\perp$ for several values of $m$. An order parameter that does not vanish in one of the two phases it separates can still be used in the Binder cumulant to locate the critical point by finding the intersecting point of $U_4(m,J_\perp)$ for the different values of $m$. This can be seen in the bottom figure at $J_\perp = 2.21$. Inset of top figure: Binder cumulant in the vicinity of $J_\perp = 2.16$ shows that not all $U_4(m,J_\perp)$ of the different values of $m$ cross simultaneously, thus it is a  spurious crossing point. Inset of bottom figure: Closeup of $J_\perp = 2.22$ reveals that it is not an intersection between the different values of $m$. A larger version of these two insets are shown in Figs.~\ref{fig:binder_cumulant_s2_s5_2896_a_inset_tki} and \ref{fig:binder_cumulant_s2_s5_2896_c_inset_tki}, respectively.}
  \label{fig:binder_cumulants_s2_s5_tki}
\end{figure}

Even though $O^2_\text{string}$ does not vanish in one of the phases being separated by the critical point, it can still be used in the Binder cumulant to locate the critical point. The only drawback to this is that the Binder cumulant formed out of $O^2_\text{string}$ does not form a crossing point between the different values of $m$ for any $s$. Nonetheless, an intersecting point between the different $m$'s still forms and this suffices in locating the critical point. To illustrate this, $U_4(m,J_\perp)$ is tabulated in Fig.~\ref{fig:binder_cumulants_s2_s5_tki} for $s = 2$ (top) and $s = s^* = 5.29$ (bottom), where the latter was obtained using a linear solver. While it is noticeable that the top figure appears to have a crossing point at $J_\perp \sim 2.16$, this point is far from the critical point that one would expect from observing the cumulants in Fig.~\ref{fig:cumulants_tki}. Upon close inspection of the vicinity $J_\perp \sim 2.16$, one sees that this is in fact not a good crossing point since not all the different values $m$ cross each other simultaneously as shown in the inset of the top figure. The peak however coincides with the critical point $J_\perp^c = 2.21$, but the separation between the different $m$'s does not make it a good indicator of the critical point. By increasing $s$, this peak gradually vanishes as can be seen by comparing the top and bottom figures. When $s = s^*$, as in the bottom figure, $U_4(m,J_\perp) \forall m$ intersect at $J_\perp = 2.21$, which is taken as the critical point $J_\perp^c$, in agreement with Ref.~\cite{Pillay}. Since $s = s^*$, this intersection point satisfies $\frac{\partial U_4(m,B)}{\partial m} = 0$, which is unlike the top figure of Fig.~\ref{fig:binder_cumulant_s2_s5_ising1d} of the 1D TFI model where the point at $B = 1$ was not a true intersection. The point $J_\perp = 2.22$ appears to be an intersection point, but upon close inspection, it is not since at that point $\frac{\partial U_4(m,B)}{\partial m} \neq 0$ as can be seen in the inset of the bottom figure.

\begin{figure}[h!]
  \centering\includegraphics[width=0.475\textwidth]{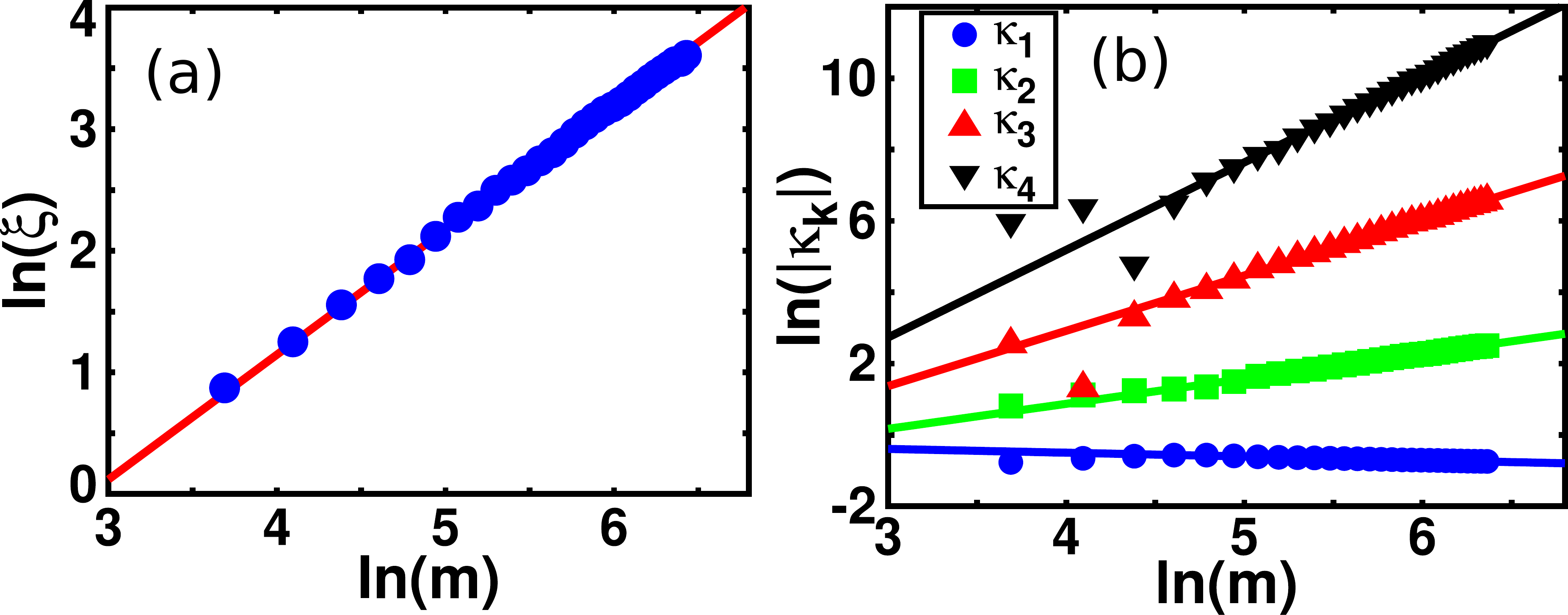}
  \caption{(Colour online) 1D TKI. Log-log plot of the (a) charge excitation correlation length $\xi$ and (b) cumulants $\kappa_i$, at the critical point $J_\perp = 2.21$ with respect to $m$. The symbols are the data points and the lines are the linear fit. The gradient of the linear fit is equal to the critical exponent.}
  \label{fig:cor_len_cumulant_m_scaling_tki}
\end{figure}

Using the critical point $J_\perp^c = 2.21$ obtained from the Binder cumulant, the critical exponents $\kappa$ and $\alpha_i$ are obtained by scaling the correlation length of the charge excitation $\xi$ and the cumulants of $O^2_\text{string}$ with respect to $m$ as shown in Figs.~\ref{fig:cor_len_cumulant_m_scaling_tki}(a) and \ref{fig:cor_len_cumulant_m_scaling_tki}(b) respectively. This gives $\kappa \sim 1.027 \pm 0.005$, $\alpha_1 = -0.1056 \pm 0.0005$, $\alpha_2 = 0.701 \pm 0.003$, $\alpha_3 = 1.552 \pm 0.008$ and $\alpha_4 = 2.46 \pm 0.01$. The central charge $c$ of this phase transition has been shown to be $c \sim 1$ \cite{Pillay} which was obtained through the relation of the entanglement entropy $S$ and $\xi$ given by \cite{Calabrese}
\begin{eqnarray}
S = \frac{c}{6} \ln \xi.
\label{eqn:entropy_central_charge_xi}
\end{eqnarray}
The gradient of the linear plot of $S$ versus $\ln \xi$ is thus equal to $c/6$ which can be obtained by a linear fit through the data. Alternatively, Eq.~\ref{eqn:kappa_central_charge} can also be used to determine $\kappa$ from the handful of known values of $c$ and vice versa. Using $c = 1$, one obtains $\kappa = 1.3441$, which significantly differs from the value obtained from the linear fit in Fig.~\ref{fig:cor_len_cumulant_m_scaling_tki}(a) by $24\%$. This demonstrates the difficulty and inaccuracy in determining $c$ from $\kappa$. The cumulant exponents obtained from the linear fits in Fig.~\ref{fig:cor_len_cumulant_m_scaling_tki}(b) can once again be checked using Eq.~\ref{eqn:higher_cumulant_exponent_relation}. Using the obtained values of $\alpha_1$ and $\alpha_2$, one gets $\alpha_3 = -\alpha_1 + 2\alpha_2 = 0.106 + 2(0.701) = 1.508$, which differs from value obtained via linear fit in Fig.~\ref{fig:cor_len_cumulant_m_scaling_tki}(b) by $\sim 2.9\%$. Similarly, $\alpha_4 = -2\alpha_1 + 3\alpha_2 = 2(0.106) + 3(0.701) = 2.315$, which differs from the value obtained via linear fit by $\sim 5.6\%$. As before, this difference of the exponents stem from the uncertainty of the linear fit used. In addition to that, the larger contributor to this difference is the narrow region of $J_\perp$ parameter space that the critical point is located in. This affects the chosen value of the critical point $J_\perp^c$ to which $\xi$ and $\kappa_i$ have their respective exponents extracted from. In other words, one would get a smaller difference between the calculated value cumulant exponent and that obtained directly from the linear fit if a finer grid of $J_\perp$ was used in detecting the critical point, or if the critical point $J_\perp^c = 2.214$ obtained in \cite{Pillay} was used directly. The inability to use a finer grid size to zoom-in closer to the location of $J^c_\perp$ is due to numerical instabilities that plague the cumulant data when the grid size is smaller than $5\times 10^{-4}$ for this particular model.

\begin{figure}[h!]
  \centering\includegraphics[width=0.475\textwidth]{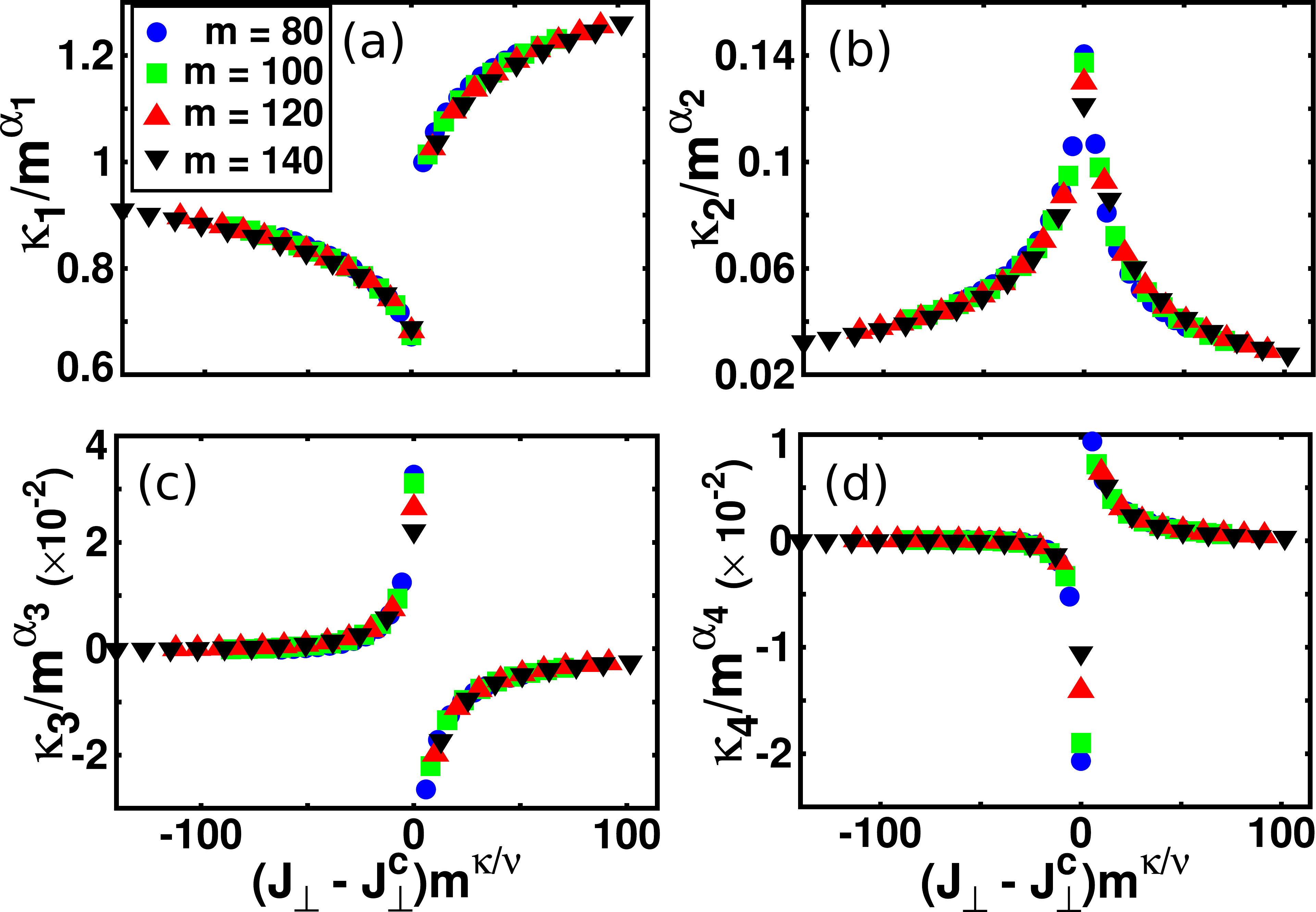}
  \caption{(Colour online) 1D TKI. Cumulants scaled according to Eq.~\ref{eqn:scaling_function_m} for several values of $m$. (a) The order parameter, (b) variance, (c) skewness and (d) kurtosis. $J_\perp^c$, $\kappa$ and $\alpha_i$ used in the scaling function are those obtained from the Binder cumulant and the linear fits of $\ln(\xi)$ and $\ln(\kappa_i)$ versus $\ln(m)$ respectively. $\nu$ is tuned such that the cumulants' sum of residual of the different values of $m$ at $J^c_\perp$ is minimized, this gives $\nu = 0.710 \pm 0.001$.}
  \label{fig:cumulants_scale_tki}
\end{figure}

\begin{figure}[h!]
  \centering\includegraphics[width=0.45\textwidth]{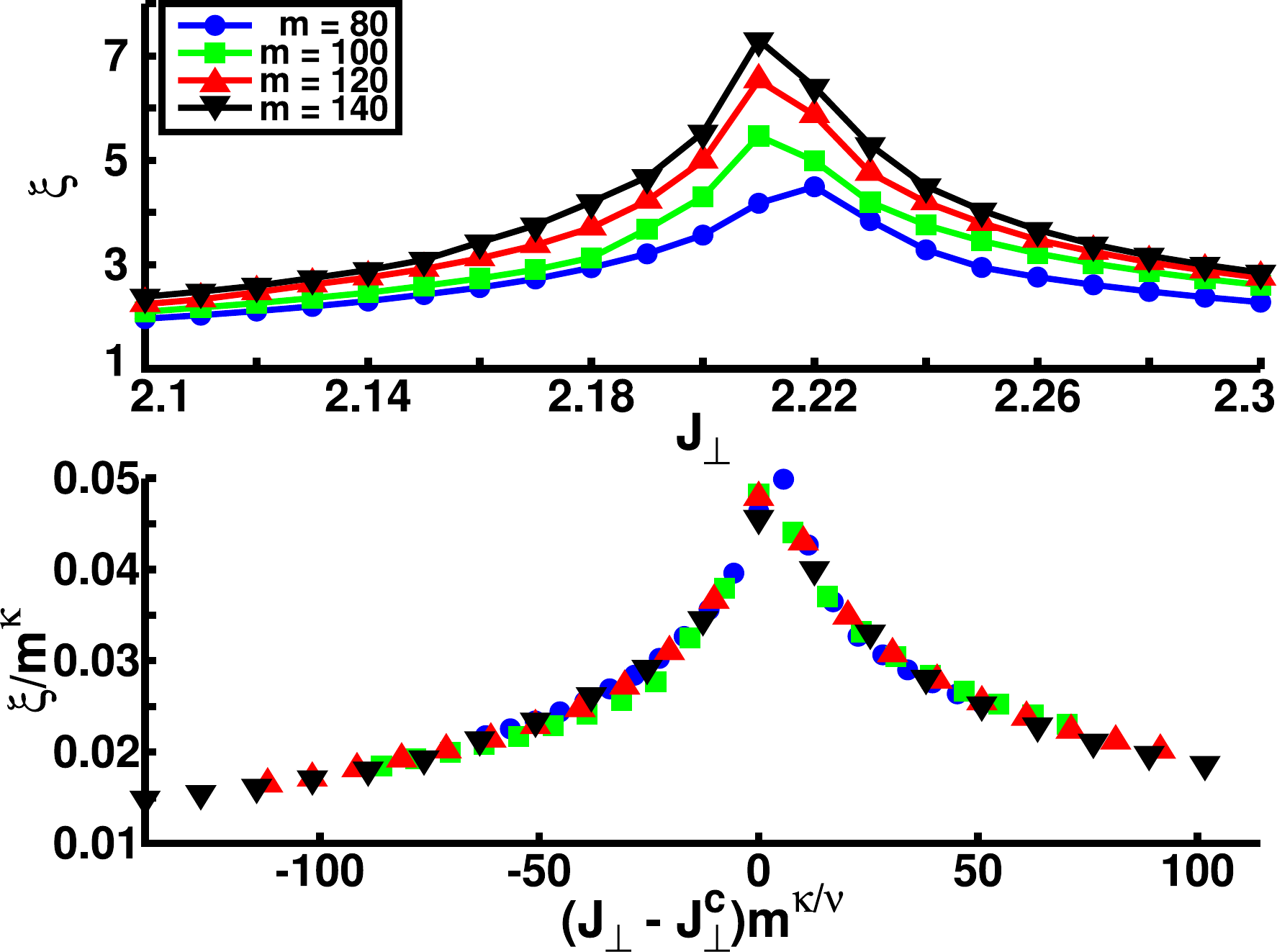}
  \caption{(Colour online) 1D TKI. Charge excitation correlation length versus $J_\perp$ (top) and charge excitation correlation length scaled according to Eq.~\ref{eqn:scaling_function_cor_len_m} (bottom) for several values of $m$. $J_\perp^c$ and $\kappa$ used in the scaling function are those obtained from the Binder cumulant and the linear fit of $\ln(\xi)$ versus $\ln(m)$ respectively. $\nu$ is tuned such that the correlation length's sum of residual of the different values of $m$ at $J^c_\perp$ is minimized, this gives $\nu = 0.710 \pm 0.001$.}
  \label{fig:cor_len_and_cor_len_scale_vs_B_tki}
\end{figure}

Using the exponents $\kappa$ and $\alpha_i$ obtained from the linear fits in Fig.~\ref{fig:cor_len_cumulant_m_scaling_tki}(a) and \ref{fig:cor_len_cumulant_m_scaling_tki}(b), and the respective cumulant and correlation length scaling functions Eqs.~\ref{eqn:scaling_function_m} and \ref{eqn:scaling_function_cor_len_m}, the value of the correlation length's critical exponent $\nu$ can be obtained. As before, this is done by tuning $\nu$ such that the respective sum of residual of the cumulants and correlation length of different values of $m$ at $J^c_\perp$ are minimized. This gives the data collapse plotted in Figs.~\ref{fig:cumulants_scale_tki} and \ref{fig:cor_len_and_cor_len_scale_vs_B_tki} with the obtained value $\nu = 0.710 \pm 0.001$. This value of $\nu$ is within the range of previously obtained values in Ref.~\cite{Pillay} of $\nu^- = 0.666$ and $\nu^+ = 0.742$, where $\nu^-$ ($\nu^+$) is the exponent obtained from fitting from below (above) the critical point.

\subsection{$S = 1$ Heisenberg chain with single-ion anisotropy}
\label{result_single_ion}
The $S = 1$ Heisenberg chain is a minimal quantum-magnetic model that demonstrates Haldane's conjecture of the existence of a spectrum gap in integer-spin antiferromagnetic chains \cite{Haldane1, Haldane2}. The Hamiltonian of this model is
\begin{eqnarray}
H &=& \sum_i J \left( S^x_i S^x_{i+1} + S^y_i S^y_{i+1} \right) + J_z S^z_i S^z_{i+1}  \nonumber \\
&& + D \sum_i \left( S^z_i \right)^2 ,
\label{eqn:Hamiltonian_single_ion}
\end{eqnarray}
where $D$ is the single-ion anisotropy term. This model is known to have several phases in the $(D, J_z)$ parameter space, namely N\'{e}el, Haldane, large-$D$, ferromagnetic and two XY phases \cite{Chen1, Chen2, Degli}.

In this work, the Hamiltonian parameters held fixed are $J = J_z = 1$. This reduces the available phases to three. When $D = 0$, the ground state is in the N\'{e}el phase which possesses a $Z_2$ symmetry. When $D < J$, it is in the Haldane phase with an incomplete $Z_2 \times Z_2$ symmetry. This phase is known to be an SPT phase protected by time-reversal ($\mathcal{T}$) or inversion ($\mathcal{I}$) symmetry (dihedral ($\mathcal{D}_2$) symmetry is broken by the single-ion anisotropy). When $D > J$, the ground state is in a topologically trivial large-$D$ phase. The transition point between the Haldane and large-$D$ phases has been determined to high precision to be $D_c/J = 0.96845(8)$ via finite DMRG with system size up to $L = 10000$ and $m = 1000$ \cite{Hu}.

\begin{figure}[h!]
  \centering\includegraphics[width=0.475\textwidth]{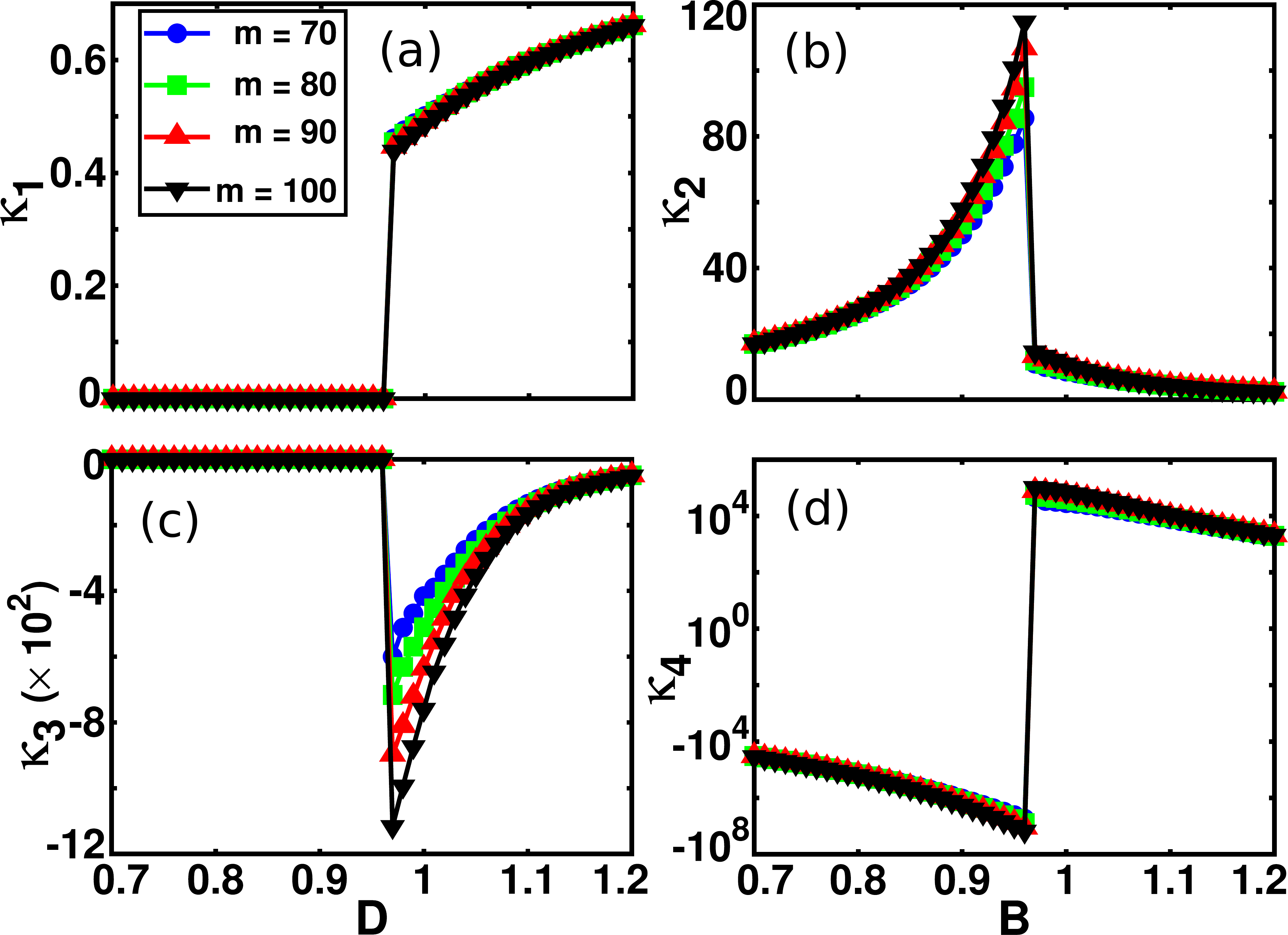}
  \caption{(Colour online) $S = 1$ Heisenberg chain with single-ion anisotropy. First four cumulants of the string order parameter $O^2_\mathcal{T}$ as a function of $D$ for several values of $m$. (a) First cumulant $\kappa_1$ is the order parameter. (b) Second cumulant $\kappa_2$ is the variance of the order parameter, which diverges at the critical point. (c) Third cumulant $\kappa_3$ is the skewness. (d) Fourth cumulant $\kappa_4$ is the kurtosis. The cumulants do not shift with increasing $m$, signifying the critical point is located in a narrow region of parameter space.}
  \label{fig:cumulants_single_ion}
\end{figure}

The key distinction between the two nonzero $D$ phases is how their ground states transform under the symmetry operation of the above mentioned symmetries. To appreciate this, one has to first look at how the iMPS transforms under symmetry operations. When an iMPS Eq.~\ref{eqn:mps} is in its canonical form, each local tensor $A_j$ can be written as a product of $m \times m$ complex matrices $\Gamma_j$ and positive, real, diagonal matrices $\Lambda$ \cite{Vidal2} which satisfies the canonical condition
\begin{eqnarray}
\sum_j \Gamma_j^\dagger \Lambda^2 \Gamma_j = \mathbbm{1}.
\label{eqn:imps_canonical_condition}
\end{eqnarray}
In the canonical form, the transfer matrix
\begin{eqnarray}
T_{\alpha \alpha' ; \beta \beta'} = \sum_j \Gamma_{j,\alpha \beta} \left( \Gamma_{j,\alpha' \beta'} \right)^* \Lambda_\beta \Lambda_{\beta'}
\label{eqn:transfer_matrix_right_eigenvec}
\end{eqnarray}
has a right eigenvector $\delta_{\beta \beta'}$ with eigenvalue 1. Similarly, the transfer matrix
\begin{eqnarray}
\tilde{T}_{\alpha \alpha' ; \beta \beta'} = \sum_j \Lambda_\alpha \Lambda_{\alpha'} \Gamma_{j,\alpha \beta} \left( \Gamma_{j,\alpha' \beta'} \right)^*
\label{eqn:transfer_matrix_left_eigenvec}
\end{eqnarray}
has a left eigenvector $\delta_{\alpha \alpha'}$ with eigenvalue 1. If this iMPS is invariant under a local symmetry $g \in G$ which is represented in the local basis as a unitary matrix $u_g$, then the $\Gamma_j$ matrices must transform under $\left(u_g\right)_{jj'}$ such that the product in Eq.~\ref{eqn:mps} does not change (up to a phase). This means that the transformed matrices satisfy \cite{Garcia}
\begin{eqnarray}
\sum_{j'} \left(u_g\right)_{jj'} \Gamma_{j'} = e^{i\theta_g} U_g^\dagger \Gamma_j U_g,
\label{eqn:imps_symmetry}
\end{eqnarray}
where $e^{i\theta_g}$ is a phase factor and $U_g$ is a unitary matrix that commutes with the $\Lambda$ matrices. $U_g$ forms an $m$-dimensional projective representation of the symmetry group
\begin{eqnarray}
U_g U_h = e^{\rho \left(g,h\right)} U_{g,h},
\end{eqnarray}
where $\rho\left(g,h\right)$ is the factor set of the representation which can be used to differentiate an SPT phase from a trivial one. Taking time-reversal symmetry as an example, $u_\mathcal{T} = U_\mathcal{T} K$, where $U_\mathcal{T}$ is a basis-dependent unitary e.g. $U_\mathcal{T} = e^{i\pi S^y}$ for the spin basis, and $K$ is the complex conjugation operation. As such, $\Gamma_j$ transforms as
\begin{eqnarray}
\Gamma_j^* = e^{i\theta_\mathcal{T}} U_\mathcal{T}^\dagger \Gamma_j U_\mathcal{T}.
\end{eqnarray}
Relating this to the canonical condition Eq.~\ref{eqn:imps_canonical_condition}, one finds
\begin{eqnarray}
\sum_j \Gamma_j^\dagger \Lambda U_\mathcal{T} U_\mathcal{T}^* \Lambda \Gamma_j = U_\mathcal{T} U_\mathcal{T}^*.
\end{eqnarray}
Thus $U_\mathcal{T} U_\mathcal{T}^*$ is an eigenvector of the transfer matrix $T$ Eq.~\ref{eqn:transfer_matrix_right_eigenvec} with eigenvalue 1. Since the only unimodular eigenvalue of $T$ is unity and this eigenvalue is unique, one gets $U_\mathcal{T} U_\mathcal{T}^* = e^{i\phi_\mathcal{T}}$. Using the unitary property $\mathbbm{1} = U_\mathcal{T} U_\mathcal{T}^\dagger$ and its transpose $\mathbbm{1} = \left( U_\mathcal{T} U_\mathcal{T}^\dagger \right)^T = U_\mathcal{T}^* U_\mathcal{T}^T$, one can eliminate the $U_\mathcal{T}$'s in $U_\mathcal{T} U_\mathcal{T}^* = e^{i\phi_\mathcal{T}}$ to get $e^{-2 i\phi_\mathcal{T}} = 1$. The latter sets the restriction $\phi_\mathcal{T} = 0, \pi$. If $\phi_\mathcal{T} = \pi$, then $U_\mathcal{T}$ is an antisymmetric matrix which causes the entanglement spectrum to be strictly even-fold degenerate i.e. the ground state is in an SPT phase \cite{Pollmann2}. Whereas if $\phi_\mathcal{T} = 0$, $U_\mathcal{T}$ is symmetric and there is no restriction on the degeneracy of the entanglement spectrum i.e. the ground state is topologically trivial. $U_\mathcal{T} U_\mathcal{T}^* = \pm 1$ thus acts as a tool to distinguish the SPT phase from the trivial one. However, in order to evaluate the cumulants, one would need an order parameter that differentiates the two phases based on their phase $e^{i\phi_\mathcal{T}}$. This can be achieved through a nonlocal order parameter \cite{Pollmann3}
\begin{eqnarray}
O^2_\mathcal{T} = \lim_{|j-k| \rightarrow \infty} \left\langle \mathbbm{1}_j \prod^k_{l=j} K_l \text{exp} \left( i \pi S^y_l \right) \mathbbm{1}_k \right\rangle.
\label{eqn:string_order_parameter2}
\end{eqnarray}
The complex conjugation operator $K_l$ acts on the local MPS tensor at site $l$ by complex-conjugating it. Unlike $U_\mathcal{T} U_\mathcal{T}^*$ which obtains the phase of the projective representation from the ancillary states of the iMPS, $O^2_\mathcal{T}$ obtains it from the physical degrees of freedom. Analogous to the evaluation of $O^2_\text{string}$ Eq.~\ref{eqn:string_order_parameter}, $O^2_\mathcal{T}$ is evaluated as the coefficient of the degree 2 component of $\braket{P^2}$, where $P = \sum_i p_\tau(i)$ but with the kink operator $p_\tau(j) = \prod_{i<j} u_\mathcal{T}(i)$. The MPO form of $P$ is given as
\begin{eqnarray}
P_i = \left[ \begin{array}{cc} e^{i\pi S_y}K & I \\
								0 & I \end{array} \right] .
\end{eqnarray}
Following the derivation in the Section \ref{appendix_variance}, this results in $O^2_\mathcal{T} = \text{Tr} \left( U_\mathcal{T} \rho_R \right)$, where $\rho_R$ is the reduced density matrix of the right bipartition. In the limit where $|j-k| \rightarrow \infty$, $U_\mathcal{T}$ is the left eigenvector of the generalized transfer matrix corresponding to $\mathcal{T}$:
\begin{eqnarray}
T^{u_\mathcal{T}}_{\alpha \alpha' ; \beta \beta'} &=& \sum_j \left( \sum_{j'} (u_\mathcal{T})_{jj'} \tilde{\Gamma}_{j',\alpha \beta} \right) \nonumber \\
&& \times \left( \Gamma_{j, \alpha' \beta'} \right)^* \Lambda_\beta \Lambda_{\beta'} ,
\end{eqnarray}
with eigenvalue 1. Since $\rho_R$ is chosen from the SVD procedure to be a diagonal matrix, and that $U_\mathcal{T}$ is antisymmetric in the SPT phase, the diagonal elements of the product $U_\mathcal{T} \rho_R$ is zero, ergo $O^2_\mathcal{T} = \text{Tr} \left( U_\mathcal{T} \rho_R \right) = 0$. In the trivial phase, $U_\mathcal{T}$ is symmetric. Hence the diagonal elements of $U_\mathcal{T} \rho_R$ is nonzero, resulting in  $O^2_\mathcal{T} = \text{Tr} \left( U_\mathcal{T} \rho_R \right) \neq 0$. Fig.~\ref{fig:cumulants_single_ion} shows the first four cumulants of $O^2_\mathcal{T}$. The first cumulant $\kappa_1$ is the nonlocal order parameter $O^2_\mathcal{T}$ itself which is zero when $D \leq 0.96$, i.e. in the SPT Haldane phase protected by $\mathcal{T}$. On the other hand, $\kappa_1 = 0$ when $D > 0.96$ which is the topologically trivial large $D$ phase. The variation with respect to $m$ of $\kappa_1$ is minute, but this is more apparent in the other cumulants $\kappa_2$ - $\kappa_4$, which all diverge at the critical point. Just as in the case of the cumulants $O^2_\text{string}$ of TKI in Fig.~\ref{fig:cumulants_tki}, the cumulants of $O^2_\mathcal{T}$ do not shift horizontally with increasing $m$, indicating that the critical point is located in narrow region of $D$ compared to the range of $D$ shown in Fig.~\ref{fig:cumulants_single_ion}. The sharp transition in the cumulants makes identifying the critical point an easy task which is taken to be $D_c = 0.96$.

\begin{figure}[h!]
  \centering\includegraphics[width=0.475\textwidth]{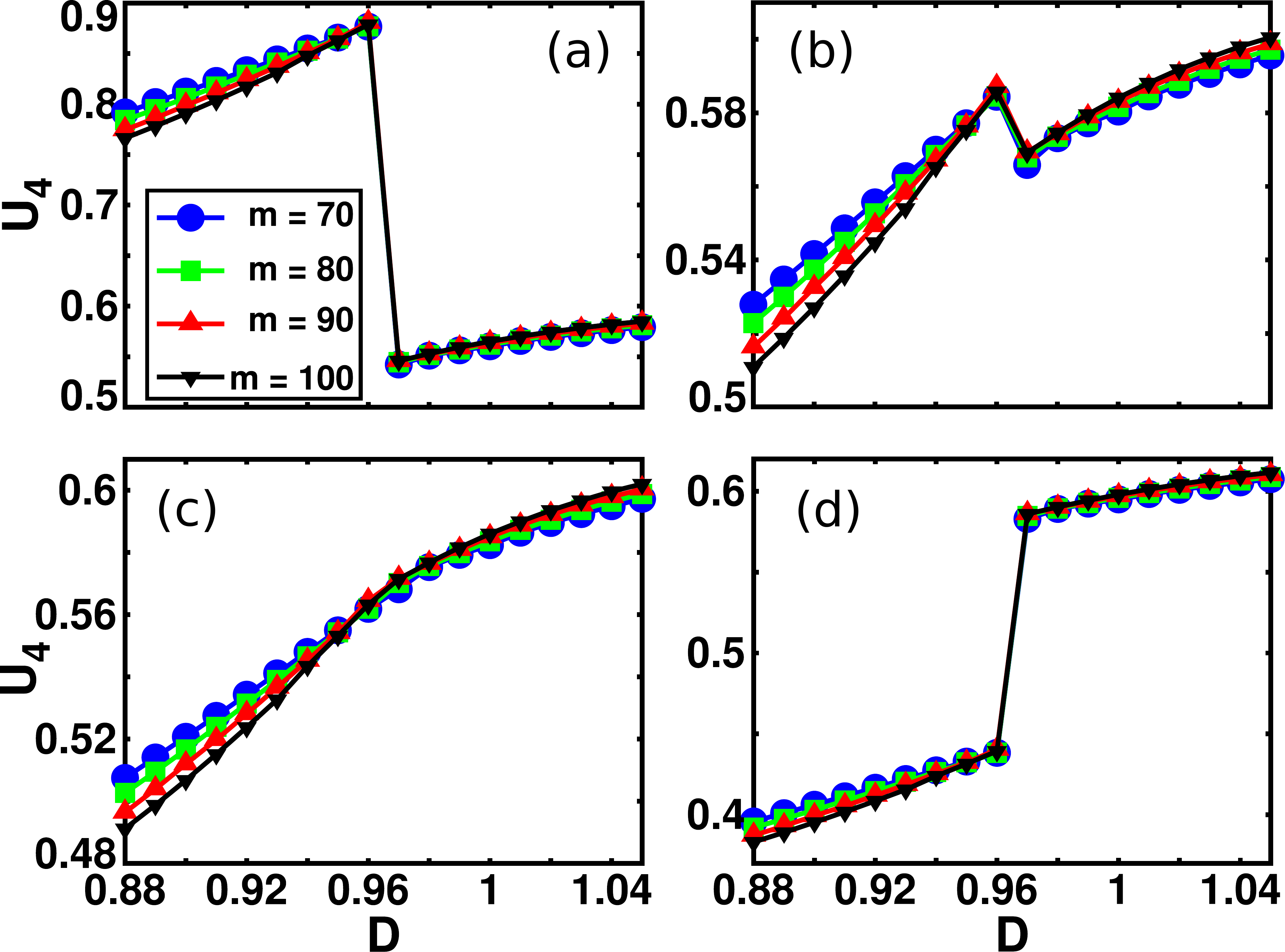}
  \caption{(Colour online) $S = 1$ Heisenberg chain with single-ion anisotropy. The Binder cumulant with (a) $s = 2$, (b) $s = 3$, (c) $s = s^* = 3.12$ and (d) $s = 4$. Increasing $s$ from (a)-(d) shifts $U_4$ until a maximum number of crossings occur between $U_4$ for the different values of $m$. This maximum number of crossings occurs in (c) where $s = 3.12$, marking the critical point $D_c = 0.96$.}
  \label{fig:binder_cumulants_s2_s5_single_ion}
\end{figure}

Even though the critical point has already been located through the use of the cumulants, the Binder cumulant can still be used as a consistency check of this critical point's location. Fig.~\ref{fig:binder_cumulants_s2_s5_single_ion} shows $U_4$ of $O^2_\mathcal{T}$ for several values of $s$. As $s$ is increased, $U_4$ in the region $D < D_c$ ($D > D_c$) shifts downwards (upwards), as can be seen in Figs.~\ref{fig:binder_cumulants_s2_s5_single_ion}(a) to \ref{fig:binder_cumulants_s2_s5_single_ion}(d). The optimum $s^*$ that gives a crossing point $U_4(m,D) \forall m$ occurs at $D = 0.96$ when $s^* = 3.12$ as shown in Fig.~\ref{fig:binder_cumulants_s2_s5_single_ion}(c). This value of $s^*$ was obtained using a linear solver to solve $\frac{\partial U_4(m,D)}{\partial m} = 0$ for $s$. The critical point is therefore taken to be $D_c = 0.96$, in agreement with that obtained from the cumulants.

\begin{figure}[h!]
  \centering\includegraphics[width=0.475\textwidth]{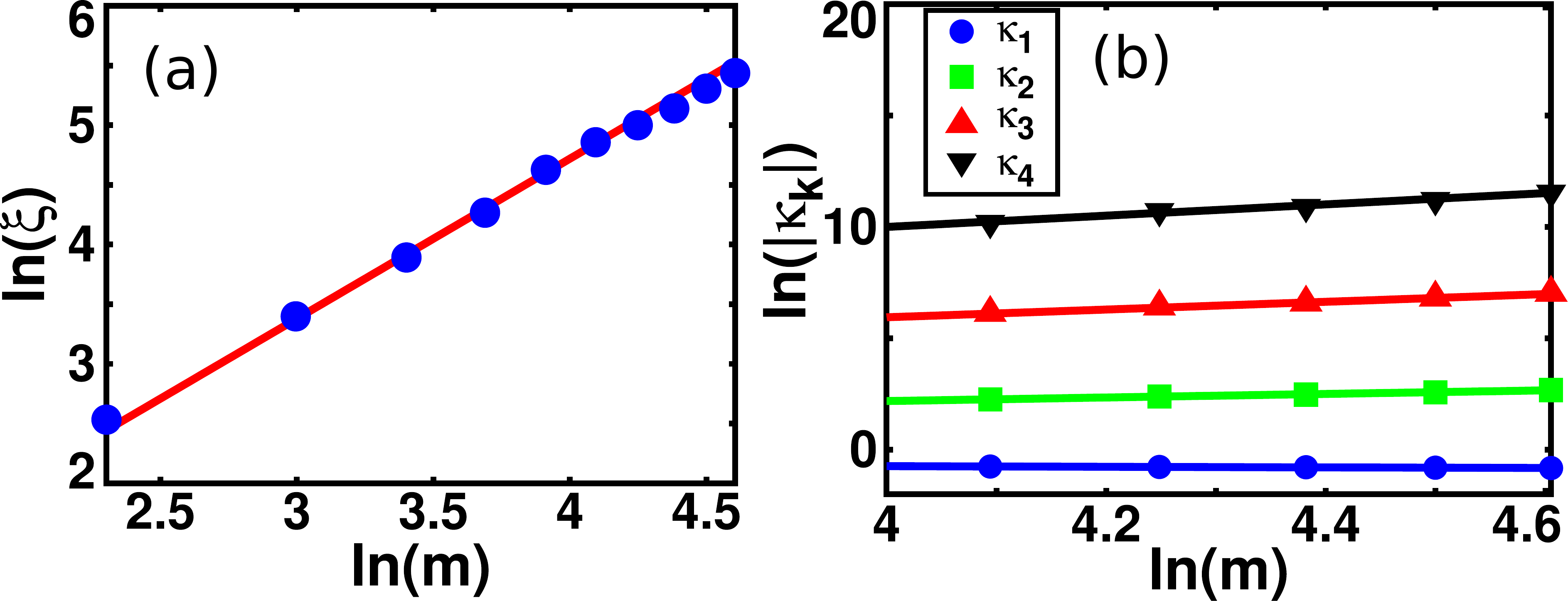}
  \caption{(Colour online) $S = 1$ Heisenberg chain with single-ion anisotropy. Log-log plot of the (a) correlation length $\xi$ and (b) cumulants, at the critical point $D = 0.97$ with respect to $m$. The symbols are the data points and the lines are the linear fit. The linear fit's gradient is equal to the critical exponent.}
  \label{fig:cor_len_cumulant_m_scaling_single_ion}
\end{figure}

Since $\kappa_1$ and $\kappa_3$ vanish when $D \leq 0.96$, it is not possible to use them to obtain their critical exponents at $D = D_c$. However, it is still possible to use the data in the vicinity of the critical point to obtain the critical exponents. Here, using the data at $D = 0.97$, $\xi$ and $\kappa_i$ are plotted against $m$ in Fig.~\ref{fig:cor_len_cumulant_m_scaling_single_ion}. By fitting these data with a linear fit and extracting their gradients, the critical exponents are obtained: $\kappa \sim 1.341 \pm 0.015$, $\alpha_1 = -0.133 \pm 0.001$, $\alpha_2 = 0.7987 \pm 0.005$, $\alpha_3 = 1.737 \pm 0.008$ and $\alpha_4 = 2.53 \pm 0.02$. From Eq.~\ref{eqn:kappa_central_charge}, the closest value of $\kappa$ to the one obtained here that gives a known value of $c$ is $\kappa = 1.3441$, corresponding to $c = 1$, in good agreement with Refs.~\cite{Hu, Chen1, Chen2, Degli}. The cumulant exponents can also be checked using Eq.~\ref{eqn:higher_cumulant_exponent_relation}. Using the obtained values of $\alpha_1$ and $\alpha_2$, one gets $\alpha_3 = -\alpha_1 + 2\alpha_2 = 0.133 + 2(0.7987) = 1.7304$, which differs from value obtained via linear fit in Fig.~\ref{fig:cor_len_cumulant_m_scaling_single_ion}(b) by $\sim 0.4\%$. Similarly, $\alpha_4 = -2\alpha_1 + 3\alpha_2 = 2(0.133) + 3(0.7987) = 2.662$, which differs from the value obtained via linear fit by $\sim 5\%$.

\begin{figure}[h!]
  \centering\includegraphics[width=0.475\textwidth]{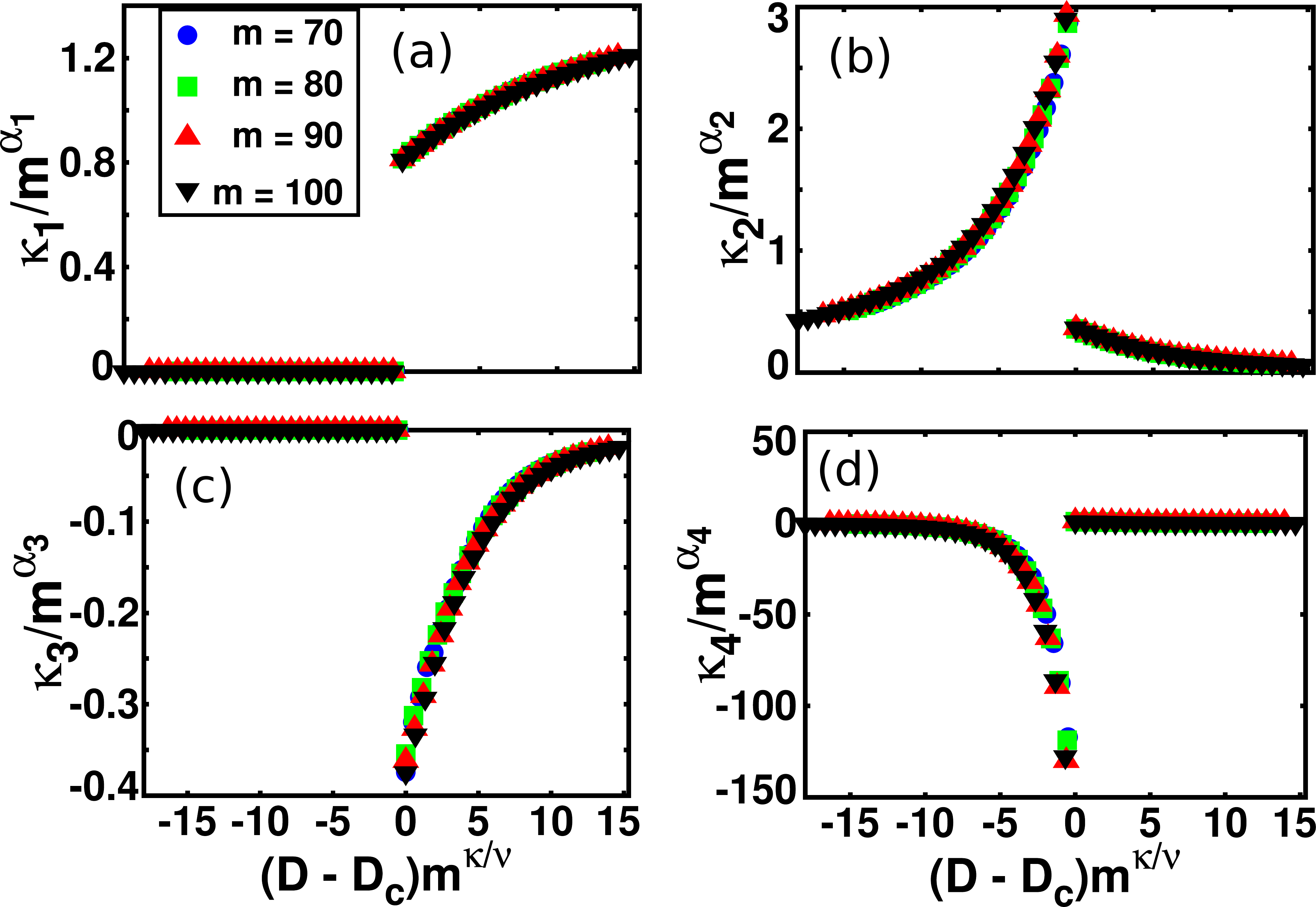}
  \caption{(Colour online) $S = 1$ Heisenberg chain with single-ion anisotropy. Cumulants scaled according to Eq.~\ref{eqn:scaling_function_m} for several values of $m$. (a) The order parameter, (b) variance, (c) skewness and (d) kurtosis. $D = 0.97$, $\kappa$ and $\alpha_i$ used in the scaling function are those obtained from the Binder cumulant and the linear fits of $\ln(\xi)$ and $\ln(\kappa_i)$ versus $\ln(m)$ respectively. $\nu$ is tuned such that the cumulants' sum of residual of the different values of $m$ at $D_c$ is minimized, this gives $\nu = 1.470 \pm 0.001$.}
  \label{fig:cumulants_scale_single_ion}
\end{figure}

\begin{figure}[h!]
  \centering\includegraphics[width=0.45\textwidth]{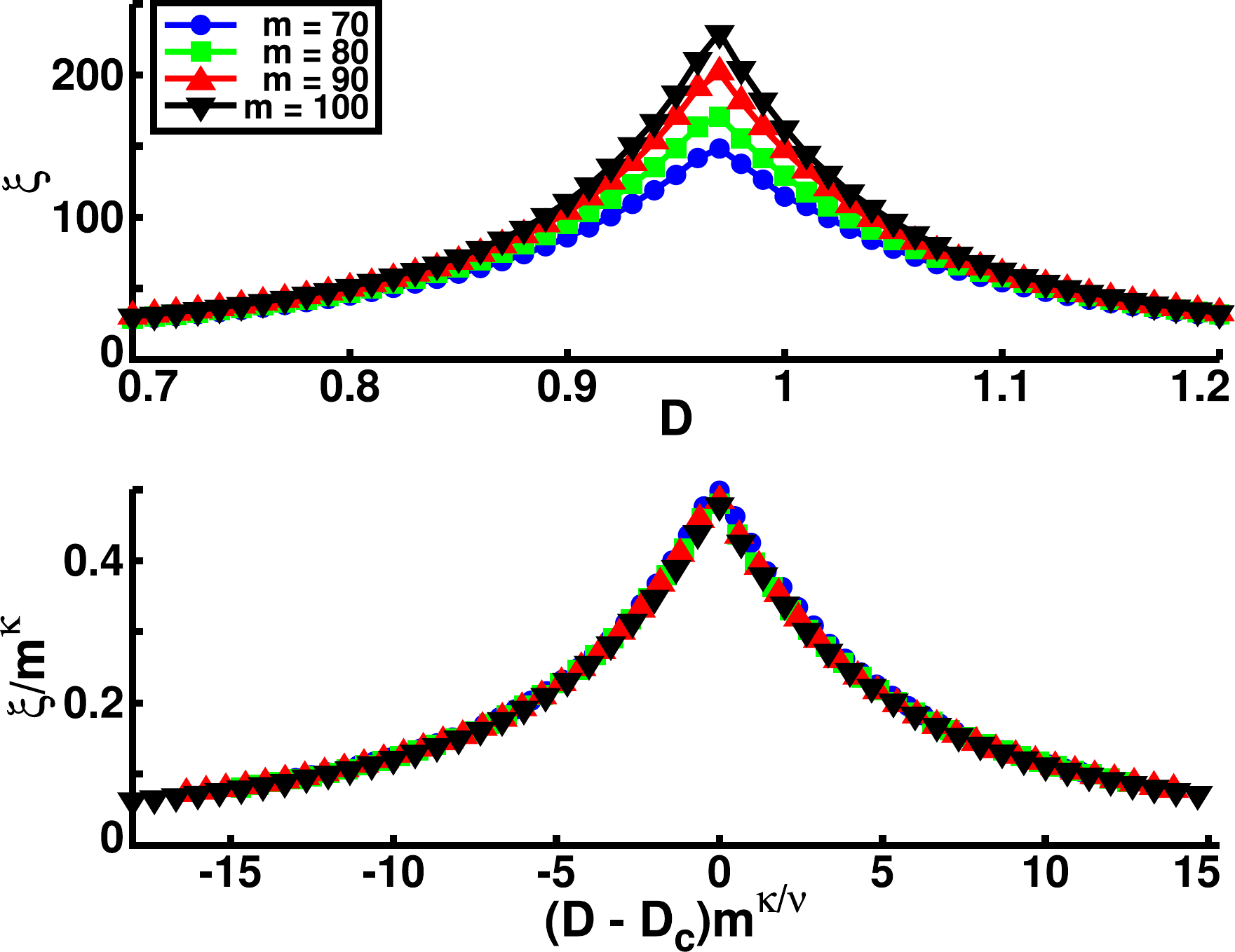}
  \caption{(Colour online) $S = 1$ Heisenberg chain with single-ion anisotropy. Correlation length versus $D$ (top) and correlation length scaled according to Eq.~\ref{eqn:scaling_function_cor_len_m} (bottom) for several values of $m$. $D = 0.97$ and $\kappa$ used in the scaling function are those obtained from the Binder cumulant and the linear fit of $\ln(\xi)$ versus $\ln(m)$ respectively. $\nu$ is tuned such that the correlation length's sum of residual of the different values of $m$ at $D_c$ is minimized, this gives $\nu = 1.470 \pm 0.001$.}
  \label{fig:cor_len_and_cor_len_scale_vs_B_single_ion}
\end{figure}

The obtained values of the exponents $\kappa$ and $\alpha_i$ are now used in the scaling functions of the cumulants and the correlation length Eqs.~\ref{eqn:scaling_function_m} and \ref{eqn:scaling_function_cor_len_m} respectively to determine the value of the exponent $\nu$. By tuning $\nu$ such that the sum of residual of the cumulants and correlation length of different values of $m$ at $D_c$ is minimized, the best data collapse is obtained which marks the optimum value of $\nu$. For the sake of consistency, the same value $D = 0.97$ used to obtain the exponents $\kappa$ and $\alpha_i$'s is used here. The data collapse of the cumulants are displayed in Fig.~\ref{fig:cumulants_scale_single_ion} and that of the correlation length is displayed in Fig.~\ref{fig:cor_len_and_cor_len_scale_vs_B_single_ion} where the best data collapse occurs when $\nu = 1.470 \pm 0.001$. This differs with the value $\nu = 1.472(4)$ obtained in Ref.~\cite{Hu} by $\sim 0.16\%$.

\subsection{Two-dimensional square lattice transverse field Ising model on an infinite cylinder}
\label{result_tfi_cylinder}
\begin{figure}[h!]
  \centering\includegraphics[width=0.475\textwidth]{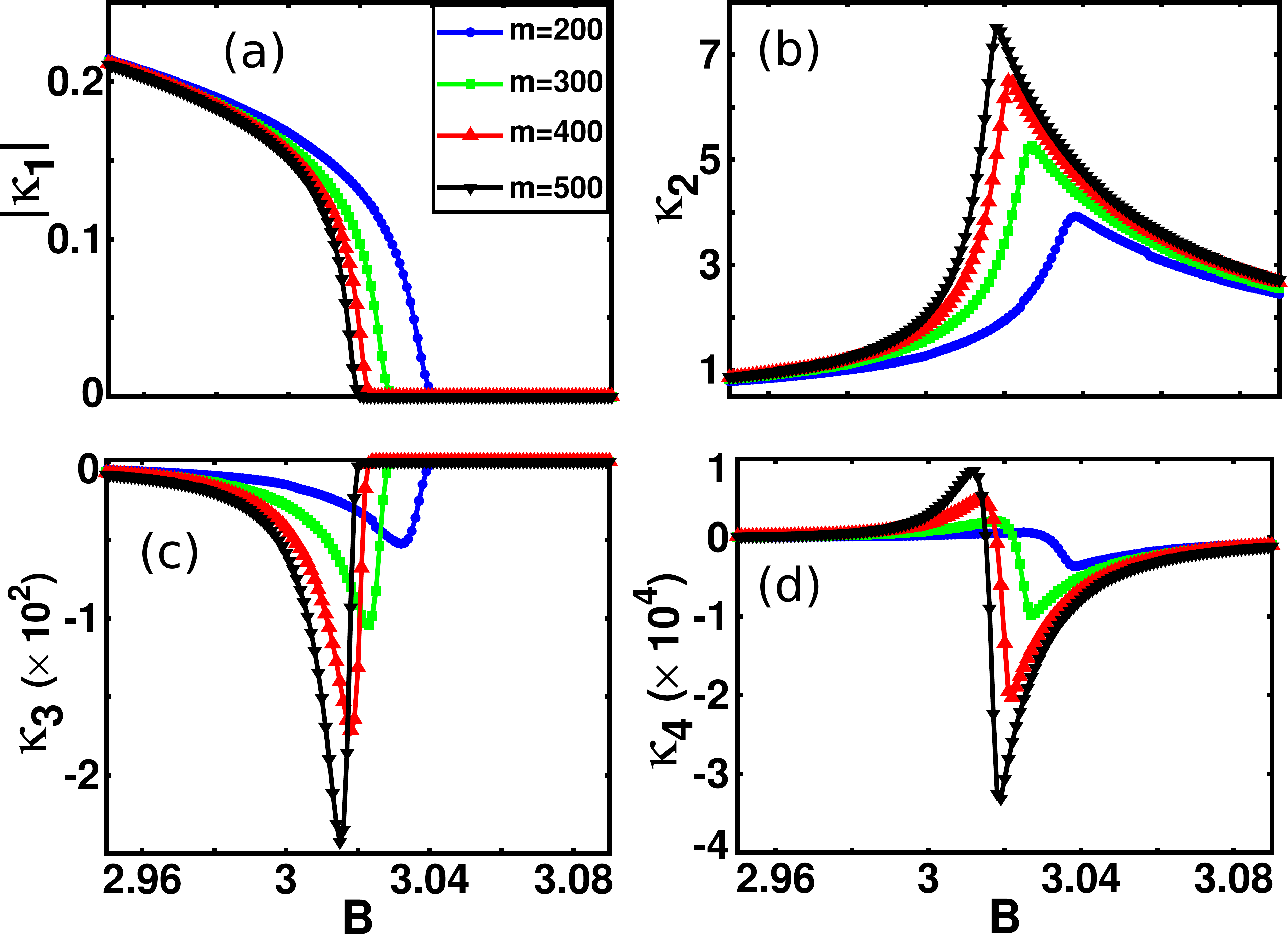}
  \caption{(Colour online) 2D TFI model on an infinite cylinder of circumference $w = 12$. First four cumulants of the order parameter $\braket{M} = \sum_{i,j} \sigma^z_{i,j}$ as a function of transverse field $B$ for several values of $m$. (a) First cumulant $\kappa_1$ is the order parameter itself which is nonzero when $B < B_c$ and zero when $B > B_c$. (b) Second cumulant $\kappa_2$ is the variance of the order parameter, which diverges at the critical point. (c) Third cumulant $\kappa_3$ is the skewness (d) Fourth cumulant $\kappa_4$ is the kurtosis. The vanishing of $\kappa_1$ and the peaks of $\kappa_2$, $\kappa_3$ and $\kappa_4$ shift towards the critical point as $m$ is increased.}
  \label{fig:cumulants_ising_w12}
\end{figure}

The 2D square lattice transverse field Ising model on an infinite cylinder is given by the Hamiltonian
\begin{eqnarray}
H &=& -\sum_i \sum_j^w \sigma^z_{i,j} \left( \sigma^z_{i+1,j} + \sigma^z_{i,j+1} \right) \nonumber
\\
&&+ B\sum_i \sum_j^w \sigma^x_{i,j},
\label{eqn:Hamiltonian_ising_2d}
\end{eqnarray}
where $B$ is the transverse field strength. This is a semi-infinite cylinder i.e. its length is infinite but possesses a finite circumference $w$ that the index $j$ sums over. The circumference chosen for this work is 12 sites. Just like in the 1D case, the order parameter is the magnetization $\braket{M} = \sum_{i,j} \sigma^z_{i,j}$, which is now summed over two indices $i$ and $j$ since the system is two dimensional. Another similarity shared between the 1D and 2D TFI models is the behavior of $\braket{M}$. When $B < B_c$, the ground state is ordered and thus $\braket{M} \neq 0$. Whereas when $B > B_c$, the ground state is disordered, hence $\braket{M} = 0$. Figure \ref{fig:cumulants_ising_w12}(a) - (d) shows the first four cumulants of $\braket{M}$ as a function of $B$. The behavior of the cumulants closely resemble that of the 1D TFI in Fig.~\ref{fig:cumulants_ising1d} where $\kappa_1$ vanishes upon approaching the critical point while the other cumulants diverge. The cumulants shift significantly with increasing $m$, where their vanishing/divergence approach the critical point as $m$ is increased.

\begin{figure}[h!]
  \centering\includegraphics[width=0.45\textwidth]{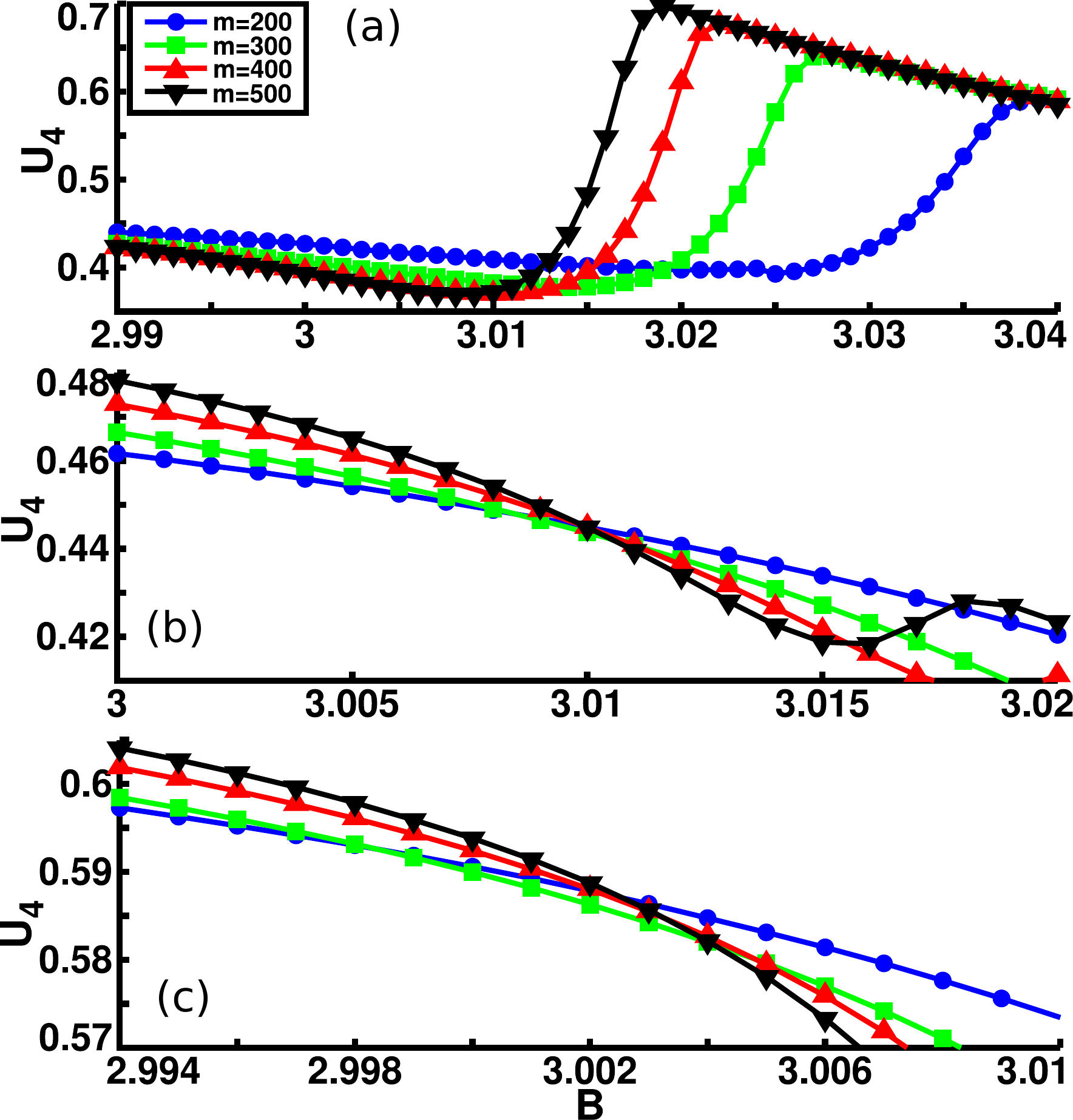}
  \caption{(Colour online) 2D TFI model on an infinite cylinder of circumference $w = 12$. The Binder cumulant for (a) $s = 2$, (b) $s = s^* = 3.27$ and (c) $s = 20$. Varying $s$ shifts $U_4(m)$, thus changes the value of $B$ where they cross. The value $s^*$ corresponds to $s$ where $U_4(m,B) \forall m$ cross. This is taken as the critical point. In this case, it is in (b) where $s = s^* = 3.27$ and $B_c = 3.01$. When $s < s^*$ and $s > s^*$ as in (a) and (c), there is no point $B$ where $U_4(m,B) \forall m$ cross simultaneously.}
  \label{fig:binder_cumulant_s2_s5_ising_w12}
\end{figure}

Since the cumulants do not easily locate the critical point, the Binder cumulant is used to do so. $U_4$ as a function of $B$ is shown in Fig.~\ref{fig:binder_cumulant_s2_s5_ising_w12} for three different values of $s$. Just as in the previous examples, varying $s$ shifts $U_4(m)$ at different rates for the different $m$ above and below the critical point. When $s < s^*$, for example in Fig.~\ref{fig:binder_cumulant_s2_s5_ising_w12}(a) where $s = 2$, there are multiple spurious crossing points formed between pairs of different $m$'s. This is analogous to the Binder cumulant for $s = 2$ in the top figure of Fig.~\ref{fig:binder_cumulant_s2_s5_ising1d} for the 1D Ising model. Increasing $s$ shifts the values of $U_4(m)$ such that $U_4(m)$ in the region below the critical point shift upwards while the values above the critical point shifts downwards - again, analogous to the increasing $s$ in the bottom figure of Fig.~\ref{fig:binder_cumulant_s2_s5_ising1d}. As a results, nearby spurious crossing points merge to form $U_4(m)^*$. This can be seen in Fig.~\ref{fig:binder_cumulant_s2_s5_ising_w12}(b) where $s = s^* = 3.27$ and $U_4(m)^*$ occurs at $B = 3.01$ which is taken as the critical point $B_c$. As before, $s^*$ was obtained by solving $\frac{\partial U_4(m,B)}{\partial m} = 0$ for $s$ using a linear solver. In the region of $3.015 < B < 3.018$, the $U_4(m = 500)$ (black inverted triangles) cross the other values of $m$ at, forming multiple spurious crossing points between pairs of $m$'s. When $s$ is further increased beyond $s^*$, $U_4(m)$ shift at different rates, thus once again forming multiple spurious crossing points between pairs of different $m$'s as can be seen in Fig.~\ref{fig:binder_cumulant_s2_s5_ising_w12}(c) where the exaggerated value of $s=20$ is chosen to clearly demonstrate this. As before, these spurious crossing points can be disregarded as candidates of the critical point.

\begin{figure}[h!]
  \centering\includegraphics[width=0.475\textwidth]{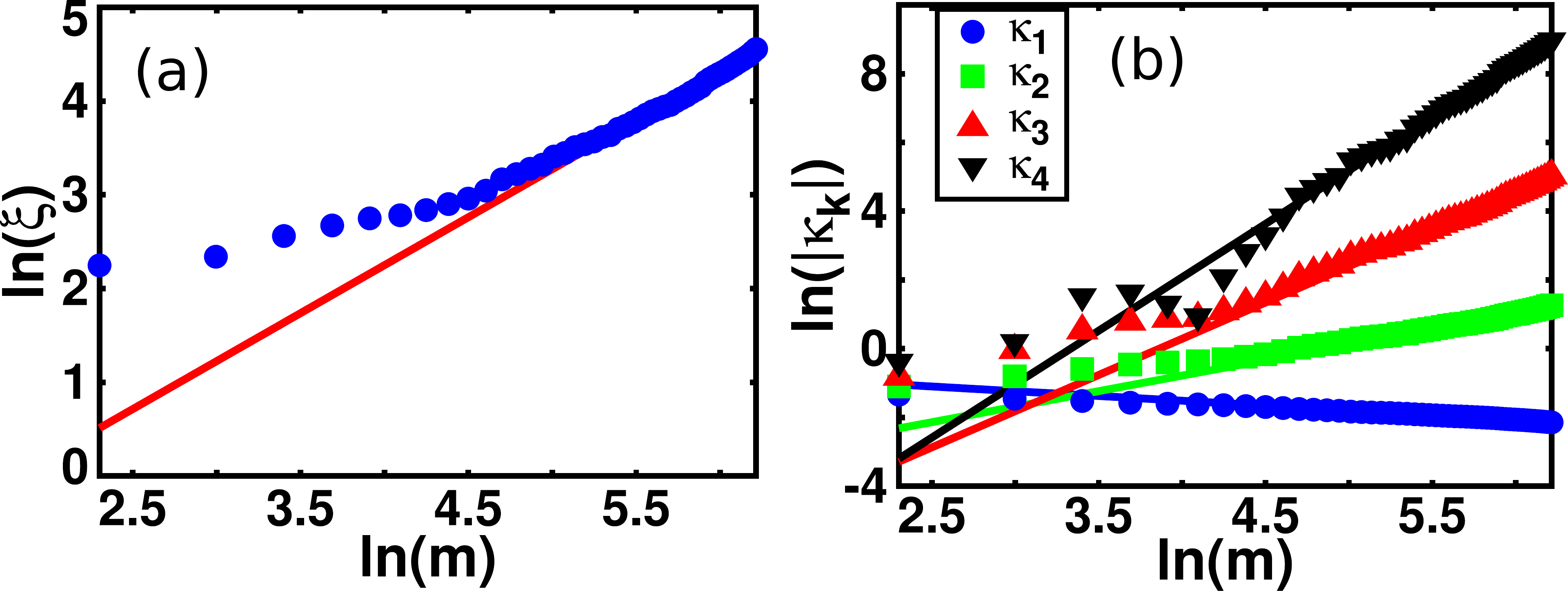}
  \caption{(Colour online) 2D TFI model on an infinite cylinder of circumference $w = 12$. Log-log plot of the (a) correlation length $\xi$ and (b) cumulants, at $B = 3.01$ with respect to $m$. The symbols are the data points and the lines are the linear fit. The gradient of the linear fit corresponds to the critical exponent.}
  \label{fig:cor_len_cumulant_m_scaling_ising_w12}
\end{figure}

Using $B_c = 3.01$ obtained from the Binder cumulant, the exponents of the correlation length and cumulants are now extracted by plotting $\xi$ and $\kappa_i$ against $m$. Fig.~\ref{fig:cor_len_cumulant_m_scaling_ising_w12}(a) shows the log-log plot of correlation length vs. $m$ for $B = 3.01$. The gradient of this linear fit gives the critical exponents $\kappa = 1.024 \pm 0.006$.  Fig.~\ref{fig:cor_len_cumulant_m_scaling_ising_w12}(b), on the other hand, is the log-log plot of the cumulants with their respective linear fits. By comparing the cumulants in this figure, one notices that the higher the cumulant order, the larger the fluctuation with respect to $m$. This makes the error in the linear fit for the higher cumulants larger, and thus the gradient of the higher cumulants is more prone to error. The gradient of the linear fits give the cumulant exponents $\alpha_1 = -0.276 \pm 0.002$, $\alpha_2 = 0.904 \pm 0.006$, $\alpha_3 = 2.11 \pm 0.01$, $\alpha_4 = 3.110 \pm 0.012$. The quality of the linear fit can be checked from the cumulant exponent relation Eq.~\ref{eqn:higher_cumulant_exponent_relation}. Using the obtained values of $\alpha_1$ and $\alpha_2$, one gets $\alpha_3 = -\alpha_1 + 2\alpha_2 = 0.276 + 2(0.904) = 2.084$, which differs from value obtained via linear fit in Fig.~\ref{fig:cor_len_cumulant_m_scaling_ising_w12}(b) by $\sim 1.2\%$. Similarly, $\alpha_4 = -2\alpha_1 + 3\alpha_2 = 2(0.276) + 3(0.904) = 3.264$, which differs from the value obtained via linear fit by $\sim 5\%$. As stated earlier, this large difference between the calculated and fitted values of the higher order cumulants' exponents stem from the fact that the higher order cumulants tend to fluctuate more. As a result, their fitted exponents are more susceptible to errors.

\begin{figure}[h!]
  \centering\includegraphics[width=0.475\textwidth]{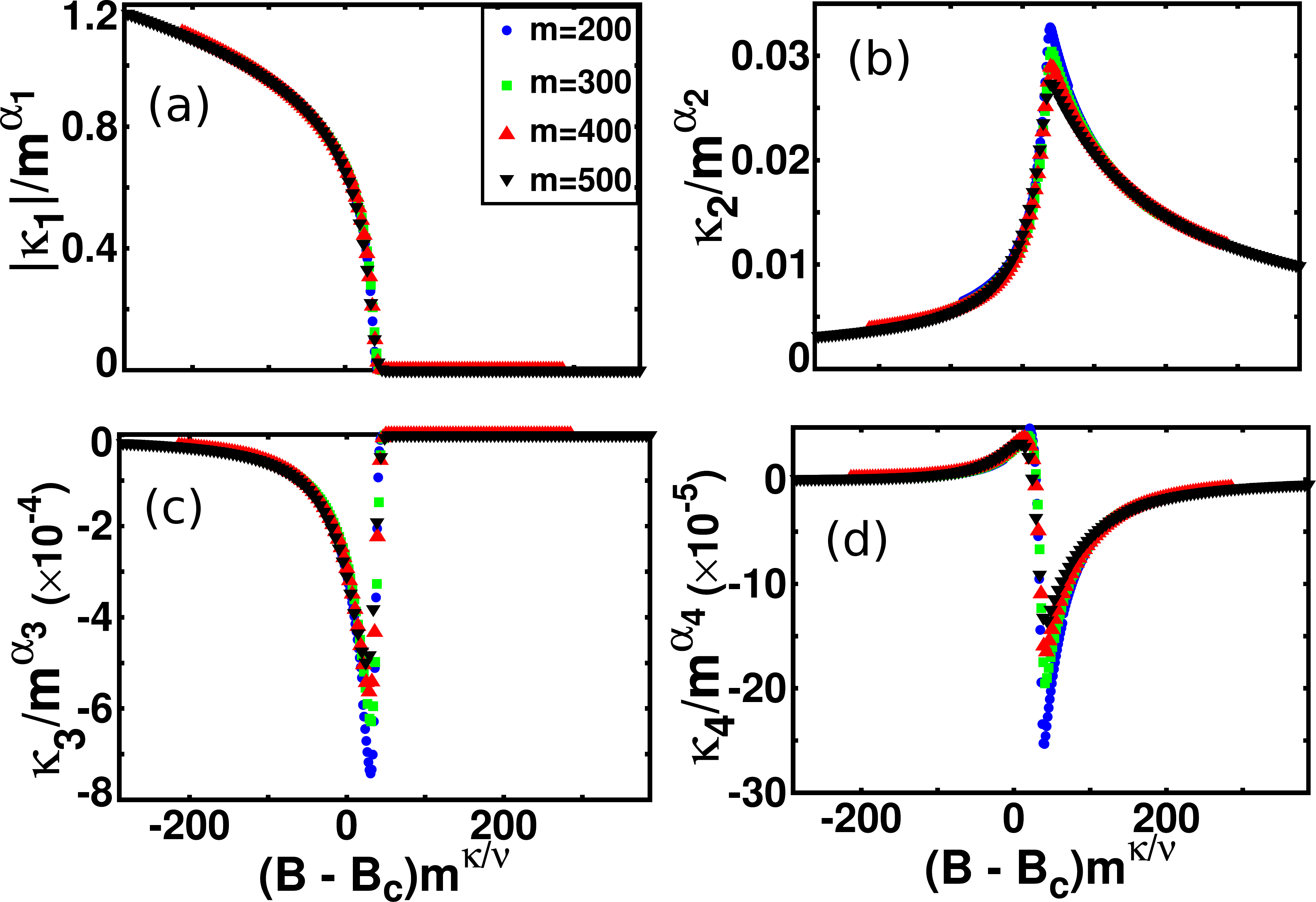}
  \caption{(Colour online) 2D TFI model on an infinite cylinder of circumference $w = 12$. Cumulants scaled according to Eq.~\ref{eqn:scaling_function_m} for several values of $m$. (a) The order parameter, (b) variance, (c) skewness and (d) kurtosis. $B_c$, $\kappa$ and $\alpha_i$ used in the scaling function are those obtained from the Binder cumulant and the linear fits of $\ln(\xi)$ and $\ln(\kappa_i)$ versus $\ln(m)$ respectively. $\nu$ is tuned such that the cumulants' sum of residual of the different values of $m$ at $B_c$ is minimized, this gives $\nu = 0.750 \pm 0.008$.}
  \label{fig:cumulants_scale_ising_w12}
\end{figure}

\begin{figure}[h!]
  \centering\includegraphics[width=0.45\textwidth]{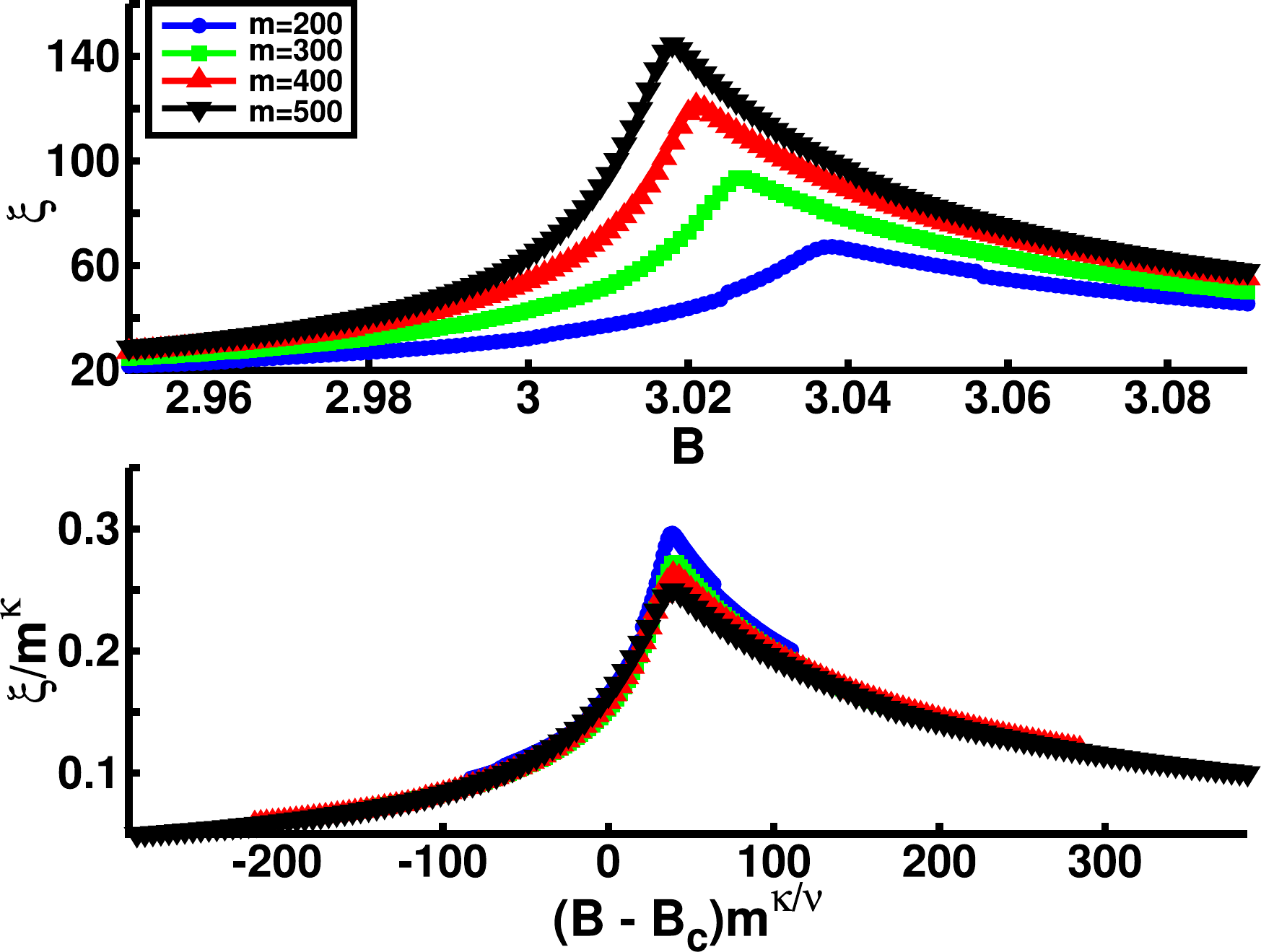}
  \caption{(Colour online) 2D TFI model on an infinite cylinder of circumference $w = 12$. Correlation length versus $B$ (top) and correlation length scaled according to Eq.~\ref{eqn:scaling_function_cor_len_m} (bottom) for several values of $m$. $B_c$ and $\kappa$ used in the scaling function are those obtained from the Binder cumulant and the linear fit of $\ln(\xi)$ versus $\ln(m)$ respectively. $\nu$ is tuned such that the correlation length's sum of residual of the different values of $m$ at $B_c$ is minimized, this gives $\nu = 0.750 \pm 0.008$.}
  \label{fig:cor_len_and_cor_len_scale_vs_B_ising_w12}
\end{figure}

Using the values of $B_c$, $\kappa$ and $\alpha_i$ obtained, together with the scaling function of the cumulants Eq.~\ref{eqn:scaling_function_m} and correlation length Eq.~\ref{eqn:scaling_function_cor_len_m}, the value of $\nu$ can now be obtained. This is done by tuning $\nu$ such that the respective sum of residual of the cumulants and correlation length of the different values of $m$ at $B_c$ are minimized. This gives the value $\nu = 0.750 \pm 0.008$ and the data collapse of the cumulant shown in Fig.~\ref{fig:cumulants_scale_ising_w12} and that of the correlation length shown in Fig.~\ref{fig:cor_len_and_cor_len_scale_vs_B_ising_w12}. This value of $\nu$ sits in between the value of the 1D TFI obtained in Section \ref{result_tfi} where $\nu = 1$, and the full 3D classical (or equivalently 2D quantum) Ising model in Ref.~\cite{Rader} where $\nu = 0.629 970(4)$. This is expected since the infinite cylinder sits geometrically in between the infinite chain (full 1D) and the infinite plane (full 2D).

\subsection{One-dimensional Bose-Hubbard model}
\label{result_bosehubbard}
The 1D Bose-Hubbard (BH) model is given by the Hamiltonian
\begin{eqnarray}
H &=& -J \sum_j \left( b^\dagger_{j+1} b_j + b^\dagger_j b_{j+1} \right) \nonumber \\
  &&+ \frac{U}{2} \sum_j n_j \left( n_j - 1 \right) ,
\label{eqn:Hamiltonian_bosehubbard}
\end{eqnarray}
where $b^\dagger_j$ ($b_j$) is the boson creation (annihilation) operator at site $j$, $n_j = b^\dagger_j b_j$ is the number operator at site $j$, and $U$ is the Hubbard repulsion term that penalizes double occupancy. For one particle per site, and fixing the energy scale $U = 1$, previous studies based on a variety of methods such as finite and infinite DMRG, quantum Monte Carlo, exact diagonalization and Bethe ansatz have shown to undergo a phase transition from the Mott insulator to a superfluid phase at $J_c \approx 0.26-0.3$ \cite{Fisher,Kuhner,Gu,Krutitsky,Carrasquilla,Rams} (and references therein). The latest work is that of Ref.~\cite{Rams} where $J_c = 0.3048(3)$ was obtained by using a maximum bond dimension of $m = 4000$ and an extrapolation of the correlation length with respect to the first two eigenvalues of the transfer matrix. This transition belongs to the Berezinskii-Kosterlitz-Thouless (BKT) universality class \cite{Fisher, Krutitsky}. Unlike other classes of phase transitions where the correlation length diverges algebraically as a power law as the critical point is approached, the BKT transition is known for its exponentially diverging correlation length \cite{Kuhner,Loison}
\begin{eqnarray}
\xi \propto \text{exp} \left( \frac{1}{|J - J_c|^\nu} \right) ,
\label{eqn:cor_len_bkt}
\end{eqnarray}
which causes all data of the critical point to be strongly plagued by finite-size effects.

\begin{figure}[h!]
  \centering\includegraphics[width=0.475\textwidth]{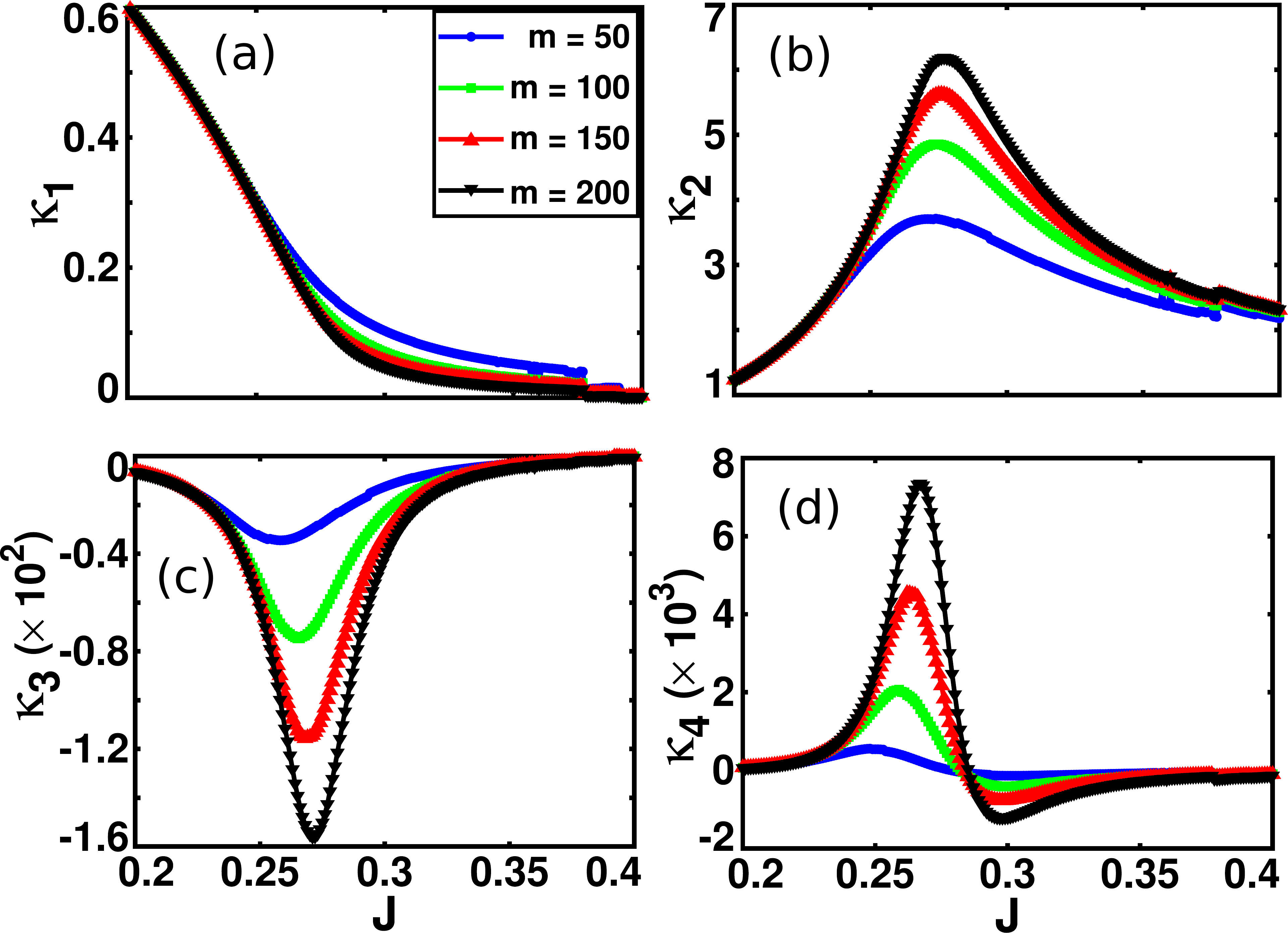}
  \caption{(Colour online) 1D BH model. First four cumulants of the string order parameter $O^2_\text{Mott}$ as a function of $J$ for several values of $m$. (a) First cumulant $\kappa_1$ is the order parameter which is zero only when the charge gap vanishes. (b) Second cumulant $\kappa_2$ is the variance of the order parameter, (c) the third cumulant $\kappa_3$ and (d) the fourth cumulant $\kappa_4$ all diverge at the critical point. The critical point is reached asymptotically with $m$ as in the BKT universality class.}
  \label{fig:cumulants_bose}
\end{figure}

In this work, the Hubbard repulsion $U$ is set to unity and the system is at half-filling. In addition, $U(1)$ particle number symmetry is enforced so that the total particle number per site is conserved. Even though the Mermin-Wagner theorem forbids the spontaneous-breaking of a continuous symmetry in 1D, an iMPS formulated without explicitly preserving $U(1)$ will spontaneously break this symmetry. For example, consider the $m = 1$ limit where the iMPS reproduces the mean-field solution, i.e. the iMPS groundstate is the exact solution of a mean-field Hamiltonian. In this limit, the iMPS explicitly breaks $U(1)$ symmetry. When $J > J_c$, this groundstate is comprised of a superfluid where the number of particles at each site is a superposition of all possible particle numbers, i.e. $\ket{\psi} = \left(a_0\ket{0} + a_1\ket{1} + a_2\ket{2} + \ldots \right)^{\otimes L}$. In the large-$m$ limit the $U(1)$ symmetry is restored, and the superfluid order parameter vanishes as required by the Mermin-Wagner theorem. By explicitly preserving $U(1)$ symmetry, the superfluid order parameter is always zero and thus is not a good choice of an order parameter. In contrast, the Mott string order parameter is zero in the superfluid phase but non-zero in the Mott-insulating phase. The latter is the region $J < J_c$ where the groundstate is dominated by the Hubbard repulsion term, therefore, double-occupancy is energetically expensive and the groundstate is comprised of a one-particle-per-site insulator. The Mott string order parameter is given as
\begin{eqnarray}
O^2_\text{Mott} = \lim_{|j-k| \rightarrow \infty} \left\langle  \mathbbm{1}_j \text{exp} \left[ i \pi \sum^k_{l=j} \left( \hat{n}_l - 1 \right) \right] \mathbbm{1}_k \right\rangle ,
\label{eqn:mott_string_order_parameter}
\end{eqnarray}
where $\hat{n}_l = b^\dagger_l b_l$ measures the number of bosons at site $l$. In the Mott insulator phase, the strong repulsion between bosons causes long-range correlations of single-site occupancy (alternatively, fluctuations in the density are short-range correlated), thus $O^2_\text{Mott} \neq 0$. On the other hand, in the gapless phase, long-range (power-law) fluctuations in the density causes $O^2_\text{Mott} = 0$. Fig.~\ref{fig:cumulants_bose}(a)-(d) show the first four cumulants of $O^2_\text{Mott}$. All the cumulants shift towards the critical point as $m$ is increased. As can be seen in Fig.~\ref{fig:cumulants_bose}(a), $\kappa_1$ decreases but does not vanish completely in the gapless superfluid phase. However, $\kappa_1$ in this phase decreases with increasing $m$, indicating that it would vanish completely in the $m \rightarrow \infty$ limit. This corresponds to the power-law fluctuations in density being restricted to a finite correlation length by the finite $m$.

The Binder cumulant in finite system studies has been known to be cumbersome in locating the critical point of a BKT transition because one does not simply obtain a single crossing point between the different system sizes \cite{Loison}. Instead, multiple crossing points between each system size is observed. As a results, one has to compare each crossing point and extrapolate them to get the final crossing point which marks the critical point. Even so, this extrapolation is not straight-forward and can give a very different critical point if one is not careful to do the correct comparison between many different system sizes. Because of this, a large number of system sizes and comparisons are required to locate the correct critical point. Projecting this fact onto the consideration of an infinite system described in this work, the additional length scaling parameter $s$ would add an extra degree of difficulty since there is no clear way to determine its optimum value $s^*$ which is defined as the crossing point of the Binder cumulant for all values of $m$. As such, the deployment of the Binder cumulant to determine the critical point is deemed impractical. Instead, the scaling functions of the cumulants and correlation length are used directly to determine the critical point and critical exponents simultaneously.

\begin{figure}[h!]
  \centering\includegraphics[width=0.475\textwidth]{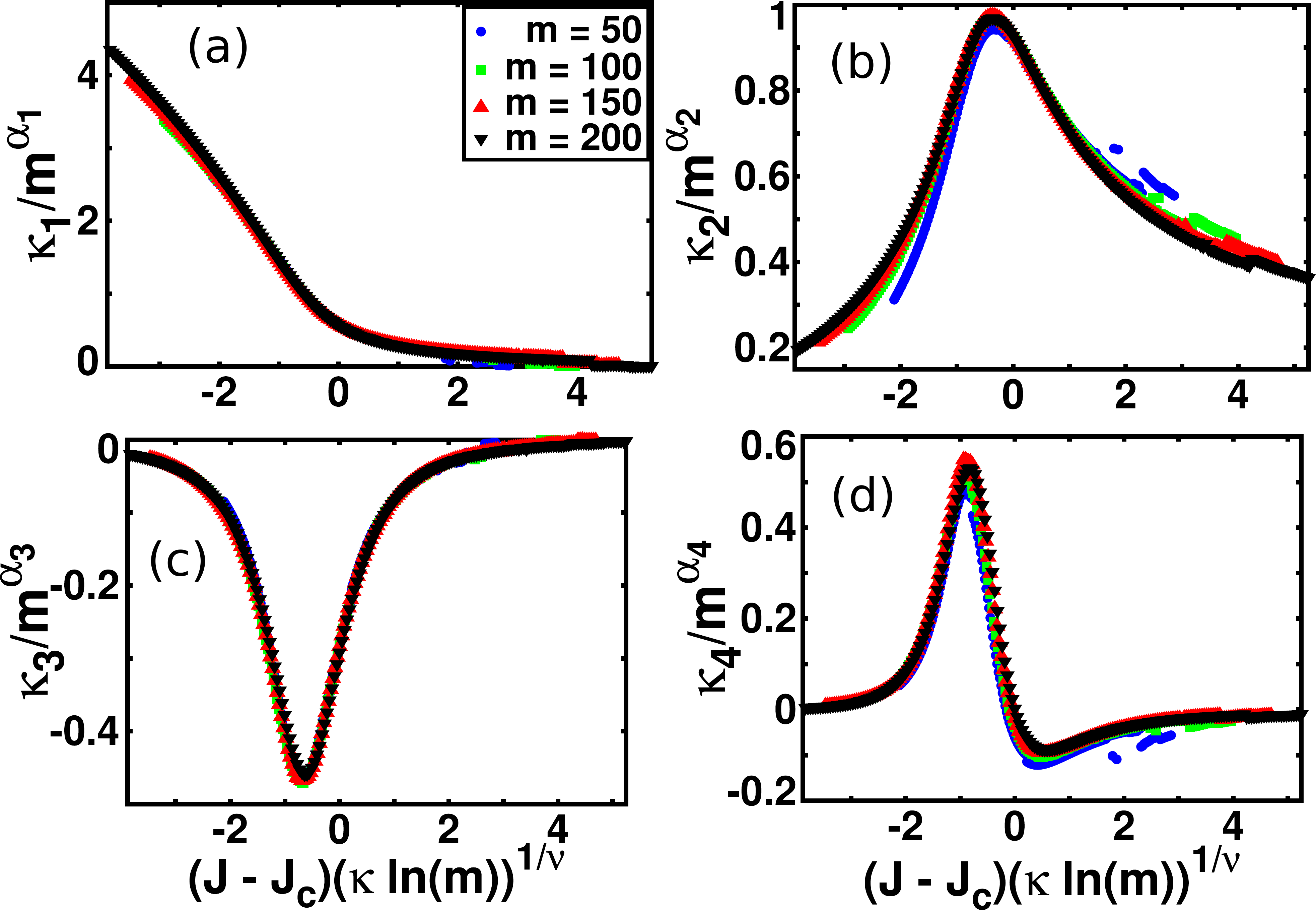}
  \caption{(Colour online) 1D BH model. Cumulants scaled according to Eq.~\ref{eqn:scaling_function_m} for several values of $m$. (a) The order parameter, (b) variance, (c) skewness and (d) kurtosis. $J_c$, $\kappa$, $\nu$ and $\alpha_i$ are tuned until the cumulants' sum of residual of the different values of $m$ over a range of $J$ is minimized. The values of these parameters are $J_c = 0.2850 \pm 0.0005$, $\kappa = 1.275 \pm 0.001$, $\nu = 0.500 \pm 0.001$, $\alpha_1 = -0.375 \pm 0.001$, $\alpha_2 = 0.350 \pm 0.005$, $\alpha_3 = 1.10 \pm 0.02$ and $\alpha_4 = 1.80 \pm 0.05$.}
  \label{fig:cumulants_scale_bose}
\end{figure}

Since $\xi$ in Eq.~\ref{eqn:cor_len_bkt} scales as an exponential instead of a power law as in Eq.~\ref{eqn:cor_len_thermo_limit}, a new form of the scaled parameter Eq.~\ref{eqn:L_indep_finite_size_effect} has to be obtained. Starting from Eq.~\ref{eqn:cor_len_bkt} for a finite system and following the same steps to obtain Eq.~\ref{eqn:L_indep_finite_size_effect} from Eq.~\ref{eqn:cor_len_thermo_limit}, one gets:
\begin{eqnarray}
\xi(L,J^*(L)) \propto \text{exp} \left( \frac{1}{|J^*(L) - J_c|^\nu} \right) \nonumber \\
L = \text{exp} \left( \frac{g}{|J^*(L) - J_c|^\nu} \right) \nonumber \\
|J^*(L) - J_c| (\ln(L))^{1/\nu} = g' .
\end{eqnarray}
Converting this to an infinite system described by an iMPS by the substitution $L \rightarrow \xi \propto m^\kappa$, gives
\begin{eqnarray}
|J^*(m) - J_c| (\kappa \ln(m))^{1/\nu} = g' ,
\label{eqn:m_indep_finite_size_effect_BKT}
\end{eqnarray}
where $g$ is a proportionality constant and $g' = g^{1/\nu}$. Using this together with the cumulant scaling function of the form Eq.~\ref{eqn:scaling_function_m} and the correlation length scaling function Eq.~\ref{eqn:scaling_function_cor_len_m}, the data collapse of the cumulants and correlation length can be obtained respectively. The data collapse of the four cumulants are plotted in Fig.~\ref{fig:cumulants_scale_bose}. By tuning each cumulant exponents $\alpha_i$ simultaneously with the critical point $J_c$ and exponents $\kappa$ and $\nu$, the best data collapse, i.e. the collapse corresponding to the cumulants' minimized sum of residual of the different values of $m$ over a range of $J$, is obtained with the critical point $J_c = 0.2850 \pm 0.0005$, and exponents $\kappa = 1.275 \pm 0.001$, $\nu = 0.500 \pm 0.001$, $\alpha_1 = -0.375 \pm 0.001$, $\alpha_2 = 0.350 \pm 0.005$, $\alpha_3 = 1.10 \pm 0.02$, $\alpha_4 = 1.80 \pm 0.05$. Using Eq.~\ref{eqn:higher_cumulant_exponent_relation} to check the consistency of the cumulant exponents, one finds that $\alpha_3 = -\alpha_1 + 2\alpha_2 = 0.375 + 2(0.35) = 1.075$, which differs by $\sim 2.3\%$ from the value obtained by directly tuning $\alpha_3$ in the data collapse scaling function. Similarly, $\alpha_4 = -2\alpha_1 + 3\alpha_2 = 2(0.375) + 3(0.35) = 1.80$, which is exactly the value obtained from the data collapse of the scaling function. The critical point obtained here however differs from the value obtained in Refs.~\cite{Krutitsky,Carrasquilla,Rams} of $J_c \approx 0.3$ by $5\%$. The exponent $\nu$ agrees exactly with that in Ref.~\cite{Fisher}.

\begin{figure}[h!]
  \centering\includegraphics[width=0.45\textwidth]{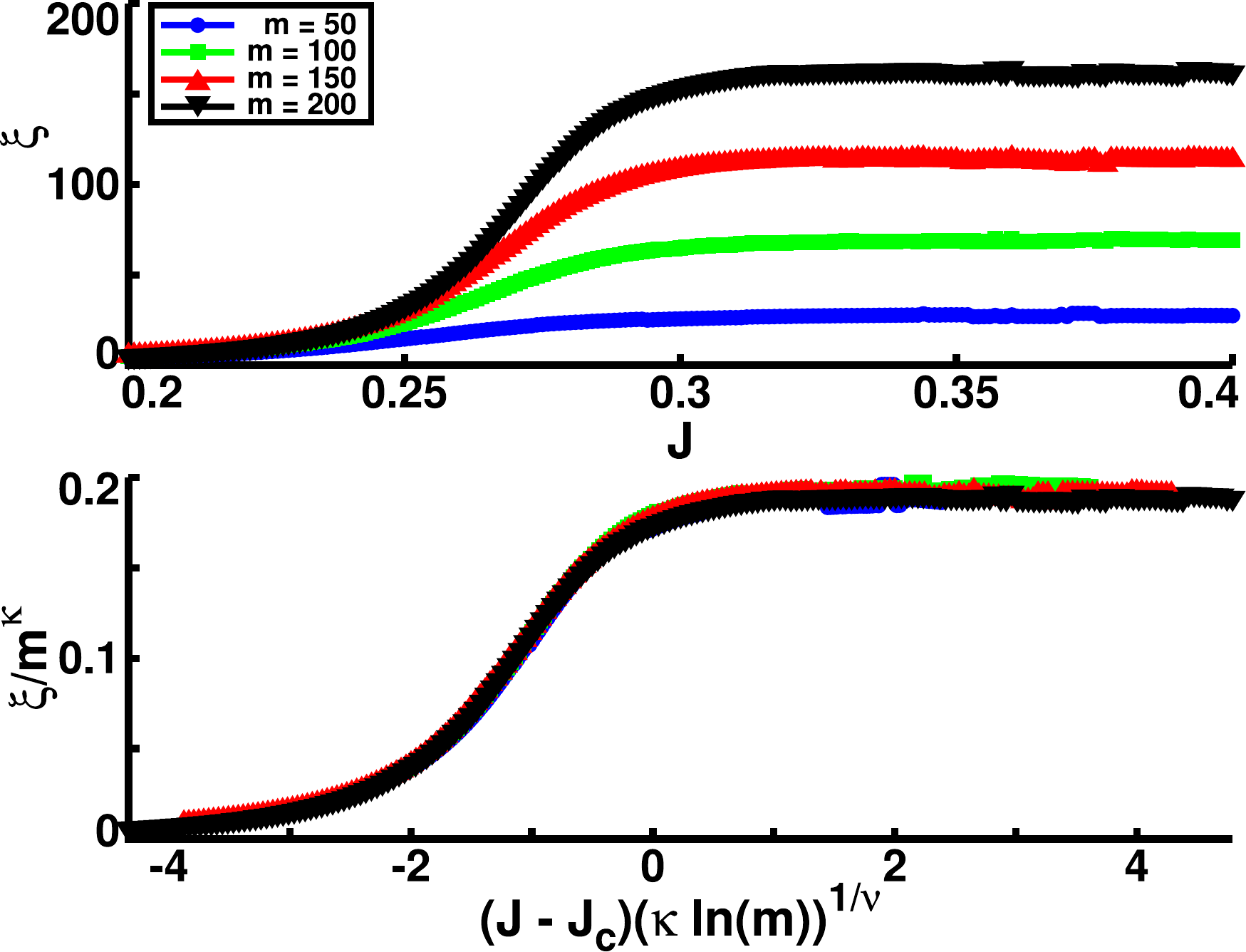}
  \caption{(Colour online) 1D BH model. Correlation length versus $J$ (top) and correlation length scaled according to Eq.~\ref{eqn:scaling_function_cor_len_m} (bottom) for several values of $m$. $J_c$, $\nu$ and $\kappa$ are tuned until correlation length's sum of residual of the different values of $m$ over a range of $J$ is minimized. The values of these parameters are $J_c = 0.295 \pm 0.001$, $\kappa = 1.275 \pm 0.001$ and $\nu = 0.500 \pm 0.001$.}
  \label{fig:cor_len_and_cor_len_scale_vs_B_bose}
\end{figure}

The data collapse of the correlation length scaling function Eq.~\ref{eqn:scaling_function_cor_len_m} is now used to check the whether the critical point and exponents $\kappa$ and $\nu$ obtained from the cumulant scaling function were correct. By minimizing the correlation length's sum of residual of different values of $m$ over a range of $J$, the data collapse obtained is shown in Fig.~\ref{fig:cor_len_and_cor_len_scale_vs_B_bose} with the values $J_c = 0.295 \pm 0.001$, $\nu = 0.500 \pm 0.001$, $\kappa = 1.275 \pm 0.001$. The critical point here is closer to that obtained in Refs.~\cite{Krutitsky,Carrasquilla,Rams} of $J_c \approx 0.3$, differing by $\sim 1.7\%$. The superfluid phase, comprising of free bosons, is described by central charge $c = 1$ \cite{Krutitsky}. Using this, Eq.~\ref{eqn:kappa_central_charge} gives $\kappa = 1.344$ which differs from the obtained value $\kappa = 1.275$ by $\sim 5.1\%$.


\section{Summary}
The order parameter cumulants were studied across the second order and BKT transition classes for several 1D and 2D exemplary systems. These cumulants were obtained using the recursive formula Eq.~\ref{eqn:fixed_pt} which offers an efficient way of computing operators of any order and unifies the procedure of calculating both local and string operators. Using the Binder cumulant, finite-entanglement scaling and scaling functions, the critical point and exponents were determined with a relatively smaller bond dimension compared to previously-known data. The procedure to obtain the critical point and exponents are summarized here:
\begin{enumerate}
\item Obtain the first four cumulants of the order parameter and correlation length $\xi$ as a function of Hamiltonian parameter $B$ across the critical point for several values of bond dimension $m$.
\item Using the four cumulants, compute the Binder cumulant $U_4(m,B)$ as a function of $B$ for all values of $m$.

\item Tune the length scale parameter $s$ relating the system size $L$ to the correlation length $\xi$ in Eq.~\ref{eqn:L_xi_relation} to obtain the crossing/intersection point of $U_4(m,B)$ for all $m$. This crossing point is taken as the critical point $B_c$. Alternatively, use a linear solver to solve $\frac{\partial U_4(m,B)}{\partial m} = 0 \,\, \forall m$ over a range of parameter $B$. This will give $B_c$ at the optimum value $s = s^*$. \label{step:binder}

\item Using $B_c$, employ finite-entanglement scaling to extract the finite-entanglement scaling exponent $\kappa$ and the cumulant exponents $\alpha_i$. This is done by plotting $\log \xi (B_c)$ and $\log \kappa_i (B_c)$ against $\log m$ respectively. Fit a linear function through these data. The critical exponent is the gradient of the linear fit. Check the consistency of the cumulant exponents by using the cumulant exponent relation Eq.~\ref{eqn:higher_cumulant_exponent_relation}. If the exponents disagree significantly, adjust the linear fit parameters. \label{step:log_kappa_alpha_linear_fit}

\item Use the cumulant scaling function Eq.~\ref{eqn:scaling_function_m}, the correlation length scaling function Eq.~\ref{eqn:scaling_function_cor_len_m} and the scaled parameter Eq.~\ref{eqn:m_indep_finite_size_effect} to obtain the data collapse of the cumulants and correlation length respectively. Tune the exponent $\nu$ to obtain the best data collapse. Fine-tune the critical point $B_c$ if necessary. \label{step:nu_lambda_c}

\item If $B_c$ is significantly different from that obtained in step \ref{step:binder}, go back to step \ref{step:log_kappa_alpha_linear_fit}, adjust the linear fit parameters, and repeat.
\end{enumerate}
Where the Binder cumulant is cumbersome in producing the critical point, such as in the case of the BKT transition class, the scaling function can be directly employed to obtain the critical point and exponents simultaneously. The consistency of the exponents can be then checked with the cumulant exponent relation.


\section{Acknowledgement}
I.P.M. acknowledges support from the ARC Future Fellowships Scheme No. FT140100625.


\section{Appendix}
\subsection{Derivation of $\braket{M^2}$}
\label{appendix_variance}
This section shows the derivation of the second order moment $\braket{M^2}$ for the operator $M$ given in Eq.~\ref{eqn:magnetization_mpo}. This is an extension of the procedure used in the calculation of the order parameter shown in Section \ref{higher_order_moments} and can be easily generalized to any higher order moments.

The operator $M^2$ is given by
\begin{eqnarray}
M^2 &=& M \otimes M \nonumber \\
	&=& \left( \begin{array}{cc} I & Z \\ & I \end{array} \right) \otimes \left( \begin{array}{cc} I & Z \\ & I \end{array} \right) \nonumber \\
	&=& \left( \begin{array}{cccc} I & Z & Z & Z^2 \\ & I & 0 & Z \\ & & I & Z \\ & & & I \end{array} \right) .
\label{appendix_eqn:magnetization_square_mpo}
\end{eqnarray}
As stated in Section \ref{higher_order_moments}, the expectation value of an operator in the form of an upper triangular MPO is calculated by the fixed-point equation of the environment matrix $E$:
\begin{eqnarray}
E_i(L+1) = T_{W_{ii}} \left(E_i(L)\right) + \sum_{j<i} T_{W_{ji}} \left(E_j(L)\right).
\label{appendix_eqn:fixed_pt}
\end{eqnarray}
For the upper triangular MPO, the operator representing the observable is $E_4(L)$ which has the expectation value of $\braket{M^2} = \text{Tr} (E_4(L) \rho_R)$. The goal here is to show that the correct polynomial degree matrix-valued coefficient Eq.~\ref{eqn:E_poly_L}:
\begin{eqnarray}
E_i(L) = \sum_{m=0}^p E_{i,m} L^m ,
\label{appendix_eqn:E_poly_L}
\end{eqnarray}
is related to $\text{Tr} (T_{Z^2} (E_4(L)) \rho_R)$ - this is done as follows. Writing out each term $E_i(L+1)$,
\begin{eqnarray}
E_1(L+1) &=& T_I(E_1(L)) \label{appendix_eqn:E1} \\
E_2(L+1) &=& T_I(E_2(L)) + T_Z(E_1(L)) \label{appendix_eqn:E2} \\
E_3(L+1) &=& T_I(E_3(L)) + T_0(E_2(L)) + T_Z(E_1(L)) \nonumber \\
		 &=& T_I(E_3(L)) + T_Z(E_1(L)) \label{appendix_eqn:E3} \\
E_4(L+1) &=& T_I(E_4(L)) + T_Z(E_3(L)) + T_Z(E_2(L)) \nonumber \\
		 &&+ T_{Z^2}(E_1(L)) \label{appendix_eqn:E4} .
\end{eqnarray}
The expectation value of interest $\text{Tr} (E_4(L) \rho_R)$ is in Eq.~\ref{appendix_eqn:E4}. To obtain this, one has to solve the Eqs.~\ref{appendix_eqn:E1}, \ref{appendix_eqn:E2} and \ref{appendix_eqn:E3} sequentially in order to obtain terms that will be substituted into the final equation Eq.~\ref{appendix_eqn:E4}.

Starting from $E_1(L+1)$, Eq.~\ref{appendix_eqn:E1} only contains the operator $T_I$, which acts trivially on $E_1(L)$ and is therefore independent of $L$. Hence the polynomial expansion of $E_1(L)$ is of polynomial order $p = 0$:
\begin{eqnarray}
E_1(L) &=& E_{1,0} L^0 \nonumber \\
&=& E_{1,0} .
\end{eqnarray}
Inserting this into Eq.~\ref{appendix_eqn:E1} gives 
\begin{eqnarray}
E_{1,0} = T_I(E_{1,0}) ,
\end{eqnarray}
which implies that $E_{1,0}$ is an eigenvector of $T_I$ with eigenvalue unity, and hence $E_{1,0} \propto \tilde{I}$, where $\tilde{I}$ is an $m \times m$ matrix. This result will be used in subsequent equations where $E_1$ is needed.

The next equation is Eq.~\ref{appendix_eqn:E2} which contains operators $T_I$ and $T_Z$. Following the same reasoning to obtain the ansatz of $E_2(L)$ via Eq.~\ref{eqn:E2_geo_series} leading to Eq.~\ref{eqn:poly_expansion_E_2}, one finds that the polynomial order of $E_2(L)$ is thus $p = 1$ and its polynomial expansion is
\begin{eqnarray}
E_2(L) = E_{2,0} + E_{2,1} L .
\label{appendix_eqn:poly_expansion_E2}
\end{eqnarray}
Inserting this and $E_{1,0} \propto \tilde{I}$ into Eq.~\ref{appendix_eqn:E2} gives
\begin{eqnarray}
E_{2,0} + E_{2,1} \times (L+1) &=& T_I(E_{2,0} + E_{2,1}L) + T_Z(\tilde{I}) \nonumber \\
E_{2,0} + E_{2,1} + E_{2,1} L &=& T_I(E_{2,0}) + T_I(E_{2,1}) L \nonumber \\
&&+ C_Z ,
\end{eqnarray}
where $C_Z \equiv T_Z(\tilde{I})$ is a constant matrix, i.e. it has no dependence on any $E_i$'s and thus its value can be calculated beforehand. Equating powers of $L$ gives
\begin{eqnarray}
L^0 &:& E_{2,0} + E_{2,1} = T_I(E_{2,0}) + C_Z \label{appendix_eqn:E2_2} \\
L^1 &:& E_{2,1} = T_I(E_{2,1}) \label{appendix_eqn:E2_3} .
\end{eqnarray}
Eq.~\ref{appendix_eqn:E2_3} implies $E_{2,1} \propto \tilde{I} = e_{2,1,1}\tilde{I}$, where $e_{2,1,1}$ is a proportionality constant. Inserting this into Eq.~\ref{appendix_eqn:E2_2} gives
\begin{eqnarray}
E_{2,0} + e_{2,1,1}\tilde{I} = T_I(E_{2,0}) + C_Z .
\label{appendix_eqn:E2_4}
\end{eqnarray}
By decomposing $E_{2,0} = \sum^{m^2}_{i=1} e_{2,0,i} \ket{\lambda_i}$, $T_I = \sum^{m^2}_{i=1} \lambda_i \ket{\lambda_i}\bra{\lambda_i}$ and $C_Z \equiv T_Z(\tilde{I}) = \sum^{m^2}_{i=1} C_{Z,i} \ket{\lambda_i}$, where $C_{Z,i}$ are elements of the constant matrix $C_Z$, Eq.~\ref{appendix_eqn:E2_4} becomes
\begin{eqnarray}
\sum^{m^2}_{i=1} e_{2,0,i} \ket{\lambda_i} + e_{2,1,1}\tilde{I}  &=& \sum^{m^2}_{i=1} \lambda_i \ket{\lambda_i}\bra{\lambda_i} \left( \sum^{m^2}_{i'=1} e_{2,0,i'} \ket{\lambda_{i'}} \right) \nonumber \\
&&+ \sum^{m^2}_{i=1} C_{Z,i} \ket{\lambda_i} \nonumber \\
&=& \sum^{m^2}_{ii'=1} \lambda_i e_{2,0,i'} \ket{\lambda_i} \left\langle \lambda_i | \lambda_{i'} \right\rangle \nonumber \\
&&+ \sum^{m^2}_{i=1} C_{Z,i} \ket{\lambda_i} \nonumber \\
&=& \sum^{m^2}_{i=1} \lambda_i e_{2,0,i} \ket{\lambda_i} + \sum^{m^2}_{i=1} C_{Z,i} \ket{\lambda_i} , \nonumber \\
\label{appendix_eqn:E2_5}
\end{eqnarray}
where the orthogonality relation $\sum_{i'} \left\langle \lambda_i | \lambda_i' \right\rangle = \delta_{ii'}$ was used in the last step. Further decomposing Eq.~\ref{appendix_eqn:E2_5} into a term parallel to the identity and terms perpendicular to the identity,
\begin{eqnarray}
e_{2,0,1}\tilde{I} + \sum^{m^2}_{i=2} e_{2,0,i} \ket{\lambda_i} &+& e_{2,1,1}\tilde{I} \nonumber \\
&=& e_{2,0,1}\tilde{I} + \sum^{m^2}_{i=2} \lambda_i e_{2,0,i} \ket{\lambda_i} \nonumber \\
&&+ C_{Z,1} \tilde{I} + \sum^{m^2}_{i=2} C_{Z,i} \ket{\lambda_i} \nonumber \\
\sum^{m^2}_{i=2} e_{2,0,i} \ket{\lambda_i} + e_{2,1,1}\tilde{I} &=& \sum^{m^2}_{i=2} \lambda_i e_{2,0,i} \ket{\lambda_i} + C_{Z,1} \tilde{I} \nonumber \\
&&+ \sum^{m^2}_{i=2} C_{Z,i} \ket{\lambda_i} .
\label{appendix_eqn:E2_6}
\end{eqnarray}
The parts that are parallel to the identity in Eq.~\ref{appendix_eqn:E2_6} are
\begin{eqnarray}
e_{2,1,1}\tilde{I} &=& C_{Z,1} \tilde{I} .
\label{appendix_eqn:E2_7}
\end{eqnarray}
Multiplying both sides by of Eq.~\ref{appendix_eqn:E2_7} by $\rho_R$ and taking the trace gives
\begin{eqnarray}
\text{Tr} \left( e_{2,1,1}\tilde{I}\rho_R \right) &=& \text{Tr} \left( C_{Z,1} \tilde{I} \rho_R \right) \nonumber \\
e_{2,1,1} &=& C_{Z,1} ,
\label{appendix_eqn:E2_8}
\end{eqnarray}
where $\tilde{I} \rho_R = 1$ was used. The terms that are perpendicular to the identity in Eq.~\ref{appendix_eqn:E2_6} will be used in the later part for calculating $E_1(L)$ where the value of $e_{2,0,i}$ will be needed. Since there is no explicit way of determining $e_{2,0,i}$, it has to be solved with a numerical solver. This is done be rewriting the parts of Eq.~\ref{appendix_eqn:E2_6} that are perpendicular to the identity, i.e.
\begin{eqnarray}
\sum^{m^2}_{i=2} \left(1 - \lambda_i \right) e_{2,0,i} \ket{\lambda_i} = \sum^{m^2}_{i=2} C_{Z,i}\ket{\lambda_i} ,
\label{appendix_eqn:E2_9}
\end{eqnarray}
as a set of linear equations
\begin{eqnarray}
\left(1 - \lambda_i \right) e_{2,0,i} = C_{Z,i} .
\label{appendix_eqn:E2_10}
\end{eqnarray}
Eq.~\ref{appendix_eqn:E2_10} can now be solved for $e_{2,0,i}$ with a linear solver such as the generalized minimal residual solver (GMRES).

Eq.~\ref{appendix_eqn:E3} is similar to that of Eq.~\ref{appendix_eqn:E2}. Applying the same procedure gives 
\begin{eqnarray}
e_{3,1,1} = C_{Z,1} ,
\label{appendix_eqn:E3_2}
\end{eqnarray}
and
\begin{eqnarray}
\left(1 - \lambda_i \right) e_{3,0,i} &= C_{Z,i} .
\label{appendix_eqn:E3_3}
\end{eqnarray}
This implies
\begin{eqnarray}
e_{3,1,i} = e_{2,1,i} \,\,\, \forall i .
\label{appendix_eqn:E3_4}
\end{eqnarray}

One can now look for the ansatz for the final term $E_4(L)$ in Eq.~\ref{appendix_eqn:E4} in the asymptotic large-$L$ limit. To do so, the similar reasoning to obtain the ansatz of $E_2(L)$ via Eq.~\ref{eqn:E2_geo_series} leading to Eq.~\ref{eqn:poly_expansion_E_2} is applied here to obtain the ansatz for $E_4(L)$. In this case, $E_4(L)$ is the form of a geometric series, and the large $L$ limit depends on the nature of the spectral decomposition of $C_Z$ and $C_{Z^2}$. Since $T_I$ only has a single eigenvalue equal to 1 and all other eigenvalues are strictly less than 1, $E_4(L)$ can diverge at most quadratically with $L$, hence $E_4(L)$ has polynomial order $p = 2$ and its polynomial expansion is
\begin{eqnarray}
E_4(L) = E_{4,0} + E_{4,1} L + E_{4,2} L^2 .
\label{appendix_eqn:poly_expansion_E4}
\end{eqnarray}
Inserting this, together with $E_1(L)$, $E_2(L)$ and $E_3(L)$, into Eq.~\ref{appendix_eqn:E4} gives
\begin{eqnarray}
E_{4,0} &+& E_{4,1} \times (L+1) + E_{4,2} \times (L+1)^2 \nonumber \\
&=& T_I(E_{4,0} + E_{4,1} L + E_{4,2} L^2) \nonumber \\
&&+ T_Z(E_{3,0} + E_{3,1} L) \nonumber \\
&&+ T_Z(E_{2,0} + E_{2,1} L) \nonumber \\
&&+ T_{Z^2}(E_{1,0}) \nonumber \\
E_{4,0} &+& E_{4,1} + E_{4,2} + (E_{4,1} + 2E_{4,2})L + E_{4,2}L^2\nonumber \\
&=& T_I(E_{4,0}) + T_Z(E_{3,0}) + T_Z(E_{2,0}) + T_{Z^2}(E_{1,0}) \nonumber \\
&&+ (T_I(E_{4,1}) + T_Z(E_{3,1}) + T_Z(E_{2,1}))L \nonumber \\
&&+ T_I(E_{4,2})L^2 .
\label{appendix_eqn:E4_2}
\end{eqnarray}
Equating powers of $L$,
\begin{eqnarray}
L^0 : E_{4,0} + E_{4,1} + E_{4,2} &=& T_I(E_{4,0}) + T_Z(E_{3,0}) \nonumber \\
&&+ T_Z(E_{2,0}) + T_{Z^2}(E_{1,0}) \nonumber \\
&=& T_I(E_{4,0}) + T_Z(E_{3,0}) \nonumber \\
&&+ T_Z(E_{2,0}) + C_{Z^2} \label{appendix_eqn:E4_3} \\
L^1 : E_{4,1} + 2E_{4,2} &=& T_I(E_{4,1}) + T_Z(E_{3,1}) \nonumber \\
&&+ T_Z(E_{2,1}) \label{appendix_eqn:E4_4} \\
L^2 : E_{4,2} &=& T_I(E_{4,2}) \label{appendix_eqn:E4_5} ,
\end{eqnarray}
where the last term on the right hand side of Eq.~\ref{appendix_eqn:E4_3} was $T_{Z^2}(E_{1,0}) = T_{Z^2}(\tilde{I}) \equiv C_{Z^2}$. Analogous to $C_Z$, $C_{Z^2}$ is a constant matrix, i.e. it has no dependence on any $E_i$'s, and thus its value can be calculated beforehand. Eq.~\ref{appendix_eqn:E4_5} implies $E_{4,2} \propto \tilde{I} = e_{4,2,1}\tilde{I}$, where $e_{4,2,1}$ is a proportionality constant. Substituting this into Eqs.~\ref{appendix_eqn:E4_4} gives
\begin{eqnarray}
E_{4,1} + 2e_{4,2,1}\tilde{I} &=& T_I(E_{4,1}) + T_Z(E_{3,1}) \nonumber \\
&&+ T_Z(E_{2,1}) \label{appendix_eqn:E4_4_2} .
\end{eqnarray}
By decomposing $T_I = \sum^{m^2}_{i=1} \lambda_i \ket{\lambda_i}\bra{\lambda_i}$, $E_{4,1} = \sum^{m^2}_{i=1} e_{4,1,i} \ket{\lambda_i}$, $T_Z = \sum^{m^2}_{i=1} C_{Z,i} \ket{\lambda_i}\bra{\lambda_i}$, $E_{3,1} = \sum^{m^2}_{i=1} e_{3,1,i} \ket{\lambda_i}$ and $E_{2,1} = \sum^{m^2}_{i=1} e_{2,1,i} \ket{\lambda_i}$, Eq.~\ref{appendix_eqn:E4_4_2} becomes
\begin{eqnarray}
\sum^{m^2}_{i=1} e_{4,1,i} \ket{\lambda_i} &+& 2e_{4,2,1}\tilde{I} \nonumber \\
&=& \sum^{m^2}_{i=1} \lambda_i \ket{\lambda_i}\bra{\lambda_i} \left( \sum^{m^2}_{i'=1} e_{4,1,i'} \ket{\lambda_{i'}} \right) \nonumber \\
&&+ \sum^{m^2}_{i=1} C_{Z,i} \ket{\lambda_i}\bra{\lambda_i} \left( \sum^{m^2}_{i'=1} e_{3,1,i'} \ket{\lambda_{i'}} \right) \nonumber \\
&&+ \sum^{m^2}_{i=1} C_{Z,i} \ket{\lambda_i}\bra{\lambda_i} \left( \sum^{m^2}_{i'=1} e_{2,1,i'} \ket{\lambda_{i'}} \right) .\nonumber \\
\label{appendix_eqn:E4_4_3}
\end{eqnarray}
Using Eq.~\ref{appendix_eqn:E3_4}, Eq.~\ref{appendix_eqn:E4_4_3} becomes
\begin{eqnarray}
\sum^{m^2}_{i=1} e_{4,1,i} \ket{\lambda_i} &+& 2e_{4,2,1}\tilde{I} \nonumber \\
&=& \sum^{m^2}_{i=1} \lambda_i \ket{\lambda_i}\bra{\lambda_i} \left( \sum^{m^2}_{i'=1} e_{4,1,i'} \ket{\lambda_{i'}} \right) \nonumber \\
&&+ 2 \sum^{m^2}_{i=1} C_{Z,i} \ket{\lambda_i}\bra{\lambda_i} \left( \sum^{m^2}_{i'=1} e_{3,1,i'} \ket{\lambda_{i'}} \right) \nonumber \\
&=& \sum^{m^2}_{ii'=1} \lambda_i e_{4,1,i'} \ket{\lambda_i} \left\langle \lambda_i | \lambda_{i'} \right\rangle \nonumber \\
&&+ 2 \sum^{m^2}_{ii'=1} C_{Z,i} e_{3,1,i'} \ket{\lambda_i} \left\langle \lambda_i | \lambda_{i'} \right\rangle \nonumber \\
&=& \sum^{m^2}_{i=1} \lambda_i e_{4,1,i} \ket{\lambda_i} + 2 \sum^{m^2}_{i=1} C_{Z,i} e_{3,1,i} \ket{\lambda_i} , \nonumber \\
\label{appendix_eqn:E4_4_4}
\end{eqnarray}
where the orthogonality relation $\sum_{i'} \left\langle \lambda_i | \lambda_{i'} \right\rangle = \delta_{ii'}$ was used in the last step. Further decomposing Eq.~\ref{appendix_eqn:E4_4_4} into a term parallel to the identity and terms perpendicular to the identity,
\begin{eqnarray}
e_{4,1,1}\tilde{I} + \sum^{m^2}_{i=2} e_{4,1,i} \ket{\lambda_i} &+& 2e_{4,2,1}\tilde{I} \nonumber \\
&=& e_{4,1,1}\tilde{I} + \sum^{m^2}_{i=2} \lambda_i e_{4,1,i} \ket{\lambda_i} \nonumber \\
&&+ 2 C_{Z,1} e_{3,1,1} \tilde{I} \nonumber \\
&&+ 2 \sum^{m^2}_{i=2} C_{Z,i} e_{3,1,i} \ket{\lambda_i} \nonumber \\
\sum^{m^2}_{i=2} e_{4,1,i} \ket{\lambda_i} + 2e_{4,2,1}\tilde{I} &=& \sum^{m^2}_{i=2} \lambda_i e_{4,1,i} \ket{\lambda_i} + 2 C_{Z,1} e_{3,1,1}\tilde{I} \nonumber \\
&&+ 2 \sum^{m^2}_{i=2} C_{Z,i} e_{3,1,i} \ket{\lambda_i} .
\label{appendix_eqn:E4_4_5}
\end{eqnarray}
The terms parallel to the identity in Eq.~\ref{appendix_eqn:E4_4_5} are
\begin{eqnarray}
e_{4,2,1}\tilde{I} &=& C_{Z,1} e_{3,1,1} \tilde{I} .
\label{appendix_eqn:E4_4_6}
\end{eqnarray}
Multiplying both side of Eq.~\ref{appendix_eqn:E4_4_6} by $\rho _R$ and taking the trace gives
\begin{eqnarray}
\text{Tr} \left( e_{4,2,1}\tilde{I} \rho_R \right) &=& \text{Tr} \left( C_{Z,1} e_{3,1,1} \tilde{I} \rho_R \right) \nonumber \\
e_{4,2,1} &=& C_{Z,1} e_{3,1,1} . 
\label{appendix_eqn:E4_4_7}
\end{eqnarray}
Substituting in Eq.~\ref{appendix_eqn:E3_2} gives
\begin{eqnarray}
e_{4,2,1} &=& C^2_{Z,1} . 
\label{appendix_eqn:E4_4_8}
\end{eqnarray}
Eq.~\ref{appendix_eqn:E4_4_8} is the contribution to the expectation value per site coming from polynomial degree $p = 2$ i.e. the degree of $L^2$ with coefficient $E_{4,2}$, cf. the last term on the right hand side of Eq.~\ref{appendix_eqn:poly_expansion_E4}. Comparing this result with the right hand side of Eq.~\ref{eqn:second_cumulant_per_site} reveals that $e_{4,2,1}$ is $\kappa_1^2$. The terms perpendicular to the identity in Eq.~\ref{appendix_eqn:E4_4_5} are not needed for the subsequent calculation of the $L^0$ terms. Thus there is no need to determine them.

The same procedure to obtain Eq.~\ref{appendix_eqn:E4_4_8} is now applied to Eq.~\ref{appendix_eqn:E4_3} to obtain the contribution to the expectation value from the $L^0$ term. Using $E_{4,2} = e_{4,2,1}\tilde{I}$, Eq.~\ref{appendix_eqn:E4_3} becomes
\begin{eqnarray}
E_{4,0} + E_{4,1} + e_{4,2,1}\tilde{I} &=& T_I(E_{4,0}) + T_Z(E_{3,0}) \nonumber \\
&&+ T_Z(E_{2,0}) + C_{Z^2} \label{appendix_eqn:E4_3_2} .
\end{eqnarray}
Using the eigenvalue decomposition above in addition to $C_{Z^2} \equiv T_{Z^2}(\tilde{I}) = \sum_{i=1}^{m^2} C_{Z^2,i} \ket{\lambda_i}$ where $C_{Z^2,i}$ are the elements of the constant matrix $C_{Z^2}$, Eq.~\ref{appendix_eqn:E4_3_2} becomes
\begin{eqnarray}
\sum^{m^2}_{i=1}e_{4,0,i}\ket{\lambda_i} &+& \sum^{m^2}_{i=1}e_{4,1,i}\ket{\lambda_i} + e_{4,2,1}\tilde{I} \nonumber \\
&=& \sum^{m^2}_{ii'=1} \lambda_i e_{4,0,i'} \ket{\lambda_i}\left\langle \lambda_i | \lambda_{i'} \right\rangle \nonumber \\
&&+ \sum^{m^2}_{ii'=1} C_{Z,i} e_{3,0,i'} \ket{\lambda_i}\left\langle \lambda_i | \lambda_{i'} \right\rangle \nonumber \\
&&+ \sum^{m^2}_{ii'=1} C_{Z,i} e_{2,0,i'} \ket{\lambda_i}\left\langle \lambda_i | \lambda_{i'} \right\rangle \nonumber \\
&&+ \sum_{i=1}^{m^2} C_{Z^2,i} \ket{\lambda_i} \nonumber \\
&=& \sum^{m^2}_{i=1} \lambda_i e_{4,0,i} \ket{\lambda_i} + \sum^{m^2}_{i=1} C_{Z,i} e_{3,0,i} \ket{\lambda_i} \nonumber \\
&&+ \sum^{m^2}_{i=1} C_{Z,i} e_{2,0,i} \ket{\lambda_i} + \sum_{i=1}^{m^2} C_{Z^2,i} \ket{\lambda_i} . \nonumber \\
\label{appendix_eqn:E4_3_3}
\end{eqnarray}
Using Eq.~\ref{appendix_eqn:E3_4}, Eq.~\ref{appendix_eqn:E4_3_3} becomes
\begin{eqnarray}
\sum^{m^2}_{i=1}e_{4,0,i}\ket{\lambda_i} &+& \sum^{m^2}_{i=1}e_{4,1,i}\ket{\lambda_i} + e_{4,2,1}\tilde{I} \nonumber \\
&=& \sum^{m^2}_{i=1} \lambda_i e_{4,0,i} \ket{\lambda_i} + 2 \sum^{m^2}_{i=1} C_{Z,i} e_{3,0,i} \ket{\lambda_i} \nonumber \\
&&+ \sum_{i=1}^{m^2} C_{Z^2,i} \ket{\lambda_i} .
\end{eqnarray}
Separating the term parallel to the identity from terms perpendicular to it,
\begin{eqnarray}
e_{4,0,1}\tilde{I} &+& \sum^{m^2}_{i=2}e_{4,0,i}\ket{\lambda_i} + e_{4,1,1}\tilde{I} \nonumber \\
&&+ \sum^{m^2}_{i=2}e_{4,1,i}\ket{\lambda_i} + e_{4,2,1}\tilde{I} \nonumber \\
&=& e_{4,0,1}\tilde{I} + \sum^{m^2}_{i=2} \lambda_i e_{4,0,i} \ket{\lambda_i} \nonumber \\
&&+ 2 C_{Z,1} e_{3,0,1} \tilde{I} + 2 \sum^{m^2}_{i=2} C_{Z,i} e_{3,0,i} \ket{\lambda_i} \nonumber \\
&&+ C_{Z^2,1} \tilde{I} + \sum_{i=2}^{m^2} C_{Z^2,i} \ket{\lambda_i} \nonumber \\
\sum^{m^2}_{i=2}e_{4,0,i}\ket{\lambda_i} &+& e_{4,1,1}\tilde{I} + \sum^{m^2}_{i=2}e_{4,1,i}\ket{\lambda_i} + e_{4,2,1}\tilde{I} \nonumber \\
&=& \sum^{m^2}_{i=2} \lambda_i e_{4,0,i} \ket{\lambda_i} + 2 C_{Z,1} e_{3,0,1} \tilde{I} \nonumber \\
&&+ 2 \sum^{m^2}_{i=2} C_{Z,i} e_{3,0,i} \ket{\lambda_i} + C_{Z^2,1} \tilde{I} \nonumber \\
&&+ \sum_{i=2}^{m^2} C_{Z^2,i} \ket{\lambda_i} .
\label{appendix_eqn:E4_3_4}
\end{eqnarray}
Multiplying both sides of Eq.~\ref{appendix_eqn:E4_3_4} by $\rho_R$ and taking the trace,
\begin{eqnarray}
&& \text{Tr} \left( \sum^{m^2}_{i=2}e_{4,0,i}\ket{\lambda_i} \rho_R \right) + \text{Tr} \left( e_{4,1,1}\tilde{I} \rho_R \right) \nonumber \\ 
&&+ \text{Tr} \left( \sum^{m^2}_{i=2}e_{4,1,i}\ket{\lambda_i} \rho_R \right) + \text{Tr} \left( e_{4,2,1}\tilde{I} \rho_R \right) \nonumber \\
&=& \text{Tr} \left( \sum^{m^2}_{i=2} \lambda_i e_{4,0,i} \ket{\lambda_i} \rho_R \right) + 2 \text{Tr} \left( C_{Z,1} e_{3,0,1} \tilde{I} \rho_R \right) \nonumber \\
&&+ 2 \text{Tr} \left( \sum^{m^2}_{i=2} C_{Z,i} e_{3,0,i} \ket{\lambda_i} \rho_R \right) + \text{Tr} \left( C_{Z^2,1} \tilde{I} \rho_R \right) \nonumber \\
&&+ \text{Tr} \left( \sum_{i=2}^{m^2} C_{Z^2,i} \ket{\lambda_i} \rho_R \right) .
\label{appendix_eqn:E4_3_5} 
\end{eqnarray}
Using $\ket{\lambda_i}\rho_R = 0$ for the terms perpendicular to the identity and $\tilde{I}\rho_R = 1$ for the terms parallel to the identity, Eq.~\ref{appendix_eqn:E4_3_5} becomes
\begin{eqnarray}
e_{4,1,1} + e_{4,2,1} &=& 2 C_{Z,1} e_{3,0,1} + C_{Z^2,1} .
\label{appendix_eqn:E4_3_6}
\end{eqnarray}
Substituting in Eq.~\ref{appendix_eqn:E4_4_8} gives
\begin{eqnarray}
e_{4,1,1} &=& -C_{Z,1}^2 + 2 C_{Z,1} e_{3,0,1} + C_{Z^2,1} .
\label{appendix_eqn:E4_3_7}
\end{eqnarray}
The elements $e_{3,0,1}$ are to be obtained from the linear solver in Eq.~\ref{appendix_eqn:E3_3}. Eq.~\ref{appendix_eqn:E4_3_7} is the contribution coming from polynomial degree $p = 1$ i.e. the degree of $L^1$ with coefficient $E_{4,1}$, cf. Eq.~\ref{appendix_eqn:poly_expansion_E4}. Comparing this result with the right hand side of Eq.~\ref{eqn:second_cumulant_per_site} reveals that $e_{4,1,1}$ is $\kappa_2$.


\subsection{Inset figures of the Binder cumulant}
\label{appendix_inset_fig_binder_cumulant}
\begin{figure}[h!]
  \centering\includegraphics[width=0.45\textwidth]{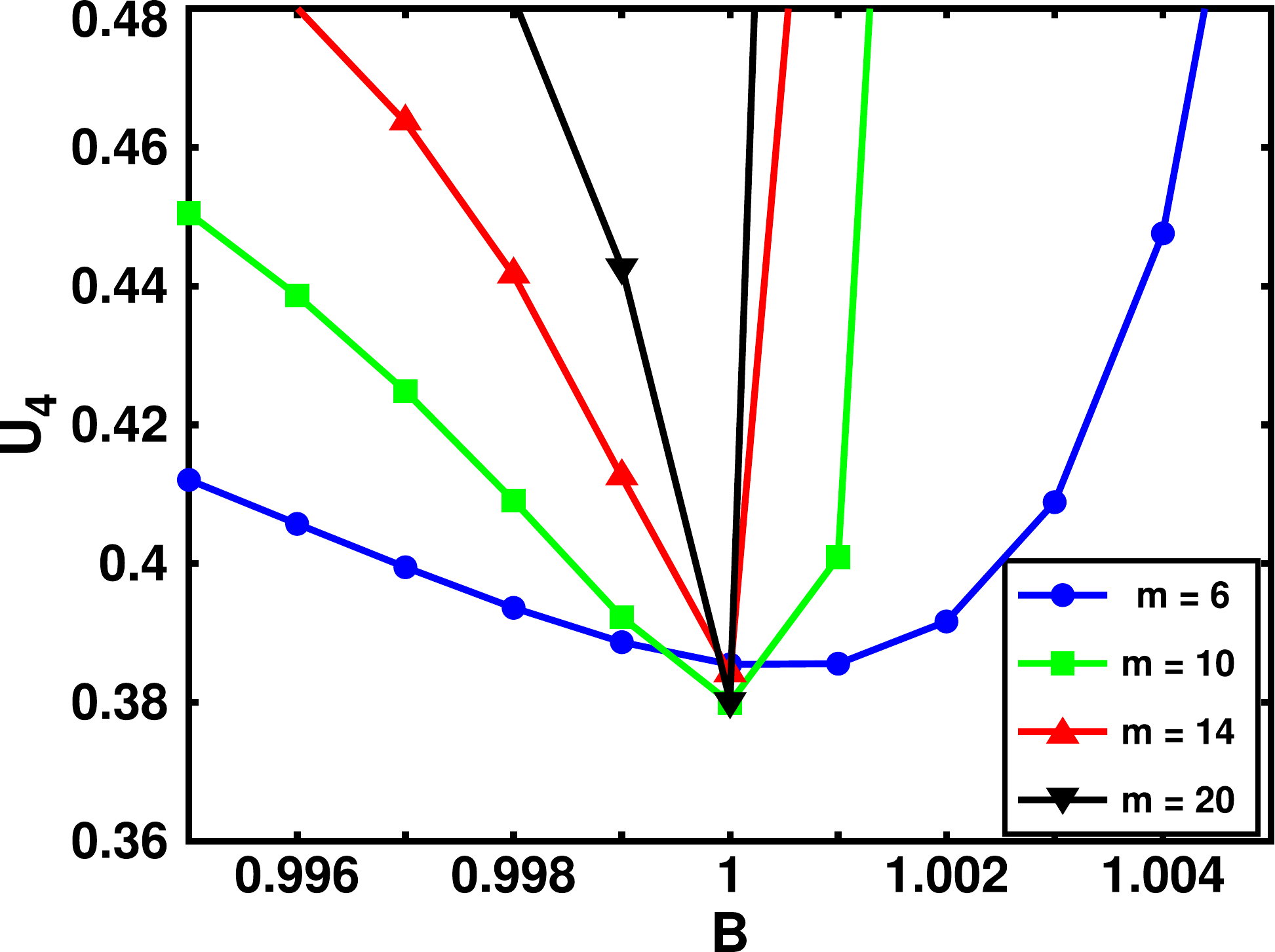}
  \caption{(Colour online) 1D TFI model. Larger version of the inset in the top figure of fig.~\ref{fig:binder_cumulant_s2_s5_ising1d} corresponding $s = 2$. An imperfect intersection at $B = 1$ is formed between $U_4$ of different values of $m$, i.e. $\frac{\partial U_4(m,B)}{\partial m}$ is a small but nonzero value.}
  \label{fig:binder_cumulant_s2_s5_3075_inset_ising1d}
\end{figure}

This section displays the larger version of the three inset figures corresponding to the Binder cumulant $U_4$ of the 1D TFI model and the 1D TKI for the purpose of clearly seeing the imperfect intersection of $U_4$ between the different values of $m$. Fig.~\ref{fig:binder_cumulant_s2_s5_3075_inset_ising1d} corresponds to the inset in the top figure of Fig.~\ref{fig:binder_cumulant_s2_s5_ising1d} corresponding to the 1D TFI model when $s = 2 \neq s^*$. At $B = 1$, an imperfect intersection of $U_4$ between the different values of $m$ is apparent.

\begin{figure}[h!]
  \centering\includegraphics[width=0.45\textwidth]{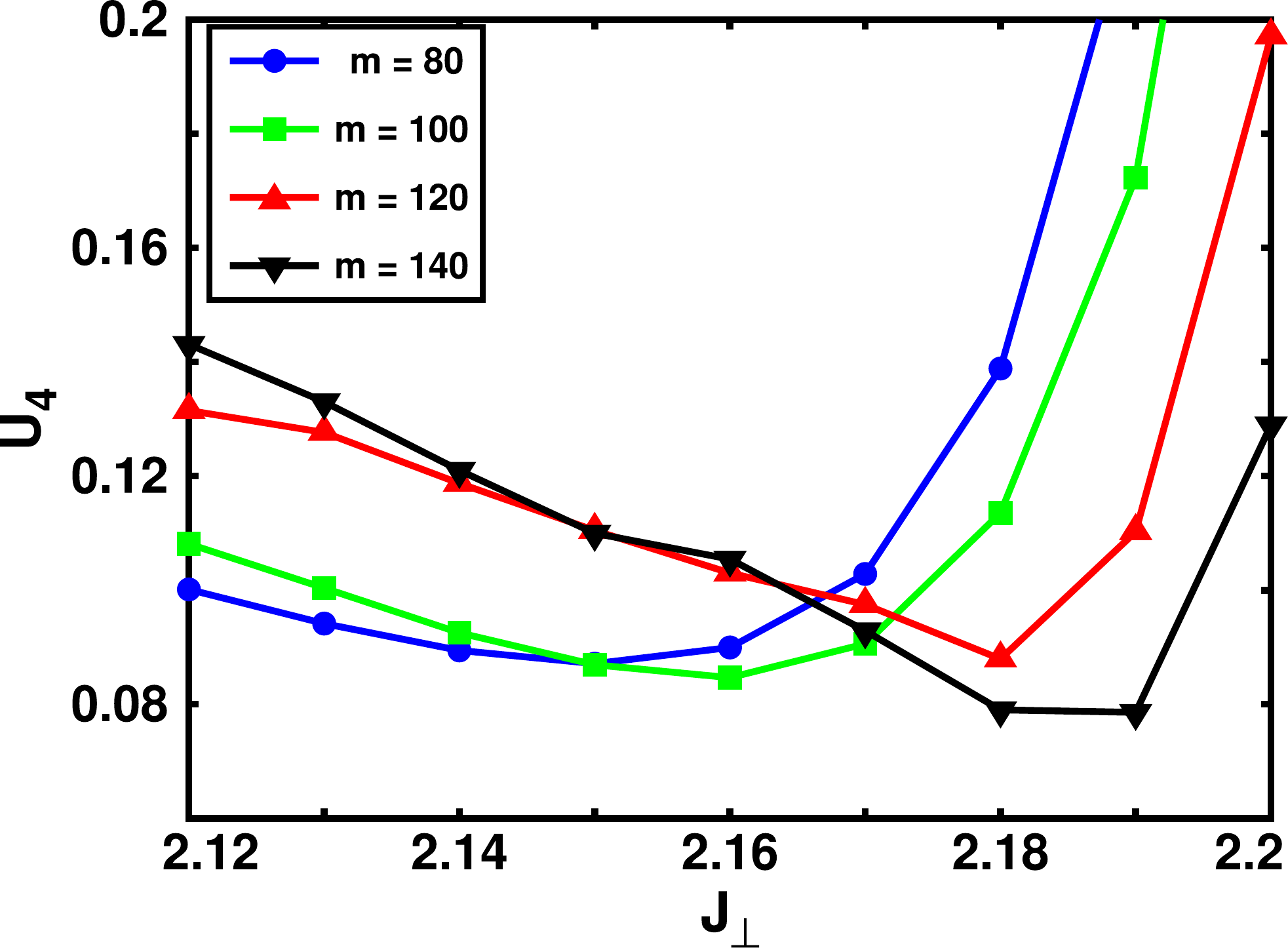}
  \caption{(Colour online) 1D TKI. Larger version of the inset in the top figure of fig.~\ref{fig:binder_cumulants_s2_s5_tki} corresponding $s = 2$. Not all $U_4(m,J_\perp)$ of the different values of $m$ cross simultaneously, thus it is a spurious crossing point.}
  \label{fig:binder_cumulant_s2_s5_2896_a_inset_tki}
\end{figure}

\begin{figure}[h!]
  \centering\includegraphics[width=0.45\textwidth]{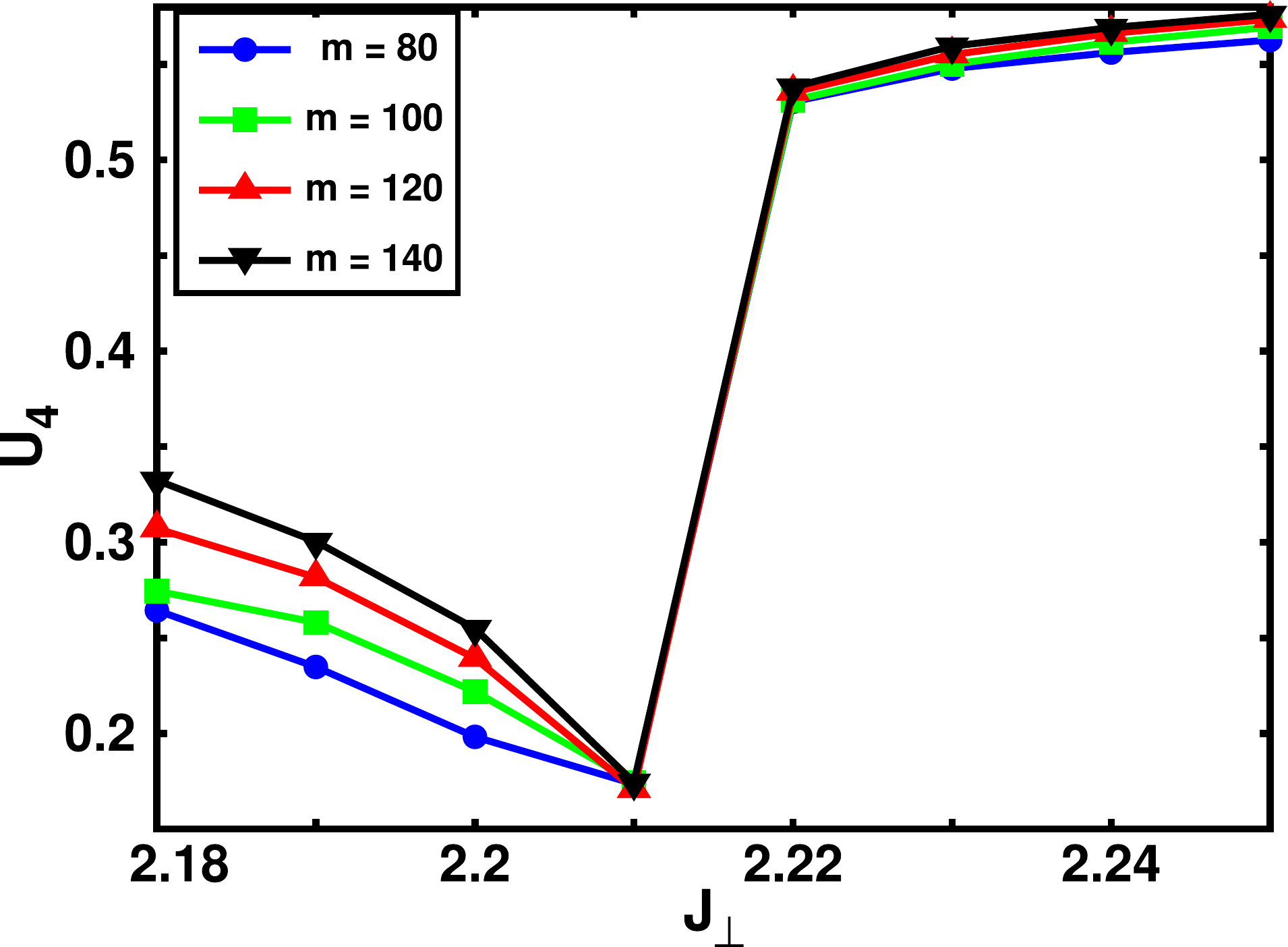}
  \caption{(Colour online) 1D TKI. Larger version of the inset in the bottom figure of fig.~\ref{fig:binder_cumulants_s2_s5_tki} corresponding $s = s^* = 5.29$. Closeup of $J_\perp = 2.22$ reveals that $U_4$ does not intersect there.}
  \label{fig:binder_cumulant_s2_s5_2896_c_inset_tki}
\end{figure}

Figs.~\ref{fig:binder_cumulant_s2_s5_2896_a_inset_tki} and \ref{fig:binder_cumulant_s2_s5_2896_c_inset_tki} correspond to the inset in the top and bottoms figures of Fig.~\ref{fig:binder_cumulants_s2_s5_tki} respectively for the 1D TKI. When $s = 2 \neq s^*$, multiple spurious crossing points occur in fig.~\ref{fig:binder_cumulant_s2_s5_2896_a_inset_tki}. When $s = s^* = 5.29$, the point at $J_\perp = 2.22$ appears to be a crossing point, however upon closer inspection as shown in fig.~\ref{fig:binder_cumulant_s2_s5_2896_c_inset_tki}, it is not - there is only one crossing point at $J_\perp = 2.21$ as shown in the bottom figure of fig.~\ref{fig:binder_cumulants_s2_s5_tki}.



\end{document}